\documentclass[journal,onecolumn,font=12]{IEEEtran}
 
\usepackage[cmex10]{amsmath}

\usepackage{amssymb}
\usepackage{color}
\usepackage{bbm}
\usepackage{mathrsfs}

\usepackage{graphicx} 
\usepackage{hyperref} 

\allowdisplaybreaks[2]

\usepackage{array} 

\usepackage{cite}
\usepackage{url}
\usepackage{epstopdf,tikz}

\newtheorem{theorem}{Theorem}
\newtheorem{definition}{Definition}
\newtheorem{lemma}{Lemma}
\newtheorem{proposition}{Proposition}
\newtheorem{corollary}{Corollary}
\newtheorem{remark}{Remark}

\newcommand{\TV}{\text{TV}}

\newcommand{\sadaf}[1]{{\color{black}#1}}

\newcommand{\T}{\text{T}}
\newcommand{\Lo}{\text{L}}
\newcommand{\Hi}{\text{H}}
\newcommand{\D}{\text{D}}

\newcommand{\V}{\text{V}}
\newcommand{\w}{\text{w}}

\def\bx{{x}^n}

\DeclareMathOperator*{\argmax}{argmax}

\begin{document}

%
% paper title
% can use linebreaks \\ within to get better formatting as desired
\title{State Masking Over a Two-State Compound Channel}

% author names and affiliations
% use a multiple column layout for up to three different
% affiliations
%\author{\IEEEauthorblockN{Sadaf Salehkalaibar\\ Department of Electrical and Computer Engineering, \\ University of %Tehran,\\ e-mail: s.saleh@ut.ac.ir}}
%\author{\IEEEauthorblockN{author1}}

\author{Sadaf Salehkalaibar, Mohammad Hossein Yassaee, Vincent Y. F. Tan, Mehrasa Ahmadipour
	\thanks{Sadaf  Salehkalaibar is  with the Department of Electrical and Computer Engineering, College of Engineering, University of Tehran, Tehran, Iran (e-mail:  {s.saleh@ut.ac.ir})}
	\thanks{Mohammad Hossein Yassaee is with  Institute for Research in Fundamental Sciences (IPM), Tehran, Iran  (e-mail: yassaee@ipm.ir).}
	\thanks{Vincent Y. F. Tan is  with Department of Electrical and Computer Engineering and the Department of Mathematics, National University of Singapore, 
		(e-mail:  {vtan@nus.edu.sg}).}
	\thanks{Mehrasa Ahmadipour is with LTCI, Telecom ParisTech, Universit\'e Paris-Saclay, Paris, France, previously, was with the Department of Electrical and Computer Engineering, College of Engineering, University of Tehran, Tehran, Iran,
		(e-mail:  {meh.ahmadipour@gmail.com}).}
	\thanks{Parts of the material in this paper was presented at the \emph{IEEE International Symposium on Information Theory (ISIT), Paris, France, 2019.}}
}

% make the title area
\maketitle

\begin{abstract}
\sadaf{We consider  fundamental limits for communicating over a compound channel when the state of the channel needs to be masked.  
 Our model is closely related to an area of study known as covert communication that is a   setting in which the    transmitter wishes to communicate to  legitimate receiver(s) while ensuring that the communication is not detected by an adversary. 
The main contribution in our two-state masking setup is the establishment of bounds on the throughput-key region when the constraint that quantifies how much the  states are masked is defined to be the total variation distance between the two channel-induced distributions.   For the scenario in which the key length is infinite, we provide  sufficient conditions for when the bounds to coincide for the throughput, which follows the square-root law. Numerical examples, including that of a Gaussian channel, are provided to illustrate our results. }
%Covert communication in the information-theoretic context has been primarily concerned with fundamental limits when the transmitter wishes to communicate to  legitimate receiver(s) while ensuring that the communication is not detected by an adversary. This paper, however, considers an alternative, and no less important, setting in which the object to be masked is the state of the compound channel. Such a communication model has applications in the prevention of malicious parties seeking to jam the communication signal when, for example, the signal-to-noise ratio of a wireless channel is found to be low. Our main contribution is the  establishment of  bounds on the throughput-key region when the covertness constraint is defined in terms of the total variation distance. In addition, for the scenario in which the key length is infinite, we provide a sufficient condition for when the bounds coincide for the scaling of the throughput, which follows the square-root law. Numerical examples, including that of a Gaussian channel, are provided to illustrate our results. 
\end{abstract}


\IEEEpeerreviewmaketitle

\section{Introduction}

%\begin{figure}[t]
%	\centering
%	\includegraphics[scale=0.3]{covert_model.eps}
%	\caption{Covert Communication over a Compound Channel.}
%	\label{figure0}
%\end{figure}  
\sadaf{This paper establishes fundamental limits for communicating over compound channels when the state of the channel needs to be masked from an adversary. Our work lies at the intersection of two major areas of study in the Shannon theory---state masking and covert communication. The goal of {\em state masking} is to reliably send a message from  a transmitter to a receiver in the presence of a masking constraint on the state of the channel. In this setting, an adversary who is observing the channel output should not be able to  reliably infer the channel's state. Fundamental limits for  state masking problem has been considered in several previous works \cite{Neri, Ligong-state, Young, Ulukus, Tian, Vishwanath, Kim2, Mael}.    In {\em covert communication}, the goal is to hide   the very act message transmission from potentially malicious parties.   Covert communication has been extensively studied in the information theory community in recent years~\cite{Bash,Ligong,Bloch}. Here, the transmitter feeds a specific symbol, called the {\em off-symbol}, to the channel when it decides not to send a message, signifying that the transmitter is in the \emph{off-transmission state}.  When the transmitter decides to send a {\em bona fide} message, the channel, together with the codeword associated to the message, induce a certain output distribution that is different from that when a message is not sent. One desires to design codes such that distinguishing between these two output distributions is difficult in the sense that the failure probability is high. In both the covert communication and state masking problems, observations from the output of the channel should not leak information about the \emph{state}. In the former, the state refers to the transmitter's off-transmission state while in the latter, it refers to the channel's state. We review related works on state masking and covet communication in Section~\ref{sec:related}.}

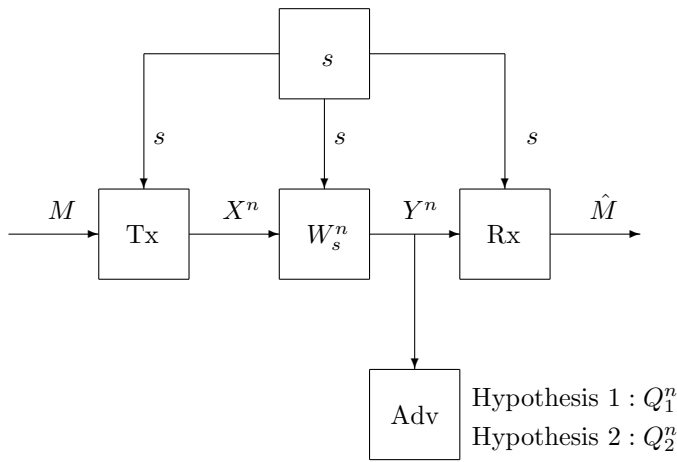
\begin{figure}[t]
	\centering
\setlength{\unitlength}{.4mm}
\begin{picture}(200, 150)
%\thicklines
\put(0, 75){\vector(1, 0){30}}
\put(60, 75){\vector(1,0){30}}
\put(120, 75){\vector(1,0){30}}
\put(180, 75){\vector(1,0){30}}
\put(30, 60){\line(1, 0){30}}
\put(30, 60){\line(0,1){30}}
\put(60, 60){\line(0,1){30}}
\put(30, 90){\line(1,0){30}}

\put(90, 60){\line(1, 0){30}}
\put(90, 60){\line(0,1){30}}
\put(120,60){\line(0,1){30}}
\put(90, 90){\line(1,0){30}}

\put(10, 80){  $M$}
\put(68, 80){  $X^n$}
%\put(51, -10){  $\bbE[\rvg(X)]\le\Gamma$}
%\put(65, 8){  $[2^{nR}]$}
\put(128, 80){  $Y^n$} 
\put(39, 72){$\mathrm{Tx}$} 
\put(99, 72){$W_s^n$} 

\put(150, 60){\line(1, 0){30}}
\put(150, 60){\line(0,1){30}}
\put(180, 60){\line(0,1){30}}
\put(150, 90){\line(1,0){30}}
\put(159, 72){$\mathrm{Rx}$} 
\put(190, 80){  $\hat{M}$} 
%\put(186, 1){  $\Pr(\hatM \ne M)$} 

\put(90, 120){\line(1, 0){30}}
\put(90, 120){\line(0,1){30}}
\put(120, 120){\line(0,1){30}}
\put(90, 150){\line(1,0){30}}
%\put(120, 75){\vector(1,0){30}}
%\put(165, 60){\vector(0,-1){30}}
\put(105, 120){\vector(0,-1){30}}
\put(120, 135){\line(1,0){45}}
\put(165,135){\vector(0,-1){45}}

\put(90, 135){\line(-1,0){45}}
\put(45, 135){\vector(0,-1){45}}

%\put(150, 60){\line(1, 0){30}}
%\put(150, 60){\line(0,1){30}}
%\put(180, 60){\line(0,1){30}}
%\put(150, 90){\line(1,0){30}}

\put(105, 105){  $s$} 
\put(169, 105){  $s$} 
\put(45, 105){  $s$} 
%\put(165, 45){  $M_{\mathrm{d}}$} 
%\put(128, 81){  $S$} 
\put(104, 131){$s$} 
%\put(157, 76){State} 
%\put(157, 67){Enc} 

\put(120, 0){\line(1, 0){30}}
\put(120, 0){\line(0,1){30}}
\put(150, 0){\line(0,1){30}}
\put(120, 30){\line(1,0){30}}

\put(135, 75){\vector(0, -1){45}}
\put(127, 12){$\mathrm{Adv}$} 

\put(154, 18){$\mathrm{Hypothesis}\,\, 1: Q_1^n$} 
\put(154, 05){$\mathrm{Hypothesis}\,\, 2: Q_2^n$} 
  \end{picture}
\caption{State Masking Over a Two-State Compound Channel}
\label{figure1b}
\end{figure}

More specifically, in this work, we consider state masking over a compound channel with two states $s\in\{1,2\}$ (see Fig.~\ref{figure1b}). In a compound channel~\cite{blackwell, dobrushin, wolfowitz}, the channel law $W_s$ depends on a certain state $s$ which remains {\em fixed} throughout the transmission (see~\cite[Section~7.2]{ElGamal}). We assume that the channel state is known at the transmitter and the receiver. \sadaf{Instead of focusing on the covert communication setting in which covertness is measured in terms of the  ability of an observer at the receiver being able to distinguish between the absence or presence of the transmission of a message, here we consider imposing a {\em state masking constraint} on communication.  The masking constraint quantifies the ability of an adversary to distinguish between the two channel states by observing the channel output over an $n$-length transmission block  denoted by $Y^n$.} More precisely, the masking constraint places a bound on  the probability of success of an adversary in a binary hypothesis test that seeks to distinguish between the channel states $s\in \{1,2\}$.
%covertness being measured in terms of the ability of an observer at the receiver being able to distinguish between the absence or presence of message transmission, here we quantify covertness as the ability of an adversary in distinguishing between the two channel states. That is, we quantify covertness in terms of the probability of success of an adversary in a binary hypothesis test that seeks to distinguish between $s\in \{1,2\}$. 
 We wish to design coding schemes and establish impossibility results such that this  probability of success  is upper bounded by a small constant,  meaning that the adversary's test is unreliable. Our main contribution is in the establishment of bounds that characterize the optimal tradeoff between the throughput of reliable communication  in the presence of the masking constraint 
%deniable (i.e., the adversary cannot learn whether communication is occurring or what the channel state is) 
%  and reliable communication 
  as well as the length of the shared key between the transmitter and receiver.

%, while we establish bounds on the throughput under the constraint that the adversary cannot learn the channel state.

% For example, consider a channel with an AWGN where at one of the states, the channel is \emph{more noisy} than that of the other state.  If the communication is not covert in the sense that the adversary finds out a channel state corresponds to the more noisy channel, then it can disrupt with the transmission by sending jamming signals with a small power consumption. Motivated by this scenario, we consider covert communication over a compound channel as the ability of the adversary in distinguishing the channel state. The compound channel without a covertness constraint has been previously studied in [],[]. Covert communication over compound binary symmetric channels was studied in \cite{Jaggi} but the setting is different from ours. Specifically, the best throughput to a deniable (i.e., the adversary cannot learn whether communication is occurring) and reliable communication was established, while we establish bounds on the throughput under the constraint that the adversary cannot learn the channel state.

\sadaf{To describe our model and techniques in more detail, we consider the following setup for the state masking problem over a two-state compound channel.  For a given blocklength $n$, the masking constraint is measured by the total variation distance between the two channel output distributions of states $ s=1$ or $s=2$, denoted by $Q_1^n$ and $Q_2^n$ respectively. We upper bound the total variation distance by a   constant $\delta \in (0,1)$. We assume that a shared key with bounded length is available between the transmitter and receiver. We provide inner and outer bounds on the throughput-key region and show that the square-root law, similarly to that in covert communication~\cite{Bash}, also holds in this setting. The bounds match for some special cases and thus, for these cases, we have completely characterized the optimal transmission throughput given a shared key with a certain length. The achievability scheme uses two codebooks where each of them is used for communicating given a  channel state and the maximum likelihood (ML) decoder  at the receiver. The proof of outer  bound  requires us to analyze the total variation distance between two {\em non-product} distributions $Q_1^n$ and $Q_2^n$. The single-letter characterization of the masking constraint requires relating the total variation distance to a binary hypothesis testing problem and using  various probability bounds  (e.g., the Berry-Esseen theorem) to control the false alarm and missed detection probabilities. This idea of relating the total variation distance to a binary hypothesis test for converse bounds was inspired by \cite{Mehrdad} where a similar technique was used in deriving converses for the second-order asymptotics in covert communication.}% A list of contributions is summarized in Section \ref{sec:contributions}. 	}

\subsection{Related Works}\label{sec:related}
In this section, we review papers in the general areas of state masking and covert communication and a more detailed discussion of how our setup differs from the  closely related works \cite{Jaggi,   Neri, Ligong-state}. 
%\begin{itemize}
%\item Related works in state masking
%\item Related works in covert communication
%\end{itemize}

\sadaf{For state masking,  a channel coding problem was studied in \cite{Neri, Ligong-state} where a minimum amount of information about the channel state should be learned from the channel output. The state masking problem is related to  state amplification \cite{Young} where the primary goal of a transmitter which has access to the channel state is to send a message while conveying some information about the state to the receiver. The problem of simultaneous message transmission and state amplification when the transmitter has an imperfect non-causal knowledge of the state has been investigated in \cite{Tian}. The work in \cite{Vishwanath} considered a state-dependent broadcast channel where the state should be conveyed to one of the receivers and  masked  from the other receiver. When the state information is available causally  at the transmitter, the optimal tradeoff between the rate and the state estimation error has been characterized in \cite{Kim2}. State amplification and state masking from the point of view of empirical coordination have been studied in \cite{Mael}. We emphasize that in our work, the state to be masked is that of the compound channel; this setup has not been studied in these previous works. }

There has been a flurry of research activity for the covert communication problem in recent years. Indeed, covert communication over a channel with an additive white Gaussian noise  \cite{Bash},  a discrete memoryless channel (DMC) \cite{Ligong,Bloch}, a classical-quantum channel \cite{Ligong2}, a channel with state \cite{Lee}, in the presence of an adversarial jammer \cite{Jaggi2},  and from the second-order perspective \cite{Mehrdad} have been considered. It is shown that the maximum amount of information that can be sent scales with the \emph{square-root} of the transmission blocklength; this is known as the {\em square-root law}. The extension to multi-user setups has been considered for multiple-access channels \cite{Bloch3}, broadcast channels \cite{Vincent} and relay channels \cite{Ligong4}.

\sadaf{
	In a   work related to the present paper~\cite{Jaggi}, covert communication over compound binary symmetric channels  was  studied but the setting is different from ours. Specifically, the optimal throughput to deniably  and reliably communicate was established, whereas our main contribution is to establish bounds on the throughput under the constraint that the adversary cannot learn the channel state.} 	\sadaf{
	The proposed setup is also different from the previous works on state masking, e.g., \cite{Neri, Ligong-state}, in the following aspects. First, in their setting, the channel state is an identically and independently distributed (i.i.d.) random variable over an $n$-length transmission block denoted by $S^n$. More precisely, the state masking constraint introduced in \cite{Neri} is given by:
	\begin{IEEEeqnarray}{rCl}
		\limsup_{n\to\infty} \frac{1}{n}I(S^n;Y^n)\leq L,
		\end{IEEEeqnarray}
	for some positive $L$.
	In a recent work \cite{Ligong-state}, a more stringent masking constraint has been imposed on the setup where it is required that
	\begin{IEEEeqnarray}{rCl}
		\lim_{n\to\infty} \frac{1}{n}I(S^n;Y^n)=0.
		\end{IEEEeqnarray}
The optimal rate of communication has been derived in \cite{Ligong-state} for different cases based on the knowledge of state at the transmitter. Among these cases, there is a DMC with a state that is random but constant throughout the course of transmission (over $n$ channel uses). Moreover, the state is generated randomly according to a known distribution $\mathsf{P}_S$.  The specific masking constraint of this setup is assumed to be as follows:
\begin{IEEEeqnarray}{rCl}
	\lim_{n\to\infty} I(S;Y^n)=0.\label{Ligong-eq}
	\end{IEEEeqnarray}
This model is related to our proposed setup in the sense that the state remains fixed during the block of transmission. However, the state distribution is not known in our compound channel. Thus, a state masking constraint in analogous to \eqref{Ligong-eq} for the compound setup involves a maximization over all possible state distributions accounting for the worst-case scenario. Adapting constraint~\eqref{Ligong-eq} for the proposed setup yields the following masking constraint:
\begin{IEEEeqnarray}{rCl}
	\lim_{n\to\infty}\max_{\mathsf{P}_S} I(S;Y^n)=0.\label{Ligong-compound}
\end{IEEEeqnarray}
The extreme case of our proposed masking constraint in which $Q_1^n=Q_2^n$ implies constraint~\eqref{Ligong-compound}. Considering the fact that designing two sets of codebooks that induce exactly the same channel output distribution is generally infeasible, we allow the total variation distance between $Q_1^n$ and $Q_2^n$ to be non-zero but smaller than  a small positive number $\delta$.
Finally, in the settings of \cite{Neri, Ligong-state}, the square-root law in the  zero-rate regime has not been investigated. The main focus of our work is to prove that the square-root law holds in the proposed setup using analytical tools from covert communication. }

%We conclude this section with a summary of the main contributions, introduction of notation, and the organization of the paper.

\subsection{Main Contributions} \label{sec:contributions}

The main contributions of this paper are as follows:
\sadaf{\begin{itemize}
	\item We consider state masking  over a compound channel and provide an inner bound on the the optimal transmission throughput-shared key region (Theorem~\ref{key-thm}). 
	\item If we assume that the key rate is sufficiently high and the error  probability of decoding the message at the legitimate receiver is vanishing, we are also able to  upper bound the optimal transmission throughput (Theorem~\ref{thm-upper}). It is shown that the upper and lower bounds match for some special cases (Theorem~\ref{opt-thm}). 
	\item Examples are provided to compare the lower and upper bounds. In particular, a Gaussian setup is also considered where we identify parameter regimes in which the conditions of the optimality result in Theorem~\ref{opt-thm} are satisfied and thus we have a complete characterization of the throughput (Corollary~\ref{Gaus-cor}). 
	\end{itemize}
}

\subsection{Notation}

We mostly use the notation of \cite{ElGamal}. Random variables are shown by capital letters, e.g., $X$, $Y$ and their realizations by lower-case letters, e.g., $x$, $y$. The alphabets of random variables are denoted by script symbols such as $\mathcal{X}$, $\mathcal{Y}$. The sequence of random variables $(X_i,\ldots,X_j)$ and its realizations $(x_i,\ldots,x_j)$ are abbreviated as  $X_i^j$ and $x_i^j$ respectively. We use $X^j$ and $x^j$ instead of $X_1^j$ and $x_1^j$. The empirical distribution of sequence $x^n$ is also known as its \emph{type}. 

The probability mass function (pmf) of a discrete random variable $X$ defined over the channel input alphabet $\mathcal{X}$ denoted using the letter $P$. The pmf of a discrete random variable $Y$ defined over the channel output alphabet $\mathcal{Y}$ is denoted using the letter $Q$. The distributions of the  sequences of random variables $X^n$ and $Y^n$ are denoted by $P^n$ and $Q^n$ respectively.    The notation $P^{\times n}$ denotes the $n$-fold product distribution, i.e., for every $x^n\in\mathcal{X}^n$, we have
\begin{align}
P^{\times n}(x^n)\triangleq\prod_{i=1}^n P(x_i).
\end{align}
The expectation and variance operators are written as $\mathbb{E}[\cdot]$ and $\mathbb{V}[\cdot]$, respectively. The notation $\mathbbm{1}\left\{ \cdot \right\}$ denotes the indicator function.  The probability of an event $\mathcal{E}$ is denoted by $\mathbb{P}(\mathcal{E})$. 

The total variation distance between two distributions $P_1$ and $P_2$ is denoted by $d_{\TV}(P_1,P_2)$ and is defined as $d_{\TV}(P_1,P_2)\triangleq \frac{1}{2}\sum_x|{P_1(x)-P_2(x)|} \in [0,1]$. The chi-squared distance between two distributions $P_1$ and $P_2$ is denoted by $\chi_2(P_1\| P_2)$ and is defined as $\chi_2(P_1\| P_2)\triangleq \mathbb{E}_{P_2}[(\frac{P_1(X)}{P_2(X)}-1)^2]$.
The KL-divergence (relative entropy) is denoted by $D(P_1\| P_2)$. The Bhattacharyya coefficient between two distributions $P_1$ and $P_2$ is denoted by $F(P_1,P_2)$ and is defined as $F(P_1,P_2)\triangleq \sum_x\sqrt{P_1(x)P_2(x)}$.

The Hamming weight of sequence $x^n$ is denoted as $\w(x^n)$. The binary entropy function is denoted as  $h_{\text{b}}(\cdot )$. The binary symmetric channel with crossover probability $\epsilon$ is denoted by $\text{BSC}(\epsilon)$. The Bernoulli distribution, i.e., one that is defined over $\{0,1\}$, is denoted by $\text{Bern}( \cdot )$. The cumulative distribution function of a standard normal distribution is denoted by $\Phi(\cdot )$. The notation $\log(\cdot )$ denotes the natural logarithm. The exponential function is written as $\exp(\cdot )$.

\subsection{Organization}
The remainder of the paper is organized as follows. Section~\ref{covert-def} describes the state masking setup over a compound channel.  Section~\ref{sec:enforce} provides the lower and upper bounds on the optimal transmission throughput and some examples to discuss the bounds. The paper is concluded in Section~\ref{sec:conclusion} and by technical appendices.

\section{System Model and Definitions}\label{covert-def}

Consider  communication in the presence of a masking constraint over a compound channel  with input and output alphabets $\mathcal{X}$ and $\mathcal{Y}$; see Fig.~\ref{figure1b}. The compound channel consists of two channel pmfs $W_s( \cdot |\cdot ), s\in\{1,2\}$ where the channel state $s$ is available at both the transmitter and the receiver. The message set is shown by $\mathcal{M}$. The transmitter and receiver share a secret key $K\in \mathcal{K}$. The transmitter  sends an $n$-length input $X^n$ over the channel using a state-dependent encoding function $f_s^{(n)}\colon\mathcal{M}\times \mathcal{K}\to \mathcal{X}^n$ where  
\begin{align}
X^n &= f_s^{(n)}(M,K).
\end{align}
The receiver uses a decoding function $g_s^{(n)}\colon \mathcal{Y}^n\times \mathcal{K}\to \mathcal{M}$ which maps the channel output to an estimated message as follows:
\begin{align}
\widehat{M} &= g_s^{(n)}(Y^n,K).
\end{align}

\begin{definition}\label{def-main} An $(n,|\mathcal{M}|, |\mathcal{K}|)$-code consists of a message set $\mathcal{M}$, a key set $\mathcal{K}$, two encoding functions $f_s^{(n)}:\mathcal{M}\times \mathcal{K}\to\mathcal{X}^n$, and two decoding functions $g_s^{(n)}:\mathcal{Y}^n\times \mathcal{K}\to\mathcal{M}$ for $s\in\{1,2\}$.
	For each state $s$, the code induces the following joint pmf on the random variables $(M,K,X^n,Y^n,\widehat{M})$:\footnote{The adversary knows the key distribution
	% which is used for generating the codebook
	 but does not know the exact  key used by the transmitter and receiver.}
	\begin{align}
	P_{MKX^nY^n\widehat{M}}^{(s)}(m,k,x^n,y^n,\hat{m})\triangleq \frac{1}{|\mathcal{M}|\cdot |\mathcal{K}|} W_s^n(y^n|x^n)\cdot \mathbbm{1}\left\{ f_s^{(n)}(m,k)=x^n \right\}\cdot \mathbbm{1}\left\{ g_s^{(n)}(y^n,k)=\hat{m} \right\}.\label{dist-induced}
	\end{align}
We also define the output distribution on ${\cal Y}^n$ induced by the joint distribution in \eqref{dist-induced} with state $s \in \{1,2\}$ as $Q_s^n$. 

We next define the performance metrics of interest given a code. First, we define the conditional probability of error conditioned on the state, key and the message as 
	\begin{IEEEeqnarray}{rCl}
		P_{\mathrm{e}}(s,k,m)\triangleq \mathbb{P}\big(\widehat{M}\neq M|S=s, K=k, M=m\big).
	\end{IEEEeqnarray}
The  conditional probability of error conditioned on the state and key is 
	\begin{IEEEeqnarray}{rCl}
		\bar{P}_{\mathrm{e}}(s,k)\triangleq \frac{1}{|\mathcal{M}|}\sum_{m=1}^{|\mathcal{M}|}P_{\mathrm{e}}(s,k,m),\qquad \forall s\in\{1,2\},\;k\in\mathcal{K}.\label{average-def}
	\end{IEEEeqnarray}
The maximum error probability is defined as 
	\begin{IEEEeqnarray}{rCl}
		P_{\mathrm{e}} \triangleq \max_{s=1,2}\;\; \max_{k\in\mathcal{K}}\;\;\max_{m\in\mathcal{M}}\;P_{\mathrm{e}}(s,k,m).\label{Pe-def}
	\end{IEEEeqnarray}	
\end{definition}

%In the following, we first present the mathematical setup for the first approach. Next, the second approach is discussed. }  

%	Evaluating the rate in \eqref{rate-positive-example} for the above example yields the following optimal rate:
%	\begin{align}
%	L_1^*(0^+,\delta_1,\delta_2) =1-h_{\text{b}}\left(\frac{1}{3}\right).
%	\end{align}

%\begin{figure}[b]
%	\centering
%	\includegraphics[scale=0.22]{case1.eps}
%	\caption{Example of Case 1.}
%	\label{figure1}
%\end{figure}

We  note that in our system model, similarly to \cite{Ligong}, we have assumed that the adversary observes the channel output through the {\em same} channel as the legitimate receiver.\medskip

%\sadaf{Defining the channel output marginal on $\mathcal{Y}^n$ induced by \eqref{dist-induced} to be $Q_s^n$, we present the following two definitions.}  \medskip

\sadaf{\begin{definition} \label{def:code} For $(\epsilon,\delta)\in (0,1)^2$, an $(n,|\mathcal{M}|,|\mathcal{K}|)$-code is said to be an $(n,|\mathcal{M}|,|\mathcal{K}|,\epsilon,\delta)$-code if  the total variation distance  $d_{\TV}(Q_1^n,Q_2^n)$ is no larger than $\delta$ and the maximum error probability is no larger than $\epsilon$.
	\end{definition}}
\medskip

\sadaf{ \begin{definition}\label{TV-def}
		For $(\epsilon,\delta)\in (0,1)^2$, a pair  $(L,S)\in\mathbb{R}_+^2$ is said to be  $(\epsilon,\delta)$-achievable if there exists a sequence of  $(n,|\mathcal{M}|,|\mathcal{K}|,\epsilon,\delta)$-codes such that
		\begin{align}
		\liminf_{n\to \infty}\; \frac{\log |\mathcal{M}|}{\sqrt{n}}&\ge L, \label{eqn:liminf_L}\\*
		\limsup_{n\to\infty}\; \frac{\log |\mathcal{K}|}{\sqrt{n}} &\leq S.\label{liminf_LK}
		\end{align}
		 The set of all $(\epsilon,\delta)$-achievable  pairs $(L ,S)$ is denoted as  $\mathcal{L}_{\TV}^*(\epsilon,\delta)$. We also define
		\begin{IEEEeqnarray}{rCl}
			L_{\TV}^*(\epsilon,\delta)=\sup\left\{ L\colon (L,S)\in \mathcal{L}_{\TV}^*(\epsilon,\delta)\;\text{for some}\; S \right\},
		\end{IEEEeqnarray}
		and 
		\begin{align}
		L_{\TV}^*(0^+,\delta)=\lim_{\epsilon\downarrow 0}L_{\TV}^*(\epsilon,\delta).
		\end{align}
\end{definition}}

	% satisfies: 
	% \begin{align}
	% P_{\mathrm{e}} \le \epsilon,
	% \end{align}

	%an $n$-length code for the covert communication over a compound DMC is said to be $(\epsilon,\delta_1,\delta_2,\beta)$-achievable if there exists a sequence of encoding and decoding functions $(f_1^{(n)},f_2^{(n)},g^{(n)})$ such that the probability of error is defined as the following:
	%	\begin{align}
	%,\label{Pe-def}
	%	\end{align}
	%	satisfies \begin{align}\displaystyle\max_{s\in\{1,2\}}P_{e_s}\leq  \epsilon,\label{Pe-bound}\end{align} 
	%	and where the following covertness conditions hold:
	%		\begin{align}
	%		\hspace{1.7cm}D\left(Q_s^{n}\|Q_0^{\times n}\right) &\leq \delta_s,\qquad s\in\{1,2\}.\label{covert_condition_kl}
	%		\end{align}  
	%We say that $L_{\beta}(\epsilon,\delta_1,\delta_2)$ is $(\epsilon,\delta_1,\delta_2,\beta)$-achievable if there exists a sequence of $(|\mathcal{M}|,n,\epsilon,\delta_1,\delta_2)$-codes such that
	%	\begin{align}
	%L_{\beta}(\epsilon,\delta_1,\delta_2)&\leq\liminf_{n\to \infty}\; \frac{\log |\mathcal{M}|}{n^{\beta}}.
	%	\end{align}

\sadaf{ The following proposition presents a necessary condition for the total variation distance between the output distributions $Q_1^n$ and $Q_2^n$ to be strictly bounded away from $1$, hence satisfying the  masking constraint in Definition~\ref{def:code}. 
	\begin{proposition}\label{feas-prop} Fix an $(n,|{\cal M}|, |{\cal K}|)$-code ${\cal C}$ as defined in  Definition~\ref{def-main}. As usual, $Q_s^n$ is the output distribution induced by   ${\cal C}$ and the channel $W_s$ for $s\in \{1,2\}$.  If the total variation distance $d_{\TV}(Q_1^n,Q_2^n)$ is strictly bounded  away from $1$, i.e., $\limsup_{n\to\infty} d_{\TV}(Q_1^n,Q_2^n)<1$,  then there exist two input distributions $P_1$ and $P_2$ over $\mathcal{X}$ such that 
		\begin{IEEEeqnarray}{rCl}
			\sum_{x\in\mathcal{X}}P_1(x)W_1(y|x)=\sum_{x\in\mathcal{X}}P_2(x)W_2(y|x), \;\;\; \forall y\,\in\mathcal{Y}.\label{prop-cons}
			\end{IEEEeqnarray}
		\end{proposition}
	\begin{IEEEproof} See Appendix~\ref{feas-prop-proof}.
		\end{IEEEproof}
	 In the following discussion, we further   assume that there exist input symbols ``$0$'' and ``$0'$''\footnote{\text{In fact, if $0\neq 0'$, we can permute the input symbols for the state $s=2$, such that the $0'$ is renamed to $0$.}} such that:
	\begin{align}
	\hspace{1.5cm}W_1(y|0)= W_2(y|0')\triangleq Q_0(y),\qquad y\in\mathcal{Y}.
	\end{align}
The above condition satisfies the condition of Proposition~\ref{feas-prop} in \eqref{prop-cons} because $P_1$ and $P_2$ therein are taken to be deterministic distributions,  placing all their masses on symbols $0$ and $0'$ respectively. It will be shown in Section \ref{sec:enforce} that the {\em square-root law} holds under this assumption; this means that the normalizations by $\sqrt{n}$ in \eqref{eqn:liminf_L} and \eqref{liminf_LK} yield finite and positive $(L,S)$ pairs.}

\sadaf{
\begin{remark}\label{rem2} In the following discussions, we  assume that all the transition probabilities $W_s(\cdot|x), s\in\{1,2\}, x\in\mathcal{X}$ are absolutely  continuous with respect to $Q_0(\cdot)$. This condition ensures   that all involved information quantities are finite. If this assumption is not required, it will be explicitly mentioned.
\end{remark} 
}

 \section{Main Results and Discussions}\label{sec:enforce} 
 
 In this section, we present our results for the proposed state masking scenario. First, inner and outer bounds to 	$\mathcal{L}_{\TV}^*(\epsilon,\delta)$ are provided.  Then, an optimality result is established. Finally, some examples are discussed.

%an $n$-length code for covert communication over a compound DMC with total variation distance as the covertness metric is said to be $(\epsilon,\delta)$-achievable if there exists a sequence of encoding and decoding functions $(f_1^{(n)},f_2^{(n)},g^{(n)})$ such that the probability of error is defined as in \eqref{Pe-def} satisfies \eqref{Pe-bound} and the total variation distance between $Q_1^n$ and $Q_2^n$ is upper bounded as:
%		\begin{align}
%		d_{\TV}\left(Q_1^n,Q_2^n\right) &\leq \delta.\label{covert_condition_PPM_kl}
%		\end{align}  
%We say that $L(\epsilon,\delta)$ is $(\epsilon,\delta)$-achievable if there exists a sequence of $(|\mathcal{M}|,n,\epsilon,\delta)$-codes such that
%	\begin{align}
%	L(\epsilon,\delta)&\leq \liminf_{n\to \infty}\; \frac{\log |\mathcal{M}|}{\sqrt{n\delta}}.
%	\end{align}
%The supremum of all $(\epsilon,\delta)$-achievable $L(\epsilon,\delta)$ is denoted by $L^*(\epsilon,\delta)$.}

\subsection{An Inner Bound to $\mathcal{L}_{\TV}^*(\epsilon,\delta)$}\label{sec:optimal}

In the following, we present an inner bound to $\mathcal{L}_{\TV}^*(\epsilon,\delta)$.  
 For positive $\gamma_1,\gamma_2$ and any three  distributions $Q_0$, ${Q}_1$ and ${Q}_2$ supported on $\mathcal{Y}$, define:
	\begin{align}
\Omega(\gamma_1,\gamma_2;Q_0,Q_1,Q_2)&\triangleq\gamma_1^2\chi_2({Q}_1\|Q_0)+\gamma_2^2\chi_2({Q}_2\|Q_0)-2\gamma_1\gamma_2\rho({Q}_1,{Q}_2,Q_0),\label{prep1} \end{align}
where
\begin{align}
\rho({Q}_1,{Q}_2,Q_0)\triangleq \mathbb{E}_{Q_0}\left[\left(\frac{{Q}_1(Y)}{Q_0(Y)}-1\right)\cdot \left(\frac{{Q}_2(Y)}{Q_0(Y)}-1\right)\right].
\label{eqn:def_rho}
\end{align}

\begin{theorem}\label{key-thm}  An inner bound to $\mathcal{L}_{\TV}^*(\epsilon,\delta)$, is given by the union of all pairs $(L,S)$ such that:
	\begin{IEEEeqnarray}{rCl}
		L\leq \min_{s\in\{1,2\}}\gamma_{s}
		%D(\bar{Q}_{s}\|Q_0)%this expression is only true for the binary channels
		D(W_s \| Q_0|\bar{P}_s),\label{key-ineq1}
	\end{IEEEeqnarray}	
	and 
	\begin{IEEEeqnarray}{rCl}
		S \geq \max_{s\in\{1,2\}}\gamma_{s}
			D(W_s \| Q_0|\bar{P}_s)
		%D(\bar{Q}_{s}\|Q_0)
		-\min_{s\in\{1,2\}}\gamma_{s}
			D(W_s \| Q_0|\bar{P}_s)
		%D(\bar{Q}_{s}\|Q_0)
		,\label{key-ineq2}
	\end{IEEEeqnarray}
%	where the union is over all input distributions $\bar{P}_s$ such that $\bar{P}_s(0)=0$ and their induced output distributions (through $W_s$)  denoted by $\bar{Q}_s$ and over all  positive $  \gamma_1,\gamma_2$ such that 
where the union is over all positive constants $\gamma_1$ and $\gamma_2 $ and input distributions $\bar{P}_s$ such that $\bar{P}_s(0)=0$ and inducing output distributions (through $W_s$) $\bar{Q}_s$ such that 
	\begin{equation}
	\Omega(\gamma_1,\gamma_2;Q_0,\bar{Q}_1,\bar{Q}_2)\le \left(2\Phi^{-1}\Big(\frac{1+\delta}{2}\Big)\right)^2.
	\label{omega-ineq}
	\end{equation}
\end{theorem}
\begin{IEEEproof} The achievable scheme is sketched as follows. Two independent codebooks are generated in which each of them is used for communicating over each channel state. The optimal decoder, namely, the ML decoder,  is used to determine the message which is sent. The analysis of error probability involves applying Shannon's achievability bound \cite[Thm~2]{Yury} to upper bound the error probability averaged over all codebooks. A key step of the analysis is the single-letterization of the  masking constraint. Finally, it is shown that there exists a ``good'' codebook which satisfies the desired properties that the maximum error probability and the masking constraint are both small. 

\underline{\textit{Preparations}}: Fix a large blocklength $n$, choose positive $\gamma_s$ for $s\in\{1,2\}$. Let $\bar{P}_{s}$ be any distribution such that $\bar{P}_{s}(0)=0$ and denote the output marginal through channel $W_s$ by $\bar{Q}_{s}$.
Choose positive $\gamma_s$ such that \eqref{omega-ineq} holds and let
\begin{equation}
\mu_{s,n}\triangleq \frac{\gamma_s}{\sqrt{n}}.\label{prep2}
\end{equation}
Define  the input distribution $P_{s,n}$ as 
\begin{align}
P_{s,n}(x) \triangleq \left\{\begin{array}{ll} \mu_{s,n}\bar{P}_{s}(x) & x\neq 0\\1-\mu_{s,n} & x=0\end{array}\right. .\label{step28}
\end{align}
Note that $n \mu_{s,n}=\Theta(\sqrt{n})$ can be thought of as the weight of each codeword in the $s$-th codebook. For $s\in\{1,2\}$, let $Q_{s,n}$ denote the output marginal of the channel $W_s$ when the input distribution is $P_{s,n}$. The distribution $Q_{s,n}$ can be written as convex combination follows:
\begin{IEEEeqnarray}{rCl}
	Q_{s,n}=(1-\mu_{s,n})Q_0+\mu_{s,n}\bar{Q}_{s}.\label{Qs-def}
\end{IEEEeqnarray}
Denote the $n$-fold products of $P_{s,n}$ and $Q_{s,n}$ by $P_s^{\times n}$ and $Q_s^{\times n}$, respectively.

\underline{\textit{Codebook Generation}}:
We generate an i.i.d.\ codebook $\mathcal{C}_s =  \left\{ x_s^n(m,k)\colon m\in\mathcal{M},k\in\mathcal{K} \right\}$ according to the input distribution $P_{s,n}$ as defined in \eqref{step28}. Denote the channel output marginal by $Q_s^n$.

\underline{\textit{Single-letter Characterization of Masking Constraint}}: The following two lemmas together provide single-letter characterizations of the masking constraint.
\begin{lemma}\label{lem-ach} We have
	\begin{align} 
	\hspace{0cm}	d_{\TV}\left(Q_1^{\times n},Q_2^{\times n}\right)\leq 2\Phi\left(\frac{1}{2}\sqrt{\Omega(\gamma_1,\gamma_2;Q_0,\bar{Q}_1,\bar{Q}_2)}\right)-1+O\left(\frac{1}{\sqrt{n}}\right). \label{eqn:dTV}
	\end{align}
	%	where 
	%	\begin{IEEEeqnarray}{rCl}
	%	\Omega(\gamma_1,\gamma_2)&\triangleq& \gamma_1^2\chi_2(\tilde{Q}_1\|Q_0)+\gamma_2^2\chi_2(\tilde{Q}_2\|Q_0)\nonumber\\&&\hspace{1.5cm}-2\gamma_1\gamma_2\rho(\tilde{Q}_1,\tilde{Q}_2,Q_0).
	%	\end{IEEEeqnarray}
\end{lemma}
\begin{IEEEproof} See Appendix \ref{lem-ach-proof}.
\end{IEEEproof}

\begin{lemma}\label{lem-key} Let $\kappa$ be an arbitrary positive number. Assume that,
	\begin{align}
	\log |\mathcal{M}|+\log |\mathcal{K}| \geq (1+\kappa)\sqrt{n} \cdot \max_{s\in\{1,2\}}\gamma_s
	D(W_s \| Q_0|\bar{P}_s)
	%D\left(\bar{Q}_s\|Q_0\right)+\log n
	.\label{lem-ineq1}
	\end{align}
	Then,  for each $s\in \{1,2\}$,  we have
	\begin{align} 
	\hspace{0cm}	\mathbb{E}\left[d_{\TV}\left(Q_s^{ n},Q_s^{\times n}\right)\right]= O\left(\frac{1}{\sqrt{n}}\right), \label{exp:dTV}
	\end{align}
	whenever $\Omega(\gamma_1,\gamma_2;Q_0,\bar{Q}_1,\bar{Q}_2)$ satisfies \eqref{omega-ineq}.
\end{lemma}
\begin{IEEEproof} See Appendix~\ref{lem-key-proof}.
\end{IEEEproof}

\underline{\textit{Encoding and Decoding}}:
If the channel is in state $s\in\{1,2\}$, the transmitter upon observing the message $m\in\mathcal{M}$ and the key $k\in\mathcal{K}$, sends the codeword $x_s^n(m,k)$ over the channel. Given  $y^n$, the decoder based on the secret key $k$ and the state $s$ uses
the optimal ML decoder
% a threshold decoder \cite{Yury}
 which chooses the message $\hat{m}$ as follows:
\begin{align}
\hat{m}=\argmax_{m\in\mathcal{M}} W^n_s(y^n|\bx_s(m,k)) 
%\frac{1}{\sqrt{n}}\log \frac{W_s\left(y^n|x_s^n(m,k)\right)}{Q_s^{\times n}(y^n)} \ge \sqrt{n} I(P_{s,n},W_s)+\xi, 
\label{eqn:rate_con}
\end{align}
%\begin{align}
%\frac{1}{\sqrt{n}}\log \frac{W_s\left(y^n|x_s^n(m,k)\right)}{Q_s^{\times n}(y^n)} \ge \sqrt{n} I(P_{s,n},W_s)+\xi, \label{eqn:rate_con}
%\end{align}
%for some arbitrarily small $\xi>0$.

%\begin{align}
%\mathcal{C}_s\triangleq,
%\end{align}
%and

%The analysis of the rate constraint is omitted.
Given a bound on the code size, the following lemma provides an upper bound on the average error probability as defined in \eqref{average-def}.
\begin{lemma}\label{PPM-thm}   If 
	\begin{IEEEeqnarray}{rCl}
		\frac{\log |\mathcal{M}|}{\sqrt{n}}\leq \min_{s\in\{1,2\}}\sqrt{n}I(P_{s,n},W_s)-n^{-1/6},\label{lem4-ineq}
		\end{IEEEeqnarray}
	Then, 
	\begin{IEEEeqnarray}{rCl}
		\mathbb{E}\left[\bar{P}_{\mathrm{e}}(s,k)\right]= O\big(n^{- {1}/{6}}\big),\qquad \forall s\in\{1,2\},\;k\in\mathcal{K}.
		\end{IEEEeqnarray}
%and 
%\begin{align}
%	\hspace{-0.2cm}\rho(\tilde{Q}_1,\tilde{Q}_2,Q_0)\!\triangleq \!\mathbb{E}_{Q_0}\left[\!\frac{(\tilde{Q}_1(Y)\!-\!Q_0(Y))  (\tilde{Q}_2(Y)\!-\! Q_0(Y))}{Q_0(Y)^2}\!\right]. \label{eqn:def_rho}
%\end{align}
\end{lemma}
\begin{IEEEproof} See Appendix~\ref{rate-analysis}.
	\end{IEEEproof}	

\underline{\textit{Existence of a Codebook with Desired Properties}}: The following lemma proves existence of a codebook with desired properties.
\begin{lemma}\label{lem4} If inequalities \eqref{lem-ineq1} and \eqref{lem4-ineq} hold for some positive $\kappa$ and $\epsilon$, then there are exists at least one ``good'' codebook such that for each $s\in \{1,2\}$, 
	\begin{align} 
	\hspace{0cm}	d_{\TV}\left(Q_s^{ n},Q_s^{\times n}\right)=
	o(1)
	% O\left(\frac{1}{\sqrt{n}}\right)
	, \label{lem4-ineq2}
	\end{align}
	and the maximum error probability as defined in \eqref{Pe-def}, satisfies:
	\begin{align}
 P_{\mathrm{e}} = o(1) \label{eqn:max_err_vanish}
 %O\left(n^{-\frac{1}{12}}\right).
	\end{align}
	\end{lemma}
\begin{IEEEproof} See Appendix~\ref{existence2-proof}.
	\end{IEEEproof}
The proof is concluded by combining Lemmas \ref{lem-ach} to \ref{lem4}. See Appendix~\ref{key-thm-proof} for details.
\end{IEEEproof}

Now, we specialize Theorem~\ref{key-thm} to the case where the length of key is infinite. Also, we restrict our attention to the binary input alphabet case in which $\mathcal{X}=\{0,1\}$ and the output distribution induced by each input symbol is denoted by
\begin{IEEEeqnarray}{rCl}
	\tilde{Q}_{1}(\cdot )\triangleq  W_1(\cdot |1),\qquad	\tilde{Q}_{2}(\cdot )\triangleq  W_2(\cdot |1).\label{output-dist-def}
\end{IEEEeqnarray}
This special case will be used later to provide an optimality result. In this case, we are able to find a closed form for the optimization problem \eqref{key-ineq1} under the constraint \eqref{omega-ineq}. The solution of this optimization is given in the following corollary.

\begin{corollary}\label{coro} Assume that the length of the key is infinite and $\mathcal{X}=\{0,1\}$. Then, $L_{\TV}^*(\epsilon,\delta)$ is lower bounded as follows:
	\begin{align}
	L_{\TV}^*(\epsilon,\delta) \geq \left\{\begin{array}{cc}
	\mathsf{L}_1&\mbox{if } \rho(\tilde{Q}_1,\tilde{Q}_2,Q_0)< \min \left(\frac{\chi_2(\tilde{Q}_1\|Q_0)D(\tilde{Q}_2\|Q_0)}{D(\tilde{Q}_1\|Q_0)},\frac{\chi_2(\tilde{Q}_2\|Q_0) D(\tilde{Q}_1\|Q_0)}{D(\tilde{Q}_2\|Q_0)}\right)\\
		\mathsf{L}_2 &\mbox{if } \rho(\tilde{Q}_1,\tilde{Q}_2,Q_0)\ge \min \left(\frac{\chi_2(\tilde{Q}_1\|Q_0)D(\tilde{Q}_2\|Q_0)}{D(\tilde{Q}_1\|Q_0)},\frac{\chi_2(\tilde{Q}_2\|Q_0) D(\tilde{Q}_1\|Q_0)}{D(\tilde{Q}_2\|Q_0)}\right)
\end{array}	  \right. \label{L*}
	\end{align}
	where
	\begin{IEEEeqnarray}{rCl}
		\mathsf{L}_1 \triangleq  2\Phi^{-1}\left(\frac{1+\delta}{2}\right)\cdot \frac{D(\tilde{Q}_1\|Q_0) D(\tilde{Q}_2\|Q_0)}{\sqrt{\chi_2(\tilde{Q}_1\|Q_0) D(\tilde{Q}_2\|Q_0)^2-2\rho(\tilde{Q}_1,\tilde{Q}_2,Q_0) D(\tilde{Q}_1\|Q_0) D(\tilde{Q}_2\|Q_0)+\chi_2(\tilde{Q}_2\|Q_0) D(\tilde{Q}_1\|Q_0)^2}},\nonumber\\\label{L-def1}
		\end{IEEEeqnarray}
%	if 
%	\begin{IEEEeqnarray}{rCl}
%		\rho(\tilde{Q}_1,\tilde{Q}_2,Q_0)< \min \left(\frac{\chi_2(\tilde{Q}_1\|Q_0)D(\tilde{Q}_2\|Q_0)}{D(\tilde{Q}_1\|Q_0)},\frac{\chi_2(\tilde{Q}_2\|Q_0) D(\tilde{Q}_1\|Q_0)}{D(\tilde{Q}_2\|Q_0)}\right),
%		\end{IEEEeqnarray}
and 
\begin{IEEEeqnarray}{rCl}
	\mathsf{L}_2 \triangleq 2\Phi^{-1}\left(\frac{1+\delta}{2}\right)\cdot \sqrt{\frac{\min\big\{\chi_2(\tilde{Q}_1\|Q_0)D(\tilde{Q}_2\|Q_0)^2,\chi_2(\tilde{Q}_2\|Q_0) D(\tilde{Q}_1\|Q_0)^2\big\}}{\chi_2(\tilde{Q}_1\|Q_0) \chi_2(\bar{Q}_2\|Q_0)-\rho^2(\tilde{Q}_1,\tilde{Q}_2,Q_0)}}.
	\end{IEEEeqnarray}	
%if
%\begin{IEEEeqnarray}{rCl}
%	 \rho(\tilde{Q}_1,\tilde{Q}_2,Q_0) \geq \min \left(\frac{\chi_2(\tilde{Q}_1\|Q_0)D(\tilde{Q}_2\|Q_0)}{D(\tilde{Q}_1\|Q_0)},\frac{\chi_2(\tilde{Q}_2\|Q_0) D(\tilde{Q}_1\|Q_0)}{D(\tilde{Q}_2\|Q_0)}\right).
%	\end{IEEEeqnarray}	
	\end{corollary}
\begin{IEEEproof} See Appendix~\ref{cor-proof}.
	\end{IEEEproof}

\subsection{An Upper Bound on $L_{\TV}^*(0^+,\delta)$ and an Optimality Result}\label{sec:upper}
 In this section, we provide an optimality  result for $L_{\TV}^*(0^+,\delta)$. As with the preceding discussion, we assume that the key is of infinite length and $\mathcal{X}=\{0,1\}$. %We consider two simplifications on the system model. First, it is assumed that the length of key is infinite. Second, we restrict to the case of $\mathcal{X}=\{0,1\}$ where the output distributions are defined in \eqref{output-dist-def}.
The following theorem presents an upper bound on $L_{\TV}^*(0^+,\delta)$ in terms of a free parameter $\varphi \in [0,1]$. Its proof is given after the statement of Theorem~\ref{opt-thm}. The proof utilizes   some ideas from Tahmasbi and Bloch~\cite{Mehrdad}. To state the theorem concisely, we recall the definitions of $\tilde{Q}_1$ and $\tilde{Q}_2$ in \eqref{output-dist-def} and additionally, define
\sadaf{\begin{IEEEeqnarray}{rCl}
	\D_{s}(\varphi)&\triangleq&\mathbb{E}_{Q_0}\left[ \frac{(\tilde{Q}_{s}(Y)-Q_0(Y)) \left(\varphi(\tilde{Q}_{1}(Y)-Q_0(Y))-(1-\varphi)(\tilde{Q}_{2}(Y)-Q_{0}(Y)) \right)}{Q_{0}(Y)^2}\right],\\
	\Delta(\varphi) &\triangleq & \mathbb{E}_{Q_0} \bigg[\bigg( \frac{\varphi(\tilde{Q}_{1}(Y)-Q_0(Y))-(1-\varphi)(\tilde{Q}_{2}(Y)-Q_0(Y))}{Q_0(Y)}\bigg)^2\bigg],
\end{IEEEeqnarray}
for some $0\leq \varphi \leq 1$, and
\sadaf{
	\begin{IEEEeqnarray}{rCl}
		\Psi \triangleq \mathbb{E}_{Q_0}\left[\left(\frac{\tilde{Q}_1(Y)-\tilde{Q}_2(Y)}{Q_0(Y)}\right)^2\right].
\end{IEEEeqnarray}
Notice that  $\D_1(\varphi)$, $\D_2(\varphi)$ and $\Delta(\varphi)$ can be equivalently written as:
\begin{IEEEeqnarray}{rCl}
	\D_1(\varphi)&=&  \varphi \chi_2(\tilde{Q}_1\|Q_0)-(1-\varphi)\rho(\tilde{Q}_1,\tilde{Q}_2,Q_0),\\
	\D_2(\varphi) &=& -(1-\varphi)\chi_2(\tilde{Q}_2\|Q_0)+\varphi \rho(\tilde{Q}_1,\tilde{Q}_2,Q_0),\label{D-definition}\\
	\Delta(\varphi) &=& \varphi^2 \chi_2(\tilde{Q}_1\|Q_0)+(1-\varphi)^2\chi_2(\tilde{Q}_2\|Q_0)-2\varphi (1-\varphi)\rho(\tilde{Q}_1,\tilde{Q}_2,Q_0).
\end{IEEEeqnarray}}
\medskip
\begin{theorem}\label{thm-upper} For any $0\le\varphi\le 1$, if 
	\begin{align}
	&\rho(\tilde{Q}_1,\tilde{Q}_2,Q_0)\leq \min\left\{\frac{\varphi}{1-\varphi}\chi_2(\tilde{Q}_1\|Q_0),\frac{1-\varphi}{\varphi}\chi_2(\tilde{Q}_2\|Q_0)\right\},\label{cons-upper}
	\end{align}
	%	 $L_{\TV}^*(0^+,\delta)$ is upper bounded as follows:
	then,
	\begin{IEEEeqnarray}{rCl}
		L_{\TV}^*(0^+,\delta) \leq   2\Phi^{-1}\left(\frac{1+\delta}{2}\right)\cdot \frac{\sqrt{\Delta(\varphi)}}{\D_1(\varphi)-\D_2(\varphi)}\cdot \max_{s\in\{1,2\}} D(\tilde{Q}_{s}\|Q_0),\label{L-upper-term}
	\end{IEEEeqnarray}
	for some $0\leq \varphi\leq 1$.

\end{theorem}} \medskip
\sadaf{The following presents an expression characterizing the minimal value of  \eqref{L-upper-term} over all possible values of $\varphi \in [0,1]$. 
	\begin{corollary}\label{cor-upper} $L^*_{\TV}(0^+,\delta)$ is upper bounded by the following:
		\begin{IEEEeqnarray}{rCl}
			L^*_{\TV}(0^+,\delta)\leq \left\{\begin{array}{cc}
				\mathsf{U}_1&\hspace{0cm}\mbox{if } \rho(\tilde{Q}_1,\tilde{Q}_2,Q_0)\leq \min \{\chi_2(\tilde{Q}_1\|Q_0),\chi_2(\tilde{Q}_2\|Q_0) \}\\[0.5ex]
				\mathsf{U}_2 &\mbox{if } \min\{\chi_2(\tilde{Q}_1\|Q_0),\chi_2(\tilde{Q}_2\|Q_0)\} <\rho(\tilde{Q}_1,\tilde{Q}_2,Q_0)
			\end{array}	  \right. \label{L-upper}
		\end{IEEEeqnarray}
		where 
		\begin{IEEEeqnarray}{rCl}
			\mathsf{U}_1&\triangleq & \frac{2\Phi^{-1}\left(\frac{1+\delta}{2}\right)}{\sqrt{\Psi}}\cdot \max_{s\in\{1,2\}} D(\tilde{Q}_{s}\|Q_0),\\
			\mathsf{U}_2&\triangleq & 2\Phi^{-1}\left(\frac{1+\delta}{2}\right)\cdot \sqrt{\frac{\min\{\chi_2(\tilde{Q}_1\|Q_0),\chi_2(\tilde{Q}_2\|Q_0)\}}{\chi_2(\tilde{Q}_1\|Q_0)\chi_2(\tilde{Q}_2\|Q_0)-\rho^2(\tilde{Q}_1,\tilde{Q}_2,Q_0)}}\cdot \max_{s\in\{1,2\}} D(\tilde{Q}_{s}\|Q_0).
		\end{IEEEeqnarray}
	\end{corollary}
	\begin{IEEEproof} See Appendix \ref{cor-upper-proof}.
	\end{IEEEproof}\medskip
}

\sadaf{The above upper bound is tight in the following special cases. 
\begin{theorem}[Optimality Results]\label{opt-thm} Assume that
	\begin{align}
D(\tilde{Q}_1\|Q_0)&=D(\tilde{Q}_2\|Q_0)\triangleq \mathbb{D}.\label{op-cons2}
\end{align}
If	
	\begin{IEEEeqnarray}{rCl}
		\rho(\tilde{Q}_1,\tilde{Q}_2,Q_0)\leq \min\big\{\chi_2(\tilde{Q}_1\|Q_0),\chi_2(\tilde{Q}_2\|Q_0)\big\},\label{conv-cons1}
		\end{IEEEeqnarray}
	then,
	\begin{align}
	L_{\TV}^*(0^+,\delta)=\frac{2\Phi^{-1}\left(\frac{1+\delta}{2}\right)}{\sqrt{\Psi}}\cdot \mathbb{D},\label{L_optimal}
	\end{align}
and if 
\begin{IEEEeqnarray}{rCl}
	\rho(\tilde{Q}_1,\tilde{Q}_2,Q_0)> \min\big\{\chi_2(\tilde{Q}_1\|Q_0),\chi_2(\tilde{Q}_2\|Q_0)\big\},\label{cond}
\end{IEEEeqnarray}
then
\begin{align}
L_{\TV}^*(0^+,\delta)=2\Phi^{-1}\left(\frac{1+\delta}{2}\right)\cdot \sqrt{\frac{\min\{\chi_2(\tilde{Q}_1\|Q_0),\chi_2(\tilde{Q}_2\|Q_0)\}}{\chi_2(\tilde{Q}_1\|Q_0)\chi_2(\tilde{Q}_2\|Q_0)-\rho^2(\tilde{Q}_1,\tilde{Q}_2,Q_0)}}\cdot \mathbb{D}.
\end{align}
\end{theorem}
\begin{IEEEproof} We first provide the proof when condition \eqref{conv-cons1} holds. The achievability follows from Corollary~\ref{coro} and considering the fact that assumptions in \eqref{op-cons2} and \eqref{conv-cons1} imply the first clause of the lower bound on $L_{\TV}^*(\epsilon,\delta)$ in~\eqref{L*} is in effect. The converse follows from Corollary~\ref{cor-upper} where the first clause of the upper bound on $L^*_{\TV}(0^+,\delta)$ in \eqref{L-upper} holds and also from the assumption in~\eqref{op-cons2}. Now, assume that the condition \eqref{cond} holds. The proofs of achievability and converse follow from the second clauses in \eqref{L*} and \eqref{L-upper}, respectively.
\end{IEEEproof}\medskip
}

\begin{IEEEproof}[Proof of Theorem~\ref{thm-upper}] The proof is sketched as follows. First, given any codebook, we relate the total variation distance to the false alarm and missed detection probabilities of a hypothesis testing problem. Then, we upper bound these probabilities which  then yields a lower bound on the total variation distance. Finally, the proof is concluded by an expurgation argument showing that there exists at least one sub-codebook of a large enough size that satisfies the desired property on the total variation distance and also by applying Fano's inequality.

\underline{\textit{Preparations}}: For any two codebooks 
\begin{align}\mathcal{C}_s=\left\{x_s^n(m,k)\colon m\in\mathcal{M},k\in\mathcal{K}\right\},\hspace{1cm} s\in\{1,2\},\end{align}
we define the following parameters:
\begin{IEEEeqnarray}{rCl}
	\mu_{\Lo,n}&\triangleq&\min_{s\in\{1,2\}}\;\displaystyle\min_{\substack{m\in\mathcal{M}}}\;\displaystyle\min_{\substack{k\in\mathcal{K}}}\; \frac{1}{n}\w\left(x_s^n(m,k)\right),\\
	\mu_{\Hi,n}&\triangleq&\max_{s\in\{1,2\}}\;\displaystyle\max_{\substack{m\in\mathcal{M}}}\;\displaystyle\max_{\substack{k\in\mathcal{K}}}\;\frac{1}{n}\w\left(x_s^n(m,k)\right),\label{mu-def}
\end{IEEEeqnarray}
and 
\begin{IEEEeqnarray}{rCl}
	\omega_{\Lo,n}\triangleq n\mu_{\Lo,n},\qquad \omega_{\Hi,n}\triangleq n\mu_{\Hi,n}.
	\end{IEEEeqnarray}
Let
\begin{subequations}\label{def-pri}
	\begin{IEEEeqnarray}{rCl}
			\Gamma_s(\varphi) &\triangleq &\mathbb{E}_{Q_0}\left[ \frac{(\tilde{Q}_{s}(Y)-Q_0(Y))\left(\varphi(\tilde{Q}_{1}(Y)-Q_0(Y))-(1-\varphi)(\tilde{Q}_{2}(Y)-Q_{0}(Y)) \right)^2}{Q_0(Y)^3}\right],\\
		\V_{s}^*(\varphi) &\triangleq&  \Delta+\mu_{\Hi,n}\cdot |\Gamma_s(\varphi)|,
	\end{IEEEeqnarray}
\end{subequations}
for some $0\leq \varphi\leq 1$ and define
\begin{IEEEeqnarray}{rCl}
	\tau(\varphi) \triangleq \frac{n\mu_{\Lo,n}}{2}\left(\D_{2}(\varphi)+\D_{1}(\varphi)\right),\label{step15}
\end{IEEEeqnarray} 
by recalling the definitions of $\D_1(\varphi)$ and $\D_2(\varphi)$ in \eqref{D-definition}.\medskip

\underline{\textit{Total Variation Distance and a Hypothesis Testing Problem}}: We relate the total variation distance to a hypothesis testing problem as follows. Consider the following inequality:
\begin{IEEEeqnarray}{rCl}
	d_{\TV}(Q_1^n,Q_2^n)\geq 1-\alpha_n-\beta_n,\label{dtv-ineqaulity}
\end{IEEEeqnarray}
where $\alpha_n$ and $\beta_n$ are the false alarm and missed detection probabilities of a possibly suboptimal hypothesis test in which under the null hypothesis $\mathcal{H}=1$, $Y^n$ is distributed according to $Q_1^n$ and under the alternative hypothesis $\mathcal{H}=2$, it is distributed according to $Q_2^n $. Define the following suboptimal test:
\begin{align}
\mathcal{A}(y^n;\varphi)\triangleq\mathbbm{1}\left\{\sum_{i=1}^n\mathcal{T}(y_i;\varphi)> \tau\right\},
\end{align}
where
\sadaf{\begin{align}
\mathcal{T}(y_i;\varphi)\triangleq  \frac{\varphi(\tilde{Q}_{1}(y_i)-Q_0(y_i))-(1-\varphi)(\tilde{Q}_{2}(y_i)-Q_0(y_i))}{Q_{0}(y_i)}.\label{step44}
\end{align}

}

The following lemma provides upper bounds on the false alarm and missed detection probabilities of this hypothesis test.

\sadaf{\begin{lemma}\label{lem2} For any $0\le\varphi\le 1$, we have:
	%	\begin{IEEEeqnarray}{rCl}
	%		\alpha_n
	%		&\leq & 1-\Phi\left(\frac{1}{2}\mu_{\Lo,n}\sqrt{n\Delta}  \right)+\frac{\sqrt{n}\mu_{\Lo,n}\mu_{\Hi,n}|\Gamma_1|}{4\sqrt{2\pi\Delta}}\nonumber\\*&&\hspace{0.5cm}+O\left(\frac{1}{\sqrt{n}}\right),\label{step160}
	%	\end{IEEEeqnarray}
	\begin{align}
	\alpha_n \le 1-\Phi\left(\frac{1}{2}\mu_{\Lo,n}\sqrt{n}\frac{\D_1(\varphi)-\D_2(\varphi)}{\sqrt{\Delta(\varphi)}}  \right)+\frac{\sqrt{n}\mu_{\Lo,n}\mu_{\Hi,n}|\Gamma_1(\varphi)|(\D_1(\varphi)-\D_2(\varphi))}{4\sqrt{2\pi\Delta(\varphi)^3}} + O\Big(\frac{1}{\sqrt{n}}\Big).\label{step160}
	\end{align}
	and 
	%\begin{IEEEeqnarray}{rCl}
	%	\beta_n
	%	&\leq & 1-\Phi\left(\frac{1}{2}\mu_{\Lo,n}\sqrt{n\Delta}  \right)+\frac{\sqrt{n}\mu_{\Lo,n}\mu_{\Hi,n}|\Gamma_2|}{4\sqrt{2\pi\Delta}}\nonumber\\*&&\hspace{0.5cm}+O\left(\frac{1}{\sqrt{n}}\right).\label{step170}
	%\end{IEEEeqnarray}
	\begin{align}
	\beta_n \le 1-\Phi\left(\frac{1}{2}\mu_{\Lo,n}\sqrt{n}\frac{\D_1(\varphi)-\D_2(\varphi)}{\sqrt{\Delta(\varphi)}}  \right)+\frac{\sqrt{n}\mu_{\Lo,n}\mu_{\Hi,n}|\Gamma_2(\varphi)|(\D_1(\varphi)-\D_2(\varphi))}{4\sqrt{2\pi\Delta(\varphi)^3}}  +O\Big(\frac{1}{\sqrt{n}}\Big).\label{step170}
	\end{align}
\end{lemma}
\begin{IEEEproof} See Appendix~\ref{lem2-proof}.
\end{IEEEproof}}\medskip

Combining \eqref{step160} and \eqref{step170} with \eqref{dtv-ineqaulity}, we get
\sadaf{\begin{IEEEeqnarray}{rCl}
	&&d_{\TV}\left(Q_1^n,Q_2^n\right)\geq 2\Phi\left(\frac{1}{2}\mu_{\Lo,n}\sqrt{n}\frac{\D_1(\varphi)-\D_2(\varphi)}{\sqrt{\Delta(\varphi)}}  \right)-1-\frac{\sqrt{n}\mu_{\Lo,n}\mu_{\Hi,n}(|\Gamma_1(\varphi)|+|\Gamma_2(\varphi)|)(\D_1(\varphi)-\D_2(\varphi))}{4\sqrt{2\pi\Delta(\varphi)^3}}+O\left(\frac{1}{\sqrt{n}}\right),\nonumber\\\label{step42b}
\end{IEEEeqnarray}}
which can be equivalently written as follows
\sadaf{\begin{IEEEeqnarray}{rCl}
	&&d_{\TV}\left(Q_1^n,Q_2^n\right)\geq 2\Phi\left(\frac{1}{2}\omega_{\Lo,n}\frac{\D_1(\varphi)-\D_2(\varphi)}{\sqrt{n\Delta(\varphi)}}   \right)-1-\frac{\omega_{\Lo,n}\omega_{\Hi,n}(|\Gamma_1(\varphi)|+|\Gamma_2(\varphi)|)(\D_1(\varphi)-\D_2(\varphi))}{4n\sqrt{2n\pi\Delta(\varphi)^3}}+O\left(\frac{1}{\sqrt{n}}\right).\label{step42}
\end{IEEEeqnarray}}
The proof is continued by showing that there exists a sub-codebook which satisfies \eqref{step42} and its size is at least $\frac{|\mathcal{M}|\cdot |\mathcal{K}|}{\sqrt{n}}$. The proof  is concluded by applying Fano's inequality. See Appendix~\ref{existence-proof} for details.
\end{IEEEproof}

\begin{remark} Assume that \eqref{op-cons2} and \eqref{conv-cons1} hold. Then, the choice of $\gamma_1=\gamma_2$ minimizes the RHS of \eqref{key-ineq1} and maximizes the RHS of \eqref{key-ineq2} in Theorem~\ref{key-thm}. Thus, it can be seen a small key length suffices under the conditions of Theorem~\ref{opt-thm}.
\end{remark}

%\medskip

\subsection{Examples}\label{example-section}

\begin{figure}[t]
	\centering
	\includegraphics[scale=1.1]{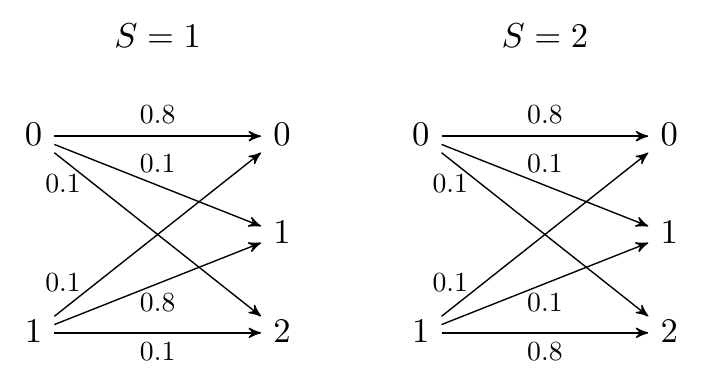}
	\caption{Example with discrete input and output alphabets}
	\label{figure2}
\end{figure}

\begin{figure}[t]
	\centering
	\includegraphics[width=.6\columnwidth]{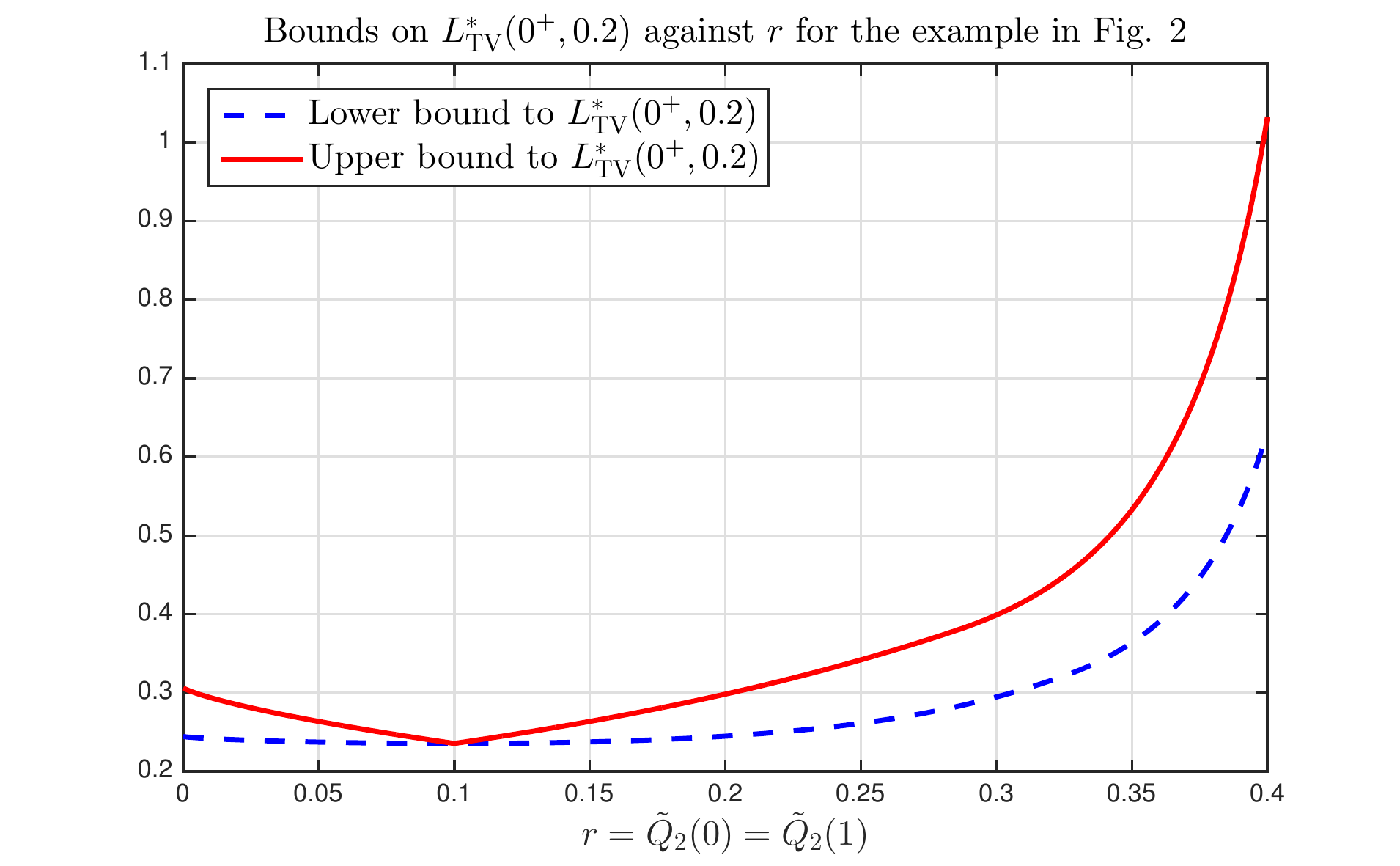}
	\caption{Lower and upper bounds on $L_{\TV}^*(0,0.2)$ for the example in Fig.~\ref{figure2}}
	\label{figure3}
\end{figure}

In this section, we provide examples that satisfy both conditions of Theorem~\ref{opt-thm} in \eqref{op-cons2} and \eqref{conv-cons1} and discuss scenarios in which we can establish the optimal communication throughput.

First, consider the setup of  Fig.~\ref{figure2}. This example satisfies the constraints~\eqref{op-cons2} and~\eqref{conv-cons1}. Choosing $\delta=0.2$ for this example, Theorem~\ref{opt-thm} yields $L_{\TV}^*(0,0.2)=0.2356$. For other values of $r\triangleq\tilde{Q}_2(0)=\tilde{Q}_2(1)$, the upper and lower bounds on $L_{\TV}^*(0,0.2)$ are plotted in Fig.~\ref{figure3}.  It can also be verified from the figure that the two bounds match for $r=0.1$.

Next, we consider a model in which the input alphabet $\mathcal{X} = \{0,1\}$ and the noise is Gaussian as follows:
\begin{align}
Y= \left\{ \begin{array}{cc}
X+Z & S=1\\
-X+Z & S=2
\end{array} \right.  \label{eqn:gauss_setup} .
\end{align}
% Under $S=1$,
%\begin{IEEEeqnarray}{rCl}
%	Y=X+Z,
%	\end{IEEEeqnarray}
Here, $Z\sim\mathcal{N}(0,\sigma^2)$ represents white Gaussian noise with variance $\sigma^2$. The channel in \eqref{eqn:gauss_setup} is analogous to the Binary Phase Shift Keying (BPSK) modulation scheme which is used widely in communication systems.  % and under $S=2$,
%\begin{IEEEeqnarray}{rCl}
%	Y=-X+Z.
%\end{IEEEeqnarray}
%where $\mathcal{X}=\{0,1\}$.

Notice that for this Gaussian setup,   communication at positive rates is not achievable since if the message $X=1$ is sent over the channel, the mean of the channel output at $S=1$ is different from that of $S=2$. Thus, an adversary observing the channel output can infer the channel state by simply estimating the mean of the output over a block of transmission. This is because when $S=2$, the channel flips the ``phase'' of the transmitted signal. Although, achieving positive rates of communication is not possible, masking the channel state is certainly feasible since the transmission of the symbol $X=0$ results in the same channel output distribution for both states (by symmetry of the model). 

For this setup, $\tilde{Q}_1$, $\tilde{Q}_2$ and $Q_0$ evaluate to $\mathcal{N}(1,\sigma^2)$, $\mathcal{N}(-1,\sigma^2)$ and $\mathcal{N}(0,\sigma^2)$, respectively.  Due to symmetry, the conditions~\eqref{op-cons2} and~\eqref{conv-cons1} of Theorem~\ref{opt-thm} are trivially satisfied in this case. The following corollary establishes the optimal throughput in this Gaussian setup.\medskip
\begin{corollary}\label{Gaus-cor} For the model in~\eqref{eqn:gauss_setup}, we have
\begin{align}
L_{\TV}^*(0^+,\delta)=\frac{\Phi^{-1}\left(\frac{1+\delta}{2}\right)}{\sqrt{2\left(\exp\left(\frac{1}{\sigma^2}\right)-\exp\left(-\frac{1}{\sigma^2}\right)\right)}}\cdot \frac{1}{\sigma^2}.\label{L_optimal-Gaus}
\end{align}
\end{corollary}
\begin{IEEEproof} See Appendix~\ref{Gaus-proof}.
	\end{IEEEproof}

\begin{figure}[t]
	\centering
	\includegraphics[width=.6\columnwidth]{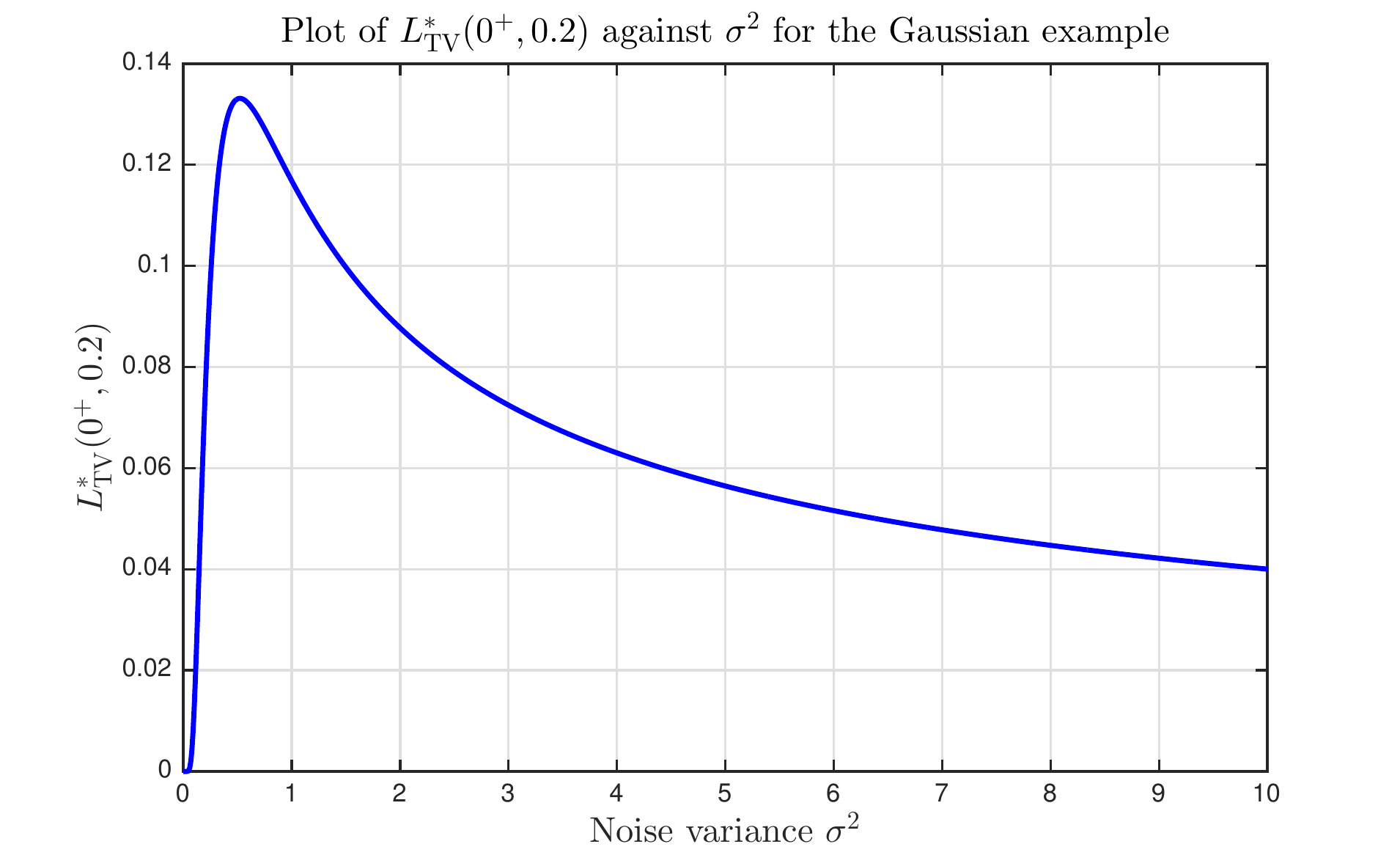}
	\caption{$L_{\TV}^*(0^+,0.2)$ for the Gaussian example}
	\label{figure4}
\end{figure}

For the proposed Gaussian model in~\eqref{eqn:gauss_setup}, Fig.~\ref{figure4} shows the optimal $L_{\TV}^*(0^+,0.2)$ plotted against $\sigma^2$. For large values of $\sigma^2$, $L^*_{\mathrm{TV}}(0^+,\delta)=\Theta\left(\frac{1}{\sigma}\right)$ and so the throughput vanishes as $\sigma$ tends to infinity.  \medskip

\subsection{Discussion}\label{subsec:concealing}
 
We now provide some remarks on the results and assumptions by an example.  In this discussion, for simplicity, we assume that the length of the key is infinite. 
\begin{figure}[t]
	\centering
	\includegraphics[scale=1.1]{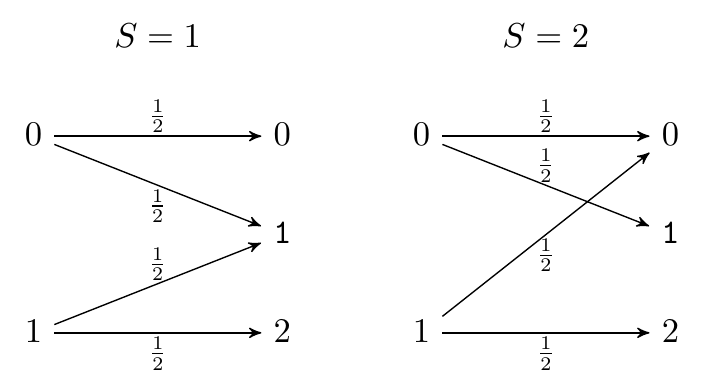}
	\caption{Example without the absolute continuity assumption}
		%	covert communication is not feasible by KL-divergence as the covertness metric.
	\label{figure5}
\end{figure} 

 Consider the example of Fig.~\ref{figure5}. The following observations are in order:
 \begin{itemize}
 \item As mentioned, the assumption of Remark~\ref{rem2} does not hold, because $W_1(2|1)\neq 0$, while $W_1(2|0)=0$ and also $W_2(2|1)\neq 0$, while $W_2(2|0)=0$. Thus our results from the previous Subsections~\ref{sec:optimal} and~\ref{sec:upper}, do not hold as the absolutely continuity assumption is violated.
 \item   Consider the achievable scheme sketched in Section~\ref{sec:optimal} with the choice $\mu_{1,n}=\mu_{2,n}\triangleq \mu_{n}$. As shown in Appendix~\ref{new-ex-proof}, under the total variation constraint for masking the state, if the number of messages $|\mathcal{M}|$ satisfies
\begin{align}
\log |\mathcal{M}| \leq \frac{\sqrt{\log \frac{1}{1-\delta}}}{2}\cdot \sqrt{n}\log n+o\left(\sqrt{n}\log n\right),\label{step3000}
\end{align}
then the desired thresholds on the maximum error probability and masking constraint are satisfied.
The above result shows that  it is possible to  communicate $\Omega(\sqrt{n}\log n)$ nats over $n$ channel uses while masking the state. %\footnote{A similar observation holds for protecting the transmission using TV distance instead of KL, in the sense that TV distance is away from zero, whenever any transmission occurred.}. 
 \item   As mentioned in the previous item, the communication rate is at least of the order of $\Omega\big(\frac{\log n}{\sqrt{n}}\big)$, which is somewhat surprising,  because the communication rate is usually of order $\Theta\big(\frac{1}{\sqrt{n}}\big)$ for covert communication problems~\cite{Bloch,Ligong}, as we have also observed in our results in the previous subsections. This again shows the importance of the absolute continuity assumption in Remark \ref{rem2}.
% \item Although we will not discuss the details, it is worth  mentioning that unlike the achievable scheme of the previous subsection, the choice $\mu_{1,n}=\mu_{2,n}$ is the unique choice which enables sending a positive number of \sadaf{nats} under the covert communication constraint.
 \end{itemize}

\section{Conclusion}\label{sec:conclusion}

In this paper, we considered  the fundamental limits of communicating over a compound channel in the presence of a masking constraint. In this setting, the constraint quantifies the ability of a malicious party in distinguishing the channel state. We provided inner and outer bounds to the optimal throughput-key region and showed that under the most general setting, the square-root law---just as in covert communications---holds for this setup. We identified regimes in which  the bounds match and in this case, the optimal throughput is established. Some examples, including that of a Gaussian setup, are provided to illustrate the bounds. 

A promising avenue for future work is to study the fundamental limits of communication over a channel with an i.i.d.\ state sequence $S^n$ available at  the encoder or the decoder or both. The distribution that generates $S^n$ is known to be either   $P_{S_1}$ or $P_{S_2}$ but otherwise it is unknown. The main goal in this {\em stochastic}---as opposed to {\em deterministic} in this paper---setting is to design a code to ensure that an adversary who is observing the channel output is unable to infer the channel state distribution, i.e.,  $P_{S_1}$ or $P_{S_2}$. The problem of covert communication over a channel with state (e.g., in the Gel'fand-Pinsker setting) has been previously studied in Lee {\em et al.}~\cite{Lee}. One can consider an alternative formulation to mask the distribution of state instead of the presence or absence of message transmission as was done in Lee {\em et al.}~\cite{Lee}.

\appendices

\sadaf{\section{Proof of Proposition~\ref{feas-prop}}\label{feas-prop-proof}
We prove the contrapositive -- that if there are no input distributions $P_1$ and $P_2$ that satisfy condition~\eqref{prop-cons}, the total variation distance between any  channel-induced output distributions $Q_1^n$ and $Q_2^n$ tends to one as $n\to\infty$. Thus, it cannot be bounded by any $\delta<1$. 	Let $\mathcal{G}_i$ be the set of all possible output distribution induced by one use of channel $W_i$. More precisely, we define the  sets $\mathcal{G}_1$ and $\mathcal{G}_2$ as follows:
\begin{IEEEeqnarray}{rCl}
	\mathcal{G}_s & \triangleq & \left\{Q_s\colon \sum_{x\in\mathcal{X}}P_s(x)W_s(\cdot |x)=Q_s(\cdot )\;\text{for some distribution}\; P_s\;\text{over}\;\mathcal{X}  \right\},\quad s\in\{1,2\}.
%	\\
%		\mathcal{G}_2 & \triangleq & \left\{Q_2 \colon \sum_{x\in\mathcal{X}}P_2(x)W_2(\cdot |x)=Q_2(\cdot )\;\text{for some distribution}\; P_2\;\text{over}\;\mathcal{X} \right\}.
	\end{IEEEeqnarray} In other words, $\mathcal{G}_s$ is the convex hull of $\{W_s(\cdot | x): x\in\mathcal{X}\}$. 
Since we have assumed that there are no distributions $P_1$ and $P_2$ which satisfy condition~\eqref{prop-cons}, then we have $\mathcal{G}_1\cap \mathcal{G}_2=\emptyset$.   We define the following parameter which is the maximum of the Bhattacharyya coefficient over the two sets of distributions $\mathcal{G}_1$ and $\mathcal{G}_2$:
\begin{IEEEeqnarray}{rCl}
	{F}(\mathcal{G}_1,\mathcal{G}_2)\triangleq \max_{Q_1\in \mathcal{G}_1, Q_2\in \mathcal{G}_2}F(Q_1,Q_2). \label{F-def}
	\end{IEEEeqnarray}
Since $\mathcal{G}_1\cap \mathcal{G}_2=\emptyset$ and  the output alphabet $\mathcal{Y}$ is finite, then there exists a positive $\delta'$ such that ${F}(\mathcal{G}_1,\mathcal{G}_2)\leq 1-\delta'$ (equivalently, there is some $\eta>0$ such that $d_{\TV}(Q_1,Q_2)>\eta$ for any $Q_1\in\mathcal{G}_1$ and $Q_2\in\mathcal{G}_2$). 

Now given the code ${\cal C}$, we again let $Q_s^n$ be the output distribution induced by ${\cal C}$ and $W_s$. Equivalently, $Q_s^n (y^n) = \sum_{x^n} \widetilde{P}_s^n(x^n)W_s(y^n|x^n)$ for some $\{\widetilde{P}_s^n(x^n): x^n\in {\cal X}^n\}$, which represents the code-induced {\em input} distribution.
First consider the    following equalities:
\begin{IEEEeqnarray}{rCl}
	Q_1(y_n|y^{n-1}) &=& \sum_{x^n}\frac{\widetilde{P}_1(x^n)W_1(y^n|x^n)}{Q_1^{n-1}(y^{n-1})} \\
	&=& \sum_{x_n}\left(\sum_{x^{n-1}}\frac{\widetilde{P}_1(x^{n-1})W_1(y^{n-1}|x^{n-1})}{Q_1^{n-1}(y^{n-1})} \widetilde{P}_1(x_n|x^{n-1})\right)W_1(y_n|x_n)\\
	&=&\sum_{x_n} \widetilde{P}_1(x_n|y^{n-1})W_1(y_n|x_n).
	\end{IEEEeqnarray}
Thus, for every $y^{n-1}$,  $Q_1(\cdot |y^{n-1})$ is induced by the (conditional) input distribution\footnote{The  distribution $\widetilde{P}_1(\cdot |y^{n-1})$ plays the role of $P_1$ in the definition of ${\cal G}_1$.} $\widetilde{P}_1(\cdot |y^{n-1})$  through the channel $W_1$. Hence, $Q_1(\cdot |y^{n-1})$  belongs to $\mathcal{G}_1$. Similarly,  for every $y^{n-1}$,  $Q_2(\cdot |y^{n-1})\in\mathcal{G}_2$. }

\sadaf{
To show that the total variation distance between any two possible output distributions $Q_1^n$ and $Q_2^n$ tends to one, it suffices to show that Bhattacharyya coefficient between them vanishes. So  consider %the Bhattacharyya coefficient between $Q_1^n$ and $Q_2^n$ as follows:
\begin{IEEEeqnarray}{rCl}
	F(Q_1^n,Q_2^n)&=&\sum_{y^n}\sqrt{Q_1^n(y^n)Q_2^n(y^n)}\\
&=&\sum_{y_n}\sqrt{Q_1(y_n|y^{n-1})Q_2(y_n|y^{n-1})}\sum_{y^{n-1}}\sqrt{Q_1^{n-1}(y^{n-1})Q_2^{n-1}(y^{n-1})}\label{step801}\\
&\leq &(1-\delta') F(Q_1^{n-1},Q_2^{n-1})\label{step800}%\\
%&\leq & (1-\delta')^n,
	\end{IEEEeqnarray}
where \eqref{step800} follows  for some $\delta'>0$ from the fact that  for every $y^{n-1}$,  $Q_1(\cdot |y^{n-1})\in \mathcal{G}_1$ and  $Q_2(\cdot |y^{n-1})\in \mathcal{G}_2$ and thus the first summation in \eqref{step801} is not larger than $1-\delta'$ by the definition of ${F}(\mathcal{G}_1,\mathcal{G}_2)$.

%To see this, consider the following sets of equalities:
%\begin{IEEEeqnarray}{rCl}
%	Q_1(y_n|y^{n-1}) &=& \sum_{x^n}\frac{P_1(x^n)W_1(y^n|x^n)}{Q_1^{n-1}(y^{n-1})} \\
%	&=& \sum_{x_n}\left(\sum_{x^{n-1}}\frac{P_1(x^{n-1})W_1(y^{n-1}|x^{n-1})}{Q_1^{n-1}(y^{n-1})} P_1(x_n|x^{n-1})\right)W_1(y_n|x_n)\\
%	&=&\sum_{x_n} P_1(x_n|y^{n-1})W_1(y_n|x_n)
%	\end{IEEEeqnarray}
%Thus $Q_1(.|y^{n-1})$ is induced by the input distribution $P_1(.|y^{n-1})$, hence it belongs to $\mathcal{G}_1$.
	
Finally, the inductive use of \eqref{step800} implies that $F(Q_1^n,Q_2^n)\leq (1-\delta')^n$, so $F(Q_1^n,Q_2^n)$ vanishes as desired.\footnote{A similar argument can be used to show that the relative entropy between any output distributions $Q_1^n$ and $Q_2^n$ grows (at least) linearly with $n$. } }

\section{Proof of Lemma \ref{lem-ach}}\label{lem-ach-proof}

First, we present the following theorem, called Berry-Esseen Theorem, which will be used repeatedly in the proofs, later. 
\begin{theorem}[Thm~1.6 in \cite{Vincent-book}]\label{BE-thm} Let $X^n\triangleq (X_1,\ldots,X_n)$ be a collection of $n$ independent random variables where each random variable has a zero mean, variance $\sigma_i^2\triangleq \mathbb{E}\left[X_i^2\right]>0$ and third absolute moment $T_i\triangleq \mathbb{E}\left[|X_i|^3\right]<\infty$. Define the average variance and average third absolute moment as $\sigma^2\triangleq \frac{1}{n}\sum_{i=1}^n\sigma_i^2$ and $T\triangleq \frac{1}{n}\sum_{i=1}^nT_i$, respectively. Then, we have:
	\begin{IEEEeqnarray}{rCl}
		\sup_{a\in\mathbb{R}}\Bigg|\mathbb{P}\left(\frac{1}{\sigma\sqrt{n}}\sum_{i=1}^nX_i<a \right)-\Phi(a)\Bigg|\leq \frac{6T}{\sigma^3\sqrt{n}}.
	\end{IEEEeqnarray}
\end{theorem}
Now, consider the following identity for the total variation distance:
\begin{IEEEeqnarray}{rCl}
	\hspace{-0.5cm}	d_{\TV}\left(Q_1^{\times n},Q_2^{\times n}\right)&= &\mathbb{P}_{Q_1^{\times n}}\left(\log \frac{Q_1^{\times n}(Y^n)}{Q_2^{\times n}(Y^n)} \geq 0\right)-\mathbb{P}_{Q_2^{\times n}}\left(\log \frac{Q_1^{\times n}(Y^n)}{Q_2^{\times n}(Y^n)} \geq 0\right)\\[2ex]
	&= &\mathbb{P}_{Q_1^{\times n}}\left(\sum_{i=1}^n\log\frac{Q_{1,n}(Y_i)}{Q_{2,n}(Y_i)}\geq 0\right)-\mathbb{P}_{Q_2^{\times n}}\left(\sum_{i=1}^n\log\frac{Q_{1,n}(Y_i)}{Q_{2,n}(Y_i)}\geq 0\right).
	\label{step18}
\end{IEEEeqnarray}
Now, we bound each of the probabilities in \eqref{step18}. First, we analyze the first probability. The mean and variance of  $\sum_{i=1}^n\log\frac{Q_{1,n}(Y_i)}{Q_{2,n}(Y_i)}$ are given by the following:
\begin{IEEEeqnarray}{rCl}
	&&\mathbb{E}\left[\sum_{i=1}^n\log\frac{Q_{1,n}(Y_i)}{Q_{2,n}(Y_i)}\right]=nD(Q_{1,n} \| Q_{2,n})\triangleq n\text{d}_1,\end{IEEEeqnarray}
and
\begin{IEEEeqnarray}{rCl}
	\mathbb{V}\left[\sum_{i=1}^n\log\frac{Q_{1,n}(Y_i)}{Q_{2,n}(Y_i)}\right]&=&n \left(\sum_yQ_{1,n}(y)\left(\log\frac{Q_{1,n}(y)}{Q_{2,n}(y)}\right)^2-D^2(Q_{1,n} \| Q_{2,n})\right)\\
	&\triangleq & n\text{v}_1.\label{step38}
\end{IEEEeqnarray}
Employing  Theorem~\ref{BE-thm}, we obtain
\begin{IEEEeqnarray}{rCl}
	\mathbb{P}_{Q_1^{\times n}}\left(\sum_{i=1}^n\log\frac{Q_{1,n}(Y_i)}{Q_{2,n}(Y_i)}\geq 0\right) \leq \Phi\left(\frac{\sqrt{n}\text{d}_1}{\sqrt{\text{v}_1}}\right)+\frac{6\text{t}_1}{\sqrt{n}\text{v}_1^{3/2}},\label{step19}
\end{IEEEeqnarray}
where
\begin{align}
\text{t}_1\triangleq \mathbb{E}_{Q_{1,n}}\bigg[\Big|\log \frac{Q_{1,n}(Y)}{Q_{2,n}(Y)}-D(Q_{1,n}\|Q_{2,n})\Big|^3\bigg].
\end{align}
Similarly, we get:
\begin{IEEEeqnarray}{rCl}
	\mathbb{P}_{Q_2^{\times n}}\left(\sum_{i=1}^n\log\frac{Q_{1,n}(Y_i)}{Q_{2,n}(Y_i)}\geq 0\right) \geq 1-\Phi\left(\frac{\sqrt{n}\text{d}_2}{\sqrt{\text{v}_2}}\right)-\frac{6\text{t}_2}{\sqrt{n}\text{v}_2^{3/2}},\nonumber\\
	\label{step20}
\end{IEEEeqnarray}
where we define:
\begin{IEEEeqnarray}{rCl}
	\text{d}_2&\triangleq & D(Q_{2,n}\|Q_{1,n}),\\
	\text{v}_2 &\triangleq & \sum_yQ_{2,n}(y)\left(\log\frac{Q_{1,n}(y)}{Q_{2,n}(y)}\right)^2-\big(D(Q_{2,n} \| Q_{1,n})\big)^2,\\
	\text{t}_2&\triangleq &\mathbb{E}_{Q_{2,n}}\bigg[\Big|\log \frac{Q_{1,n}(Y)}{Q_{2,n}(Y)}+D(Q_{2,n}\|Q_{1,n})\Big|^3\bigg].
\end{IEEEeqnarray}
Now, consider the following set of approximations:
\begin{IEEEeqnarray}{rCl}
	&&\hspace{-0.3cm}\text{d}_1 = D(Q_{1,n}\|Q_{2,n})\\
	&=&\sum_y Q_{1,n}(y)\log \frac{Q_{1,n}(y)}{Q_{2,n}(y)}\\
	&=&\mathbb{E}_{Q_{2,n}}\left[\frac{Q_{1,n}}{Q_{2,n}}\log \frac{Q_{1,n}}{Q_{2,n}}\right]\\
	&=&\mathbb{E}_{Q_{2,n}}\left[\frac{(1-\mu_{1,n})Q_0+\mu_{1,n}\bar{Q}_1}{(1-\mu_{2,n})Q_0+\mu_{2,n}\bar{Q}_2}\log \frac{(1-\mu_{1,n})Q_0+\mu_{1,n}\bar{Q}_1}{(1-\mu_{2,n})Q_0+\mu_{2,n}\bar{Q}_2}\right]\\
	&=&\mathbb{E}_{Q_{2,n}}\left[\left(1+\frac{\mu_{1,n}\big(\bar{Q}_1-Q_0\big)-\mu_{2,n} \big(\bar{Q}_{2}-Q_0\big)}{(1-\mu_{2,n})Q_0+\mu_{2,n}\bar{Q}_2}\right)\log \left(1+\frac{\mu_{1,n}\big(\bar{Q}_1-Q_0\big)-\mu_{2,n} \big(\bar{Q}_{2}-Q_0\big)}{(1-\mu_{2,n})Q_0+\mu_{2,n}\bar{Q}_2}\right)\right]\label{step24}\\
		&=&\mathbb{E}_{Q_{2,n}}\Bigg[\left(1+\frac{\mu_{1,n}\big(\bar{Q}_1-Q_0\big)-\mu_{2,n} \big(\bar{Q}_{2}-Q_0\big)}{(1-\mu_{2,n})Q_0+\mu_{2,n}\bar{Q}_2}\right) \nonumber\\&&\hspace{.5cm}\times \left(\frac{\mu_{1,n}\big(\bar{Q}_1-Q_0\big)-\mu_{2,n} \big(\bar{Q}_{2}-Q_0\big)}{(1-\mu_{2,n})Q_0+\mu_{2,n}\bar{Q}_2}-\frac{1}{2}\left(\frac{\mu_{1,n}\big(\bar{Q}_1-Q_0\big)-\mu_{2,n} \big(\bar{Q}_{2}-Q_0\big)}{(1-\mu_{2,n})Q_0+\mu_{2,n}\bar{Q}_2}\right)^2\right)\Bigg]+o\left(\frac{1}{n}\right)\\
	&=&\frac{1}{2} \mathbb{E}_{Q_{2,n}}\left[\left(\frac{\mu_{1,n}\big(\bar{Q}_1-Q_0\big)-\mu_{2,n} \big(\bar{Q}_{2}-Q_0\big)}{(1-\mu_{2,n})Q_0+\mu_{2,n}\bar{Q}_{2}}\right)^2\right]+o\left(\frac{1}{n}\right)\\[2ex]
	&=&\frac{1}{2} \mathbb{E}_{Q_0}\left[\left(\frac{\mu_{1,n}\big(\bar{Q}_1-Q_0\big)-\mu_{2,n} \big(\bar{Q}_{2}-Q_0\big)}{Q_0}\right)^2\right]+o\left(\frac{1}{n}\right)\label{step25b}\\[1ex]
	&=&\frac{1}{2}\mu_{1,n}^2\chi_2\left(\bar{Q}_1\|Q_0\right)+\frac{1}{2}\mu_{2,n}^2\chi_2\left(\bar{Q}_2\|Q_0\right)-\mu_{1,n}\mu_{2,n}\rho\left(\bar{Q}_1,\bar{Q}_2,Q_0\right)+o\left(\frac{1}{n}\right),\label{step21}
\end{IEEEeqnarray}
where \eqref{step24} follows from the Taylor expansion $\log(1+x)=x-\frac{1}{2}x^2+o(x^2)$ at $x=0$; \eqref{step25b} follows because \begin{align}Q_{2,n}=(1-\mu_{2,n})Q_0+\mu_{2,n}\bar{Q}_{2}\to Q_0.\end{align}
Now, define:
\begin{IEEEeqnarray}{rCl}
	\bar{\Omega}\left(\mu_{1,n},\mu_{2,n}\right)&\triangleq& \mu_{1,n}^2\chi_2\left(\bar{Q}_1\|Q_0\right)+\mu_{2,n}^2\chi_2\left(\bar{Q}_2\|Q_0\right)-2\mu_{1,n}\mu_{2,n}\rho\left(\bar{Q}_1,\bar{Q}_2,Q_0\right).
\end{IEEEeqnarray}
The equality \eqref{step21} can be expressed in terms of $\bar{\Omega}\left(\mu_{1,n},\mu_{2,n}\right)$ as the following:
\begin{IEEEeqnarray}{rCl}
	\text{d}_1=\frac{1}{2}\bar{\Omega}\left(\mu_{1,n},\mu_{2,n}\right)+o\left(\frac{1}{n}\right).
\end{IEEEeqnarray}
Next, we consider the following set of approximations:
\begin{IEEEeqnarray}{rCl}
	\sum_yQ_1(y)\left(\log\frac{Q_1(y)}{Q_2(y)}\right)^2
	&=&\sum_y \left((1-\mu_{1,n})Q_0(y)+\mu_{1,n}\bar{Q}_{1}(y)\right)\left(\log \frac{(1-\mu_{1,n})Q_0(y)+\mu_{1,n}\bar{Q}_{1}(y)}{(1-\mu_{2,n})Q_0(y)+\mu_{2,n}\bar{Q}_{2}(y)}\right)^2\\[1.5ex]
	&=&\sum_y \left((1-\mu_{1,n})Q_0(y)+\mu_{1,n}\bar{Q}_{1}(y)\right)\left(\log \left(1+\frac{\mu_{1,n}\big(\bar{Q}_1-Q_0 \big)-\mu_{2,n} \big(\bar{Q}_{2}-Q_0\big)}{(1-\mu_{2,n})Q_0(y)+\mu_{2,n}\bar{Q}_{2}(y)}\right)\right)^2\\[1.5ex]
	&=&\sum_y \left((1-\mu_{1,n})Q_0(y)+\mu_{1,n}\bar{Q}_{1}(y)\right)\left(\frac{\mu_{1,n}\big(\bar{Q}_1-Q_0 \big)-\mu_{2,n} \big(\bar{Q}_{2}-Q_0\big)}{(1-\mu_{2,n})Q_0(y)+\mu_{2,n}\bar{Q}_{2}(y)}\right)^2+o\left(\frac{1}{n}\right)\label{step37b}\\
	&=&\sum_y Q_0(y)\left(\frac{\mu_{1,n}\big(\bar{Q}_1-Q_0 \big)-\mu_{2,n} \big(\bar{Q}_{2}-Q_0\big)}{Q_0(y)}\right)^2+o\left(\frac{1}{n}\right)\label{step37}\\[1.5ex]
	&=& \mathbb{E}_{Q_0}\left[\left(\frac{\mu_{1,n}\big(\bar{Q}_1-Q_0 \big)-\mu_{2,n} \big(\bar{Q}_{2}-Q_0\big)}{Q_0}\right)^2\right]+o\left(\frac{1}{n}\right)\\[1.5ex]
	&=&\mu_{1,n}^2\chi_2\left(\bar{Q}_1\|Q_0\right)+\mu_{2,n}^2\chi_2\left(\bar{Q}_2\|Q_0\right)-2\mu_{1,n}\mu_{2,n}\rho\left(\bar{Q}_1,\bar{Q}_2,Q_0\right)+o\left(\frac{1}{n}\right)\\[1.5ex]
	&=&\bar{\Omega}\left(\mu_{1,n},\mu_{2,n}\right)+o\left(\frac{1}{n}\right),\label{step22}
\end{IEEEeqnarray}
where \eqref{step37b} follows from Taylor expansion $\log(1+x)=x+o(x)$ as $x\to 0$, \eqref{step37} follows because $$(1-\mu_{s,n})Q_0(y)+\mu_{s,n}\bar{Q}_{s}(y)\to Q_0(y),$$
as $\mu_{s,n}\to 0$.
Combining \eqref{step19}, \eqref{step21}, \eqref{step22} and definition of $\text{v}_1$ in \eqref{step38}, we have: 
\begin{align}
\Phi\left(\frac{\sqrt{n}\text{d}_1}{\sqrt{\text{v}_1}}\right)&= \Phi\left(\frac{1}{2}\sqrt{n\bar{\Omega}\left(\mu_{1,n},\mu_{2,n}\right)}\right)+o\left(\frac{1}{n}\right).\label{step39}
\end{align}
Similarly, 
\begin{align}
\Phi\left(\frac{\sqrt{n}\text{d}_2}{\sqrt{\text{v}_2}}\right)&= \Phi\left(\frac{1}{2}\sqrt{n\bar{\Omega}\left(\mu_{1,n},\mu_{2,n}\right)}\right)+o\left(\frac{1}{n}\right).\label{step40}
\end{align}
Following similar steps leading to \eqref{step22}, one can show that $\text{t}_s=O\big(n^{-3/2}\big)$.
Combining \eqref{step18}, \eqref{step19}, \eqref{step20}, \eqref{step39} and \eqref{step40}, and considering the fact that $\text{v}_s=\Theta\left(1/n\right)$, we obtain the following:
\begin{IEEEeqnarray}{rCl}
	d_{\TV}\left(Q_1^{\times n},Q_2^{\times n}\right) &\leq& \Phi\left(\frac{\sqrt{n}\text{d}_1}{\sqrt{\text{v}_1}}\right)+\Phi\left(\frac{\sqrt{n}\text{d}_2}{\sqrt{\text{v}_2}}\right)-1+\frac{6\text{t}_1}{\sqrt{n}\text{v}_1^{3/2}}+\frac{6\text{t}_2}{\sqrt{n}\text{v}_2^{3/2}} \label{eqn:81}
	\\&\leq& 2\Phi\left(\frac{1}{2}\sqrt{n\bar{\Omega}\left(\mu_{1,n},\mu_{2,n}\right)}\right)-1+O\left(\frac{1}{\sqrt{n}}\right).\label{step25}
\end{IEEEeqnarray}
Considering the fact that $\gamma_s=\sqrt{n}\mu_{s,n}$, completes the proof of the lemma.

\section{Proof of Lemma~\ref{lem-key}}\label{lem-key-proof}

First, we state the following lemma.
\begin{lemma} [Corollary VII.2 in \cite{Cuff}] For all positive $\eta_s$, we have:
	\begin{IEEEeqnarray}{rCl}
		\mathbb{E}_{\mathcal{C}_s}\left[d_{\TV}\left(Q_s^n,Q_s^{\times n}\right)\right]\leq \frac{1}{2}\sqrt{\frac{\exp(\eta_s)}{|\mathcal{M}|\cdot |\mathcal{K}|}}+%\frac{1}{2}
		\mathbb{P}_{W_s^nP_s^{\times n}}\left( \sum_{i=1}^n\log \frac{W_s(Y_i|X_i)}{Q_{s,n}(Y_i)} \geq \eta_s \right).\label{lem-exp}
	\end{IEEEeqnarray}
\end{lemma}
Choose 
\begin{IEEEeqnarray}{rCl}
	\eta_s = n\left(1+\frac{\kappa}{2}\right)I(P_{s,n},W_s),
	\end{IEEEeqnarray}
for some positive $\kappa$.
Notice that from \cite{Ligong}, the following approximations hold:
\begin{IEEEeqnarray}{rCl}
	I(P_s,W_s)&=&\frac{1}{\sqrt{n}}\gamma_s
	D(W_s \| Q_0|\bar{P}_s)
	%D(\bar{Q}_s\|Q_0)
	+o\left(\frac{1}{\sqrt{n}}\right)\\
	&\triangleq &\frac{\gamma_s}{\sqrt{n}}\bar{\mathbb{D}}_s+o\left(\frac{1}{\sqrt{n}}\right),\label{step1000}
	\end{IEEEeqnarray}
Thus, 
\begin{IEEEeqnarray}{rCl}
	\eta_s = \gamma_s\sqrt{n}\left(1+\frac{\kappa}{2}\right)\bar{\mathbb{D}}_s+o(\sqrt{n}).
	\end{IEEEeqnarray}
Also, we have
\begin{IEEEeqnarray}{rCl}
	\sum_{x}\sum_y P_{s,n}(x)W_s(y|x)\left(\log \frac{W_s(y|x)}{Q_{s,n}(y)}\right)^2
	&=& \frac{\gamma_s}{\sqrt{n}}\sum_{x\neq 0}\sum_y \bar{P}_{s}(x)W_s(y|x)\left(\log \frac{W_s(y|x)}{Q_0(y)}\right)^2+o\left( \frac{1}{\sqrt{n}}\right)\\*
	%&=& \frac{\gamma_s}{\sqrt{n}}\sum_y \bar{Q}_{s}(y)\left(\log \frac{\bar{Q}_s(y)}{Q_0(y)}\right)^2+O\left( \frac{1}{\sqrt{n}}\right)\\
	&\triangleq & \frac{\gamma_s}{\sqrt{n}}\bar{\mathbb{V}}_s+o\left( \frac{1}{\sqrt{n}}\right).\label{step1001}
	\end{IEEEeqnarray}
Now, consider the probability term in \eqref{lem-exp} as follows:

\begin{IEEEeqnarray}{rCl}
	&&\hspace{-1cm}\mathbb{P}_{W_s^nP_s^{\times n}}\left(\sum_{i=1}^n\log \frac{W_s(Y_i|X_i)}{Q_{s,n}(Y_i)} \geq \eta_s \right) \\
	&=&\mathbb{P}_{W_s^nP_s^{\times n}}\left(\sum_{i=1}^n\log \frac{W_s(Y_i|X_i)}{Q_{s,n}(Y_i)} \geq n(1+\frac{\kappa}{2})I(P_{s,n},W_s) \right)\\
	&=&\mathbb{P}_{W_s^nP_s^{\times n}}\left( \sum_{i=1}^n\log \frac{W_s(Y_i|X_i)}{Q_{s,n}(Y_i)}-nI(P_{s,n},W_s) \geq \frac{n\kappa}{2} I(P_{s,n},W_s) \right)\\
	&=&\mathbb{P}_{W_s^nP_s^{\times n}}\left(\sum_{i=1}^n\log \frac{W_s(Y_i|X_i)}{Q_{s,n}(Y_i)}-nI(P_{s,n},W_s) \geq \frac{\sqrt{n}\kappa \gamma_s\bar{\mathbb{D}}_s}{2}+o(\sqrt{n}) \right) \\
	&\leq & \mathbb{P}_{W_s^nP_s^{\times n}}\left( \Big|\sum_{i=1}^n\log \frac{W_s(Y_i|X_i)}{Q_{s,n}(Y_i)}-nI(P_{s,n},W_s)\Big| \geq \frac{\sqrt{n}\kappa \gamma_s\bar{\mathbb{D}}_s}{2}+o(\sqrt{n}) \right)\\
	&\leq & \frac{\sqrt{n}\gamma_s\bar{\mathbb{V}}_s}{4n\kappa^2\gamma_s^2\bar{\mathbb{D}}_s^2}+o\left(\frac{1}{\sqrt{n}}\right)\\&=& O\left(\frac{1}{\sqrt{n}}\right),
	\label{ineq2}
\end{IEEEeqnarray}
where the last inequality follows from Chebyshev's inequality and the fact that 
\begin{IEEEeqnarray}{rCl}
	\mathbb{V}\left[\sum_{i=1}^n\log \frac{W_s(Y_i|X_i)}{Q_{s,n}(Y_i)}-nI(P_{s,n},W_s)\right]&\leq & n\left(	\sum_{x}\sum_y P_{s,n}(x)W_s(y|x)\left(\log \frac{W_s(y|x)}{Q_{s,n}(y)}\right)^2\right)\\
	&\leq & \sqrt{n}\gamma_s\bar{\mathbb{V}}_s+o(\sqrt{n}). \label{step200}
	\end{IEEEeqnarray}
Combining \eqref{lem-exp} and \eqref{ineq2}, we get:
\begin{IEEEeqnarray}{rCl}
	\mathbb{E}_{\mathcal{C}_s}\left[d_{\TV}\left(Q_s^n,Q_s^{\times n}\right)\right] 
	&\leq &\frac{1}{2}\sqrt{\frac{\exp(\eta_s)}{|\mathcal{M}|\cdot |\mathcal{K}|}}+O\left(\frac{1}{\sqrt{n}}\right).\label{total-distance}
\end{IEEEeqnarray}
Now, we choose:
\begin{IEEEeqnarray}{rCl}
	\log|\mathcal{M}|+\log|\mathcal{K}|&\geq& \gamma_s\sqrt{n}\left(1+{\kappa}\right)\bar{\mathbb{D}}_s.\label{ineq600}
\end{IEEEeqnarray}
With the above choice, we can further upper bound \eqref{total-distance} as follows:
\begin{IEEEeqnarray}{rCl}
	\mathbb{E}_{\mathcal{C}_s}\left[d_{\TV}\left(Q_s^n,Q_s^{\times n}\right)\right] 
	&= & O\left(\frac{1}{\sqrt{n}}\right).
	\label{total-distance2}
\end{IEEEeqnarray}
This completes the proof of the lemma.

\section{Proof of Lemma~\ref{PPM-thm}}\label{rate-analysis}

We provide the analysis of the rate constraint in the following. Since the ML decoder results in the smallest probability of error for a given codebook, we can bound the average probability of error using Shannon's achievability bound \cite[Theorem~2]{Yury} as follows. For every $s\in\{1,2\}$ and $k\in\mathcal{K}$, the average error probability satisfies:
\begin{IEEEeqnarray}{rCl}
	\mathbb{E}\left[\bar{P}_{\mathrm{e}}(s,k)\right] \leq \mathbb{P}\left( \log \frac{W_s(Y^n|X^n)}{Q_s^{\times n}(Y^n)}\leq \log |\mathcal{M}|+\frac{1}{6}\log n \right)+n^{-1/6}.\label{Pe}
	\end{IEEEeqnarray}
Recall the definition of $\bar{\mathbb{V}}_s$ from \eqref{step1001} and consider the above probability:
\begin{IEEEeqnarray}{rCl}
	\mathbb{P}\left( \log \frac{W_s(Y^n|X^n)}{Q_s^{\times n}(Y^n)}\leq \log |\mathcal{M}|+\frac{1}{6}\log n \right)&\leq &	\mathbb{P}\left(\log \frac{W_s(Y^n|X^n)}{Q_s^{\times n}(Y^n)}\leq  nI(P_{s,n},W_s)-n^{1/3}+\frac{1}{6}\log n \right) \\
	&=&\mathbb{P}\left( \log \frac{W_s(Y^n|X^n)}{Q_s^{\times n}(Y^n)}-nI(P_{s,n},W_s)\leq  -n^{1/3}+\frac{1}{6}\log n \right)\\
	&\leq & \mathbb{P}\left(\Big|\log \frac{W_s(Y^n|X^n)}{Q_s^{\times n}(Y^n)}-nI(P_{s,n},W_s)\Big|\geq  n^{1/3}-\frac{1}{6}\log n\right) \\
	&\leq & \frac{\sqrt{n}\gamma_s\bar{\mathbb{V}}_s}{(n^{1/3}-\frac{1}{6}\log n)^2}\\
	&=& O\big(n^{-1/6}\big),
	\end{IEEEeqnarray}
where the last inequality follows from Chebyshev's inequality and steps similar to those leading to \eqref{step200}.  This completes the proof of the lemma.

\section{Proof of Existence of a Codebook With Desired Properties}\label{existence2-proof}
First, we state the following theorem which will be used later in the proof.
\begin{theorem}[McDiarmid's theorem\cite{Mc}]\label{Mc-thm} Let $\{X_k\}_{k=1}^n$ be independent random variables defined on the set $\mathcal{X}$. Consider a random variable $U=f(X^n)$ where $f$ is a function satisfying the following bounded difference property:
	\begin{IEEEeqnarray}{rCl}
		\sup_{x_1,\ldots,x_n,x'_i\in\mathcal{X}}f(x_1,\ldots,x_i,\ldots,x_n)-f(x_1,\ldots,x'_i,\ldots,x_n)\leq d_i,\qquad \forall i\in\{1,\ldots,n\},\label{eqn:bnd-diff}
	\end{IEEEeqnarray}
	for some positive numbers $d_i, 1\le i\le n$. Then, for every $r>0$, 
	\begin{IEEEeqnarray}{rCl}
		\mathbb{P}\left(U-\mathbb{E}\left[U\right]>r\right)\leq \exp\left(\frac{-2r^2}{\sum_{i=1}^n d_i^2}\right).
	\end{IEEEeqnarray}
\end{theorem}

\underline{\textit{Step 1: Probability of Satisfying the Masking Constraint}}:

Choose sequences $D_n=(\log{n})^{-1/2}$ and 
%any sufficiently slowly vanishing D_n works here!
 $a= n^{-{1}/{12}}$  and define the following events:
\begin{IEEEeqnarray}{rCl}
	\mathcal{B}_s&\triangleq& \Bigg\{ \forall k\in\mathcal{K},\; \exists \mathcal{M}_{s,k}\subset \mathcal{M},\; \mathcal{M}_s=\mathcal{M}_{s,1}\times \ldots\times \mathcal{M}_{s,k}\colon  |\mathcal{M}_{s,k}|=(1-a)|\mathcal{M}|,\;\; \nonumber\\[-1ex]
	&&\hspace{2.5cm}Q_{\mathcal{M}_s}^n(y^n)\triangleq \frac{1}{(1-a)|\mathcal{M} \| \mathcal{K}|}\;\sum_{k\in\mathcal{K}}\;\sum_{m\in\mathcal{M}_{s,k}}W_s^n(y^n|X_s^n(m,k)),\;\; d_{\TV}(Q_{\mathcal{M}_s}^n,Q_{s}^{\times n})> D_n \Bigg\},\label{event-second}
\end{IEEEeqnarray}
for $s=1,2$.

Consider the probability of this event:
\begin{IEEEeqnarray}{rCl}
	\mathbb{P}\left(\mathcal{B}_s\right) &=& \mathbb{P}\left(\forall k\in\mathcal{K}, \;\exists \mathcal{M}_{s,k}\subset \mathcal{M}\colon |\mathcal{M}_{s,k}|=(1-a)|\mathcal{M}|,\;\;d_{\TV}(Q_{\mathcal{M}_s}^n,Q_{s}^{\times n})> D_n \right)\\
	&\le & \mathbb{P}\left(\forall k\in\mathcal{K},\;\exists \mathcal{M}_{s,k}\subset \mathcal{M}\colon |\mathcal{M}_{s,k}|=(1-a)|\mathcal{M}|,\;\;d_{\TV}(Q_{\mathcal{M}_s}^n,Q_{s}^{\times n})-\mathbb{E}\left[d_{\TV}(Q_{\mathcal{M}_s}^n,Q_{s}^{\times n})\right] >\frac{D_n}{2}\right) \label{eqn:B_s1}\\
	%&\stackrel{(b)}{=}& \left(\mathbb{P}\left(\exists \mathcal{M}_{s,1}\subset \mathcal{M}\colon |\mathcal{M}_{s,1}|=(1-a_s)|\mathcal{M}|,\;\;d_{\TV}(Q_{\mathcal{M}_s}^n,Q_{s}^{\times n})> \mathbb{E}\left[d_{\TV}(Q_{\mathcal{M}_s}^n,Q_{s}^{\times n})\right] \right)\right)^{|\mathcal{K}|}\\
	&\le &% \left(
	 \sum_{\left(\mathcal{M}_{s,k}\right)_{k\in\mathcal{K}}\colon\atop |\mathcal{M}_{s,k}|=(1-a)|\mathcal{M}|}\mathbb{P}\left(d_{\TV}(Q_{\mathcal{M}_s}^n,Q_{s}^{\times n})- \mathbb{E}\left[d_{\TV}(Q_{\mathcal{M}_{s}}^n,Q_{s}^{\times n})\right]>\frac{D_n}{2}\right)%\right)%^{|\mathcal{K}|}
	,\label{step600}
\end{IEEEeqnarray}
where~\eqref{eqn:B_s1} follows because according to \eqref{exp:dTV}, the expectation of the total variation distance is upper bounded by $O(n^{-\frac{1}{2}})\le\frac{D_n}{2}$ for sufficiently large $n$, because 
\begin{equation}
\log |\mathcal{M}_s|+\log |\mathcal{K}|=\log |\mathcal{M}|+\log |\mathcal{K}|+\log(1-a)\ge  (1+\kappa')\sqrt{n} \cdot \max_{s\in\{1,2\}}\gamma_s
	D(W_s \| Q_0|\bar{P}_s),
\end{equation}
for some $\kappa'>0$ that approaches $\kappa$ as $n\to\infty$,
%$(b)$ follows
% because
%\reconsider{ from the symmetry and independence  in the code construction}
 %, we can set $k=1$
and  \eqref{step600} follows from the union bound.
%\end{itemize}

 In order to upper bound \eqref{step600}, we use the following lemma due to Liu  {\em et  al.} \cite{jingbo} (see also \cite{Mehrdad}) showing that the total variation distance of a random codebook is concentrated around its expectation. 
 \begin{lemma}[Theorem~31 in \cite{jingbo}]\label{le:jingbo}
 Consider a channel with the probability transition $W$ and an input probability distribution $P_X$. Suppose that $P_Y$ is the output distribution of the channel with the input distribution $P_X$. Let $\mathcal{C}=\{X_1,\ldots,X_{\mathsf{M}}\}$ be a random codebook with i.i.d. codewords drawn from $P_X$ and $\hat{P}_Y(\cdot )=\frac{1}{\mathsf{M}}\sum_{m=1}^\mathsf{M} W(\cdot |X_m)$ be the induced output distribution by $\mathcal{C}$. Then for any $\Delta>0$, we have
 \begin{align}
 \mathbb{P}\left(d_{\TV}(\hat{P}_Y,P_Y)-\mathbb{E}[(d_{\TV}(\hat{P}_Y,P_Y)]>\Delta\right)\le \exp\left(-2\mathsf{M}\Delta^2\right).\label{eqn:jingbo}
 \end{align}  
 \end{lemma}

 Using the above lemma with the following identifications,
 \begin{align}
 \mathcal{C}&\leftarrow \mathcal{C}_s=\bigcup_{k\in |\mathcal{K}|}\{X^n_{s}(m,k):m\in\mathcal{M}_{s,k}\},\\
 \mathsf{M}&\leftarrow (1-a)|\mathcal{M}||\mathcal{K}|,\\
 Y&\leftarrow Y^n\\
 W&\leftarrow W^n_{s},\\
 \hat{P}_{Y}&\leftarrow Q_{\mathcal{M}_s}^n,\\
 P_Y&\leftarrow Q_{s}^{\times n},\\
 \Delta&\leftarrow \frac{D_n}{2},
 \end{align}
 we get the following upper bound on
the term \eqref{step600} as follows: 
\begin{IEEEeqnarray}{rCl}
	&&\sum_{\mathcal{M}_{s,k}\subset \mathcal{M}, k\in\mathcal{K}\colon\atop |\mathcal{M}_{s,k}|=(1-a)|\mathcal{M}|}\mathbb{P}\left(d_{\TV}(Q_{\mathcal{M}_s}^n,Q_{s}^{\times n})- \mathbb{E}\left[d_{\TV}(Q_{\mathcal{M}_{s}}^n,Q_{s}^{\times n})\right]>\frac{D_n}{2}\right) \nonumber\\&&\hspace{1cm}\leq \sum_{\mathcal{M}_{s,k}\subset \mathcal{M}, k\in\mathcal{K}\colon\atop |\mathcal{M}_{s,k}|=(1-a)|\mathcal{M}|}\exp\left(-\frac{1}{2}D_n^2(1-a)|\mathcal{M}\| \mathcal{K}|\right)\\
	&&\hspace{1cm}={|\mathcal{M}| \choose (1-a)|\mathcal{M}|}^{|\mathcal{K}|}\exp\left(-\frac{1}{2}(1-a)D_n^2|\mathcal{M} \| \mathcal{K}|\right)\\
	&&\hspace{1cm}\stackrel{(d)}{\leq} \exp\left(-|\mathcal{M} \| \mathcal{K}|\left(\frac{1}{2}(1-a)D_n^2-a\log\frac{e}{a}\right)\right),\nonumber\\
\end{IEEEeqnarray}
where $(d)$ follows because 
\begin{IEEEeqnarray}{rCl}
	{|\mathcal{M}| \choose (1-a)|\mathcal{M}|}
	&\leq & \left(\frac{e}{a}\right)^{a|\mathcal{M}|},
	\end{IEEEeqnarray}
where the last inequality follows from \cite{Link2}. Thus, in summary, we get:
\begin{IEEEeqnarray}{rCl}
	\mathbb{P}\left(\mathcal{B}_1\cup\mathcal{B}_2\right) &\leq& 2\exp\left(-|\mathcal{M} \| \mathcal{K}|\left(\frac{1}{2}(1-a)D_n^2-a\log\frac{e}{a}\right)\right)\\
	%&=&\delta_{0,n}
	&\le& \exp\left(-|\mathcal{M} \| \mathcal{K}|\frac{1}{4\log n}\right)\rightarrow 0.
	\label{event0}
\end{IEEEeqnarray}
as $n\to\infty$.
%for some sequence $\delta_{0,n}$ that goes to zero as $n\to\infty$.

\underline{\textit{Step 2: Probability of Satisfying the Reliability Condition}}:

 Now, define the following event:
 \begin{IEEEeqnarray}{rCl}
	\mathcal{G} =\left\{\forall s\in\{1,2\},k\in\mathcal{K}\colon \bar{P}_{\mathrm{e}}(s,k)\leq  \mathbb{E}_{\mathcal{C}_s}\left[\bar{P}_{\mathrm{e}}(s,k)\right] +n^{-1/6}\right\}.\label{step201}
	\end{IEEEeqnarray}
%\begin{IEEEeqnarray}{rCl}
%	\mathcal{G} =\left\{\forall s\in\{1,2\},k\in\mathcal{K}\colon \bar{P}_{\mathrm{e}}(s,k)\leq 2 \mathbb{E}_{\mathcal{C}_s}\left[\bar{P}_{\mathrm{e}}(s,k)\right] \right\}.\label{step201}
%	\end{IEEEeqnarray}
Using the McDiarmid's theorem, we shall prove the following lemma, which can be thought as dual to Lemma \ref{le:jingbo},
\begin{lemma}[Concentration of average error probability around its expectation]\label{le:aep-con}
For all $k\in\mathcal{K}$ and any $\Delta>0$, we have
\begin{align}
\mathbb{P}\left(\bar{P}_e(s,k)> \mathbb{E}[\bar{P}_e(s,k)]+\Delta\right)\le \exp_e(-2|\mathcal{M}|\Delta^2).
\end{align}
\end{lemma}

This lemma with $\Delta=n^{-1/6}$ and the union bound imply
\begin{equation}
\mathbb{P}[\mathcal{G}^c]\le 2|\mathcal{K}|\exp(-2|\mathcal{M}|n^{-\frac{1}{3}})\rightarrow 0,\label{step202}
\end{equation}
for sufficiently large $n$, since $\log|\mathcal{K}|=O(\sqrt{n})$.

\begin{IEEEproof}[Proof of Lemma \ref{le:aep-con}]
Note that for a random codebook $\mathcal{C}_s$, the random variable $\bar{P}_{\mathrm{e}}(s,k)$ is a function of $|\mathcal{M}|$ independent random variables  $X_{s,1,k}^n,\ldots,X^n_{s,|\mathcal{M}|,k}$.   In this case, $\bar{P}_{\mathrm{e}}(s,k)$ can be written as follows:
\begin{align}
\bar{P}_{\mathrm{e}}(s,k)&=1-\bar{P}_{\mathrm{c}}(s,k)\\
                                       &=1-\frac{1}{|\mathcal{M}|}\sum_{y^n}\max_mW_s^n\big(y^n\big|X^n_{s,{m},k}\big)
                                       %\\
                                      % &=1-\frac{1}{|\mathcal{M}|}\sum_{y^n}W_s^n\big(y^n\big|X^n_{s,\hat{m}_{y^n},k}\big)
\end{align}
where $\bar{P}_{\mathrm{c}}(s,k)$ is the probability of correct decoding. Define:
\begin{equation}
g(\bx_{s,1,k},\cdots,\bx_{1,\mathcal{M},k})\triangleq  \frac{1}{|\mathcal{M}|}\sum_{y^n}\max_mW_s^n\left(y^n|x^n_{s,m  ,k}\right).
\end{equation}
We next show that $g$ satisfies the bounded difference property in~\eqref{eqn:bnd-diff}, so $\bar{P}_e(s,k)$   satisfies the bounded difference property as well. Let $(x^n_{s,1,k},\ldots,x^n_{s,|\mathcal{M}|,k})$ and $(\bar{x}^n_{s,1,k},\ldots,\bar{x}^n_{s,|\mathcal{M}|,k})$ be two sequences differing only in the $i$-th coordinate, that is

%For any $1\le i\le \mathcal{M}$ and any $x^n_{s,1,k},\cdots,x^n_{s,|\mathcal{M}|,k}$ let %and $\bar{x}^n_{s,i,k}$, 
%let    For any $1\le i\le \mathcal{M}$ and any $x^n_{s,1,k},\cdots,x^n_{s,|\mathcal{M}|,k}$, define the ML solution as 
%\begin{equation}
%\hat{m}_{y^n}\triangleq\argmax_m\; W^n_s(y^n|x^n_{s,m,k}),
%\end{equation}

\begin{align}
\bar{x}^n_{s,m,k} = \left\{ \begin{array}{cc}
x^n_{s,m,k} & \mbox{if } m\neq i \\
\mbox{arbitrary}& \mbox{if } m=  i 
\end{array}  \right.
\end{align}
Let the ML solution be defined as
\begin{equation}
\hat{m}_{y^n}\triangleq\argmax_m\; W^n_s(y^n|x^n_{s,m,k}). \label{eqn:ML}
\end{equation}
We have to  show that 
%Then we have, 
$g(\bx_{s,1,k},\ldots,\bx_{s,|\mathcal{M}|,k})-g(\bar{x}^n_{s,1,k},\ldots,\bar{x}^n_{s,|\mathcal{M}|,k})$ is bounded. Consider,
\begin{IEEEeqnarray}{rCl}
	&& g(\bx_{s,1,k},\ldots,\bx_{s,|\mathcal{M}|,k})-g(\bar{x}^n_{s,1,k},\ldots,\bar{x}^n_{s,|\mathcal{M}|,k})  \nonumber\\
	&&\hspace{1cm}= \frac{1}{|\mathcal{M}|}\sum_{y^n}\max_{m}W_s^n\left(y^n|x^n_{s,{m},k}\right)- \frac{1}{|\mathcal{M}|}\sum_{y^n}\max_mW_s^n\left(y^n|\bar{x}^n_{s,{m},k}\right)\\
	&&\hspace{1cm}= \frac{1}{|\mathcal{M}|}\sum_{y^n}W_s^n\left(y^n|x^n_{s,\hat{m}_{y^n},k}\right)- \frac{1}{|\mathcal{M}|}\sum_{y^n}\max_mW_s^n\left(y^n|\bar{x}^n_{s,{m},k}\right)\label{eqn:y-0}\\
	&&\hspace{1cm} \le\frac{1}{|\mathcal{M}|}\sum_{y^n}\left(W_s^n\left(y^n|x^n_{s,\hat{m}_{y^n},k}\right)-W_s^n\left(y^n|\bar{x}^n_{s,\hat{m}_{y^n},k}\right)\right)\label{eqn:y-1}\\
	&&\hspace{1cm} \le\frac{1}{|\mathcal{M}|}\sum_{y^n:\hat{m}_{y^n}=i}W_s^n\left(y^n|x^n_{s,\hat{m}_{y^n},k}\right)\label{eqn:y-2}\\
	&&\hspace{1cm}\leq \frac{1}{|\mathcal{M}|},\label{eqn:y-3}
\end{IEEEeqnarray}
where
\begin{itemize}
\item in~\eqref{eqn:y-0}, we used the definition of $\hat{m}_{y^n}$ in \eqref{eqn:ML};
\item in~\eqref{eqn:y-1} follows from the trivial inequality  $\max_m W_s^n (y^n|\bar{x}^n_{s,{m},k})\ge W_s^n (y^n|\bar{x}^n_{s,\hat{m}_{y^n},k})  $;
\item in~\eqref{eqn:y-2} follows because for $\hat{m}_{y^n}\neq i$, $\bx_{s,\hat{m}_{y^n},k}=\bar{x}^n_{s,\hat{m}_{y^n},k}$. 
\end{itemize}

Inequality \eqref{eqn:y-3} implies that $\bar{P}_e(s,k)$ has the bounded difference property with $d_i=\frac{1}{|\mathcal{M}|}$. Finally, the implication of the McDiarmid's theorem completes the proof.
\end{IEEEproof}
%Note that for a random codebook $\mathcal{C}_s$, the quantities $\bar{P}_{\mathrm{e}}(s,k), k\in\mathcal{K}$ are i.i.d. random variables. Thus, we have:
%\begin{IEEEeqnarray}{rCl}
%	\mathbb{P}(\mathcal{G})&=&\left(\mathbb{P}\left(\bar{P}_{\mathrm{e}}(1,k)\leq 2 \mathbb{E}_{\mathcal{C}_1}\left[\bar{P}_{\mathrm{e}}(1,k)\right]\right)\right)^{|\mathcal{K}|}\cdot \left(\mathbb{P}\left(\bar{P}_{\mathrm{e}}(2,k)\leq 2 \mathbb{E}_{\mathcal{C}_2}\left[\bar{P}_{\mathrm{e}}(2,k)\right]\right)\right)^{|\mathcal{K}|}\\
%	&\geq & \frac{1}{2^{2|\mathcal{K}|}}>0.\label{step202}
%	\end{IEEEeqnarray}
%Combining \eqref{event0} and \eqref{step202}, we get:
%\begin{IEEEeqnarray}{rCl}
%	\mathbb{P}\left[\mathcal{G}\bigcap (\mathcal{B}_1\cup\mathcal{B}_2)^c\right] >\mathbb{P}(\mathcal{G})-\mathbb{P}\left(\mathcal{B}_1\cup\mathcal{B}_2\right) >0.
%	\end{IEEEeqnarray}
%Thus, there exists a codebook $\mathcal{C}_s^*\in \mathcal{G}\cap \mathcal{B}_1^c\cap  \mathcal{B}_2^c$.

\underline{\textit{Step 3: Expurgation}}:
Combining \eqref{event0} and \eqref{step202}, we get:
\begin{IEEEeqnarray}{rCl}
	\mathbb{P}\left(\mathcal{G}^c \cup\mathcal{B}_1\cup\mathcal{B}_2\right) \le \mathbb{P}(\mathcal{G}^c)+\mathbb{P}\left(\mathcal{B}_1\cup\mathcal{B}_2\right) \rightarrow 0.
	\end{IEEEeqnarray}
Thus, there exists a codebook $\mathcal{C}_s^*\in \mathcal{G}\cap \mathcal{B}_1^c\cap  \mathcal{B}_2^c$.

For the codebook $\mathcal{C}_s^*$, for all $s\in\{1,2\}$ and $k\in\mathcal{K}$, we have:
\begin{IEEEeqnarray}{rCl}
	\bar{P}_{\mathrm{e}}(s,k)\leq \mathbb{E}\left[ \bar{P}_{\mathrm{e}}(s,k) \right]+n^{-1/6}=O\big(n^{-1/6}\big),
\end{IEEEeqnarray} 
where we used the bound in Lemma~\ref{PPM-thm}.

 We then remove $a|\mathcal{M}|$ codewords from the codebook $\mathcal{C}_s^*$ with largest $P_{\mathrm{e}}(s,m,k)$ for all $k\in\mathcal{K}$ and $s\in\{1,2\}$ from the subcodebook $\{ X_s^n(m,k)\colon m\in\mathcal{M} \}$ to get  new codebook $\{ X_s^n(m,k)\colon m\in\mathcal{M}_{s,k} \}$ such that $|\mathcal{M}_{s,k}|=(1-a)|\mathcal{M}|$. For these codebooks, the maximum error probability is given by (recall that $a=n^{-{1}/{12}}$):
 \begin{IEEEeqnarray}{rCl}
 	\max_m \;P_{\mathrm{e}}(m,s,k)\le \frac{1}{a}O\big(n^{-1/6}\big)=O\big(n^{-1/12}\big)=o(1).%>0.
 	\end{IEEEeqnarray}	

Further, \eqref{event0} implies that
\[
d_{\TV}(Q_{\mathcal{M}_s}^n,Q_{s}^{\times n})\le D_n=o(1).
\]
This completes the proof of existence of a codebook with desired properties. 

\section{Conclusion of Proof of Theorem~\ref{key-thm}}\label{key-thm-proof}

Using Lemma \ref{lem4}, if the constraints \eqref{lem-ineq1} and \eqref{lem4-ineq} hold, then there exists a codebook such that the maximum error probability   vanishes  as in~\eqref{eqn:max_err_vanish} and the following is satisfied:
\begin{align}
d_{\TV}(Q_1^n,Q_2^n)&\leq d_{\TV}(Q_1^n,Q_1^{\times n})+d_{\TV}(Q_1^{\times n},Q_2^{\times n})+d_{\TV}(Q_2^n,Q_2^{\times n})\\
&\stackrel{(a)}{\leq} 2\Phi\left(\frac{1}{2}\sqrt{\Omega(\gamma_1,\gamma_2;Q_0,Q_1,Q_2)}\right)-1+O\left(\frac{1}{\sqrt{n}}\right)+o(1)\\
&\stackrel{(b)}{\leq} \delta +
o(1)
%O\left(\frac{1}{\sqrt{n}}\right)
,
\end{align}
where $(a)$ follows from \eqref{lem4-ineq2}  and Lemma~\ref{lem-ach} and $(b)$ follows from \eqref{omega-ineq}. 

Dividing both sides of \eqref{lem-ineq1} and \eqref{lem4-ineq} by $\sqrt{n}$ and letting $n\to \infty$, we get:
\begin{align}
L+S\geq (1+\kappa)\cdot \max_{s\in\{1,2\}}\gamma_s
%D\left(\bar{Q}_s\|Q_0\right)
D(W_s \| Q_0|\bar{P}_s),\label{step210}
\end{align}
and 
\begin{IEEEeqnarray}{rCl}
L\leq \min_{s\in\{1,2\}}\gamma_s
%D(\bar{Q}_s\|Q_0)
D(W_s \| Q_0|\bar{P}_s),\label{step211}
\end{IEEEeqnarray}
where we have used the approximation $I(P_{s,n},W_s)=\frac{\gamma_s}{\sqrt{n}}D(W_s \| Q_0|\bar{P}_s)+o\left(\frac{1}{\sqrt{n}}\right)$. Performing Fourier-Motzkin elimination in \eqref{step210}--\eqref{step211} and letting $\kappa\to 0$ completes  the proof of the theorem.

\section{Solution of the Optimization Problem \eqref{key-ineq1}}\label{cor-proof}
Define:
$
 \bar{\chi}_{2,s}\triangleq  \chi_2(\tilde{Q}_s\|Q_0)$,  for $s\in\{1,2\}$, 
 $\bar{\rho}\triangleq-2\rho(\tilde{Q}_1,\tilde{Q}_2,Q_0)$ and $\bar{\delta}\triangleq \left(2\Phi^{-1}\left(\frac{1+\delta}{2}\right)\right)^2$. Also, recall the definition of $\bar{\mathbb{D}}_s$ from \eqref{step1000}.
Inequality \eqref{omega-ineq} can be written as follows:
\begin{align}
\bar{\chi}_{2,1}\gamma_1^2+\bar{\chi}_{2,2}\gamma_2^2+\bar{\rho}\gamma_1\gamma_2\leq \bar{\delta}.
\end{align}
The optimization problem in \eqref{key-ineq1} is given by the following:
\begin{align}
&\max_{\gamma_1,\gamma_2}\;\;\min_{s} \gamma_s\bar{\mathbb{D}}_s,\\
&\hspace{0cm}\text{s.t.}:\hspace{0.5cm}\bar{\chi}_{2,1}\gamma_1^2+\bar{\chi}_{2,2}\gamma_2^2+\bar{\rho}\gamma_1\gamma_2\leq \bar{\delta}.
\end{align}
The above optimization problem is equivalently written as:
\begin{IEEEeqnarray}{rCl}
&&	\max_{\gamma_1,\gamma_2}\;\;\min_{0\leq \xi\leq 1}\;\; \xi \gamma_1\bar{\mathbb{D}}_1+(1-\xi)\gamma_2\bar{\mathbb{D}}_2 \\
&&\hspace{0cm}\text{s.t.}:\hspace{0.5cm}\bar{\chi}_{2,1}\gamma_1^2+\bar{\chi}_{2,2}\gamma_2^2+\bar{\rho}\gamma_1\gamma_2\leq \bar{\delta}.\label{cons-min}
	\end{IEEEeqnarray}
Since the above optimization problem is convex, we have:
\begin{IEEEeqnarray}{rCl}
	&&	\min_{0\leq \xi\leq 1}\;\;\max_{\gamma_1,\gamma_2}\;\; \xi \gamma_1\bar{\mathbb{D}}_1+(1-\xi)\gamma_2\bar{\mathbb{D}}_2 \\*
	&&\hspace{0cm}\text{s.t.}:\hspace{0.5cm}\bar{\chi}_{2,1}\gamma_1^2+\bar{\chi}_{2,2}\gamma_2^2+\bar{\rho}\gamma_1\gamma_2\leq \bar{\delta}.
\end{IEEEeqnarray}
Define  
\begin{align}
 G  &\triangleq \xi \gamma_1\bar{\mathbb{D}}_1+(1-\xi)\gamma_2\bar{\mathbb{D}}_2-\lambda\left(\bar{\chi}_{2,1}\gamma_1^2+\bar{\chi}_{2,2}\gamma_2^2+\bar{\rho}\gamma_1\gamma_2- \bar{\delta}\right).
\end{align}
We calculate the derivative of $G$ with respect to $\gamma_1$ and $\gamma_2$ and let it be zero to get the optimal values $\gamma_1^*$ and $\gamma_2^*$. Thus, we obtain %get the following equations:
\begin{align}
2\bar{\chi}_{2,1}\gamma_1^*+\bar{\rho}\gamma_2^*&=\frac{\xi \bar{\mathbb{D}}_1}{\lambda},\\*
2\bar{\chi}_{2,2}\gamma_2^*+\bar{\rho}\gamma_1^*&=\frac{(1-\xi) \bar{\mathbb{D}}_2}{\lambda}.
\end{align}
This yields:
\begin{IEEEeqnarray}{rCl}
	\gamma_1^* &=& \frac{2\bar{\chi}_{2,2}\xi \bar{\mathbb{D}}_1-\bar{\rho}(1-\xi)\bar{\mathbb{D}}_2}{\lambda (4\bar{\chi}_{2,1}\bar{\chi}_{2,2}-\bar{\rho}^2)}\triangleq \frac{\bar{\gamma}_1}{\lambda},\\
	\gamma_2^* &=& \frac{2\bar{\chi}_{2,1}(1-\xi) \bar{\mathbb{D}}_2-\xi \bar{\rho}\bar{\mathbb{D}}_1}{\lambda (4\bar{\chi}_{2,1}\bar{\chi}_{2,2}-\bar{\rho}^2)}\triangleq \frac{\bar{\gamma}_2}{\lambda}.
	\end{IEEEeqnarray}
From \eqref{cons-min}, we know that
\begin{align}
\lambda \geq \sqrt{\frac{\bar{\chi}_{2,1}\bar{\gamma}_1^2+\bar{\chi}_{2,2}\bar{\gamma}_2^2+\bar{\rho}\bar{\gamma}_1\bar{\gamma}_2}{\bar{\delta}}}.
\end{align}
Thus, we have:
\begin{IEEEeqnarray}{rCl}
	\min_{0\leq \xi\leq 1}\;\max_{\gamma_1,\gamma_2}\;\; \xi \gamma_1\bar{\mathbb{D}}_1+(1-\xi)\gamma_2\bar{\mathbb{D}}_2 &&\hspace{0.3cm}=\min_{0\leq \xi\leq 1}\; \sqrt{\frac{\bar{\delta}}{\bar{\chi}_{2,1}\bar{\gamma}_1^2+\bar{\chi}_{2,2}\bar{\gamma}_2^2+\bar{\rho}\bar{\gamma}_1\bar{\gamma}_2}}\left(\xi \bar{\gamma}_1\bar{\mathbb{D}}_1+(1-\xi)\bar{\gamma}_2\bar{\mathbb{D}}_2\right).\\
	&& \hspace{0.3cm}=\min_{0\leq \xi\leq 1}\; \sqrt{\frac{4\bar{\delta}\left((1-\xi)^2\bar{\chi}_{2,1}\bar{\mathbb{D}}_2^2-\xi(1-\xi)\bar{\rho} \bar{\mathbb{D}}_1\bar{\mathbb{D}}_2+\xi ^2\bar{\chi}_{2,2} \bar{\mathbb{D}}_1^2 \right)}{\left(4\bar{\chi}_{2,1}\bar{\chi}_{2,2}-\bar{\rho}^2\right)}}\\  
	&& \hspace{0.3cm}=\left\{\begin{array}{ll}
	                               \bar{\mathbb{D}}_1\bar{\mathbb{D}}_2\sqrt{\dfrac{\bar{\delta}}{\bar{\chi}_{2,1}\bar{\mathbb{D}}_2^2+\bar{\rho}\bar{\mathbb{D}}_1\bar{\mathbb{D}}_2+\bar{\chi}_{2,2} \bar{\mathbb{D}}_1^2}};&\!\!\!\bar{\rho}>-2\min\left(\frac{\bar{\chi}_{2,1}\bar{\mathbb{D}}_2}{\bar{\mathbb{D}}_1},\frac{\bar{\chi}_{2,2}\bar{\mathbb{D}}_1}{\bar{\mathbb{D}}_2}\right)\\
	                            \sqrt{\dfrac{4\bar{\delta}\min\left(\bar{\chi}_{2,1}\bar{\mathbb{D}}_2^2,\bar{\chi}_{2,2}\bar{\mathbb{D}}_1^2\right)}{4\bar{\chi}_{2,1}\bar{\chi}_{2,2}-\bar{\rho}^2}}; &\!\!\! \bar{\rho}<-{2}\min\left(\frac{\bar{\chi}_{2,1}\bar{\mathbb{D}}_2}{\bar{\mathbb{D}}_1},\frac{\bar{\chi}_{2,2}\bar{\mathbb{D}}_1}{\bar{\mathbb{D}}_2}\right)
	       \end{array} \right.                  
\end{IEEEeqnarray}
This completes the proof.

\sadaf{\section{Proof of Corollary~\ref{cor-upper}}\label{cor-upper-proof}
We minimize the RHS of \eqref{L-upper-term} over all possible values of $0\leq \varphi \leq 1$ subject to the constraint in \eqref{cons-upper}. Recall the definitions of $\bar{\chi}_{2,s}$ and $\bar{\rho}$ from Appendix~\ref{cor-proof}. The optimization problem over $\varphi$ can be written as follows:
\begin{IEEEeqnarray}{rCl}
&&	\min_{\varphi}\; \frac{\sqrt{\varphi^2 \bar{\chi}_{2,1}+(1-\varphi)^2\bar{\chi}_{2,2}+\bar{\rho}\varphi (1-\varphi)}}{\varphi \bar{\chi}_{2,1}+(1-\varphi)\bar{\chi}_{2,2}+\frac{\bar{\rho}}{2}}\label{obj}\\
&& \hspace{0cm}\text{s.t.:}\;\;\frac{\bar{\rho}}{\bar{\rho}-2\bar{\chi}_{2,1}}\leq \varphi\leq \frac{2\bar{\chi}_{2,2}}{2\bar{\chi}_{2,2}-\bar{\rho}}.\label{opt1}
	\end{IEEEeqnarray}
Define:
\begin{IEEEeqnarray}{rCl}
	G(\varphi)\triangleq \frac{\sqrt{\varphi^2 \bar{\chi}_{2,1}+(1-\varphi)^2\bar{\chi}_{2,2}+\bar{\rho}\varphi (1-\varphi)}}{\varphi \bar{\chi}_{2,1}+(1-\varphi)\bar{\chi}_{2,2}+\frac{\bar{\rho}}{2}}.
	\end{IEEEeqnarray}
The derivative of the objective function $G$ with respect to $\varphi$ is given by the following:
\begin{IEEEeqnarray}{rCl}
\frac{\partial G(\varphi)}{\partial \varphi}=	\frac{(-1+2\varphi)(\bar{\chi}_{2,1}\bar{\chi}_{2,2}-\frac{\bar{\rho}^2}{4})}{\left(\frac{\bar{\rho}}{2}+\bar{\chi}_{2,2}(1-\varphi)+\bar{\chi}_{2,1}\varphi\right)^2\sqrt{\bar{\chi}_{2,2}(1-\varphi)^2+\varphi^2\bar{\chi}_{2,1}+\bar{\rho}\varphi(1-\varphi)}}.
	\end{IEEEeqnarray}
The above derivative term shows that the objective function $G$ has a minimum at $\varphi=\frac{1}{2}$. It is decreasing in the interval $0\leq \varphi\leq \frac{1}{2}$ and increasing in the interval $\frac{1}{2}\leq \varphi\leq 1$. It just remains to investigate the interval in \eqref{opt1} with respect to $\varphi=\frac{1}{2}$. If $\bar{\rho}\geq 0$, then the condition \eqref{cons-upper} is trivially satisfied for all $0\leq \varphi\leq 1$. Thus, in this case, the optimum value of $\varphi$ is $\frac{1}{2}$ and it yields:
\begin{IEEEeqnarray}{rCl}
	\min_{\varphi \in [0,1] } G (\varphi)= \frac{1}{\sqrt{\Psi}}.
	\end{IEEEeqnarray}
In the rest of the proof, we assume that $\bar{\rho}< 0$.
 Thus, we have the following cases:
\begin{IEEEeqnarray}{rCl}
	\min_{\varphi \in [0,1] } G (\varphi)= \left\{\begin{array}{cc}
		\frac{1}{\sqrt{\Psi}}&\hspace{2.6cm}\mbox{if } \frac{\bar{\rho}}{\bar{\rho}-2\bar{\chi}_{2,1}}\leq \frac{1}{2}\leq \frac{2\bar{\chi}_{2,2}}{2\bar{\chi}_{2,2}-\bar{\rho}}\;(\text{or} \;-\frac{\bar{\rho}}{2}\leq \min\{\bar{\chi}_{2,1},\bar{\chi}_{2,2}\})\\[1ex]
		\sqrt{\frac{\bar{\chi}_{2,1}}{\bar{\chi}_{2,1}\bar{\chi}_{2,2}-\frac{\bar{\rho}^2}{4}}} &\hspace{1.9cm}\mbox{if } \frac{2\bar{\chi}_{2,2}}{2\bar{\chi}_{2,2}-\bar{\rho}}\geq \frac{\bar{\rho}}{\bar{\rho}-2\bar{\chi}}_{2,1}\geq\frac{1}{2} \;(\text{or} \; \bar{\chi}_{2,2}\geq -\frac{\bar{\rho}}{2}\geq \bar{\chi}_{2,1})\\[1ex]
		\sqrt{\frac{\bar{\chi}_{2,2}}{\bar{\chi}_{2,1}\bar{\chi}_{2,2}-\frac{\bar{\rho}^2}{4}}} &\hspace{1.9cm} \mbox{if } \frac{\bar{\rho}}{\bar{\rho}-2\bar{\chi}}_{2,1}\leq \frac{2\bar{\chi}_{2,2}}{2\bar{\chi}_{2,2}-\bar{\rho}}\leq \frac{1}{2} \;(\text{or} \;\bar{\chi}_{2,1}\geq -\frac{\bar{\rho}}{2}\geq \bar{\chi}_{2,2})
	\end{array}	  \right.\label{conditions}
	\end{IEEEeqnarray}
Notice that the last two clauses in \eqref{conditions} can be combined together as given in the RHS of \eqref{L-upper}. This completes the proof. }

\section{Proof of \eqref{step3000} for the Example of Fig.~\ref{figure5}}\label{new-ex-proof}
First, notice that for the proposed example, we have
\begin{IEEEeqnarray}{rCl}
	Q_{1,n}(0) = \frac{1-\mu_n}{2},\;\;Q_{1,n}(1)=\frac{1}{2},\;\; Q_{1,n}(2)=\frac{\mu_n}{2},
	\end{IEEEeqnarray}
and 
\begin{IEEEeqnarray}{rCl}
	Q_{2,n}(0) = \frac{1}{2},\;\;Q_{1,n}(1)=\frac{1-\mu_n}{2},\;\; Q_{1,n}(2)=\frac{\mu_n}{2},
\end{IEEEeqnarray}
where $\mu_n=\frac{\gamma}{\sqrt{n}}$ for some positive $\gamma$ such that
\begin{align}
\gamma \leq 2\sqrt{\log \frac{1}{1-\delta}}.\label{eqn:gammadelta}
\end{align}
Now, consider the following upper bound on total variation distance in terms of Bhattacharyya coefficient \cite[Lemma 3.3.9]{Reis}
\begin{IEEEeqnarray}{rCl}
	d_{\TV}(Q_1^{\times n},Q_2^{\times n})\leq \sqrt{1-F(Q_1^{\times n},Q_2^{\times n})^2}.\label{step2000}
	\end{IEEEeqnarray}
Since $Q_1^{\times n}$ and $Q_{2}^{\times n}$ are product distributions,  one can show that:
\begin{IEEEeqnarray}{rCl}
	F(Q_1^{\times n},Q_2^{\times n}) = F(Q_{1,n},Q_{2,n})^n. \label{step2001}
	\end{IEEEeqnarray}
Calculation of the Bhattacharyya coefficient for two pmfs $Q_{1,n}$ and $Q_{2,n}$ yields the following:
\begin{IEEEeqnarray}{rCl}
F(Q_{1,n},Q_{2,n}) &=&\frac{1}{2}\left(2\sqrt{1-\mu_n}+\mu_n\right)\\	
&=& 1-\frac{1}{8}\mu_n^2+o(\mu_n^2)\label{step2002}\\
&=& 1-\frac{\gamma^2}{8n}+o\left(\frac{1}{n}\right),
	\end{IEEEeqnarray}
where \eqref{step2002} follows from Taylor series $\sqrt{1-x}= 1-\frac{x}{2}-\frac{x^2}{8}+o(x^2)$ as $x\to 0$.
Combining \eqref{step2000}, \eqref{step2001} and \eqref{step2002}, we have the following upper bound on total variation distance:
\begin{IEEEeqnarray}{rCl}
	d_{\TV}(Q_1^{\times n},Q_2^{\times n})& \leq& \sqrt{1-\left(1-\frac{\gamma^2}{8n}+o\left(\frac{1}{n}\right)\right)^{2n}}\\
	&=&\sqrt{1-\exp\left(-\frac{\gamma^2}{4}\right)}+o(1)\label{step2003}\\
	&\leq &\delta+o(1),  \label{eqn:gamma_delta}
	\end{IEEEeqnarray}
where \eqref{step2003} follows $\left(1-\frac{x}{2n}\right)^{2n}\to \exp(-x)$ as $n\to\infty$ and \eqref{eqn:gamma_delta} follows from \eqref{eqn:gammadelta}. 

To find a lower bound on the optimal throughput, one can repeat the same steps as in Appendices~\ref{lem-key-proof}, \ref{rate-analysis}, \ref{existence2-proof} and get similar results with the following approximations:
\begin{IEEEeqnarray}{rCl}
	I(P_s,W_s) &=& \frac{1}{2}h_{\text{b}}(\mu_n)\\
	&=& \frac{\sqrt{\log \frac{1}{1-\delta}}}{2}\cdot \frac{\log n}{\sqrt{n}}+o\left(\frac{\log n}{\sqrt{n}}\right),
	\end{IEEEeqnarray}
and 
\begin{IEEEeqnarray}{rCl}
	\sum_{x}\sum_y P_{s,n}(x)W_s(y|x)\left(\log \frac{W_s(y|x)}{Q_{s,n}(y)}\right)^2&=& \frac{1}{2}\left(\mu_n\left(\log \mu_n\right)^2+(1-\mu_n)\left(\log (1-\mu_n)\right)^2\right)\\
	&=& \frac{\sqrt{\log \frac{1}{1-\delta}}}{4}\cdot \frac{\log^2 n}{\sqrt{n}}+o\left(\frac{\log^2 n}{\sqrt{n}}\right).
	\end{IEEEeqnarray}
Thus, Lemmas~\ref{PPM-thm} and \ref{lem4} state that if we have
\begin{IEEEeqnarray}{rCl}
	\log |\mathcal{M}| \leq \frac{\sqrt{\log \frac{1}{1-\delta}}}{2}\cdot \sqrt{n}\log n+o\left(\sqrt{n}\log n\right),
	\end{IEEEeqnarray}
then the desired thresholds on the maximum error probability and the masking constraint are satisfied.
This completes the proof and shows that the throughout is $\Omega(\sqrt{n}\log n)$.

\section{Proof of  Lemma~\ref{lem2}}\label{lem2-proof}
\sadaf{In the following, for the ease of notation, we denote $\D_s(\varphi)$, $\Gamma_s(\varphi)$, $\Delta(\varphi)$, $\mathcal{T}(\cdot ;\varphi)$ and $\tau(\varphi)$ by $\D_s$, $\Gamma_s$, $\Delta$, $\mathcal{T}(\cdot )$ and $\tau$, respectively. That is, we drop the parameter $\varphi$ in all these functions for the sake of brevity.}

Let $\mathcal{P}^n$ be the set of all types $\pi_X$ over $\mathcal{X}^n$ and $\mathcal{C}_{s,\pi_X}$ be the sub-codebook  of  $\mathcal{C}_{s}$ consisting of all codewords with the same type $\pi_X$.  Notice that \begin{align}\mathcal{C}_s=\bigcup_{\pi_X\in\mathcal{P}^n}\mathcal{C}_{s,\pi_X}.\end{align} 
 Let $P_{X_s^n}$ be the uniform distribution over the codebook $\mathcal{C}_s= \{x_s^n(m,k):m\in\mathcal{M},k\in\mathcal{K}\}$, i.e., for every $x^n\in\mathcal{X}^n$:
%	\begin{IEEEeqnarray}{rCl}
%		P_{X_s^n}(x^n)=\left\{\begin{array}{ll} \frac{1}{|\mathcal{M}|}\;\; &\;\; x^n=x_s^n(m)\;\text{for some}\;m\in\mathcal{M}\\\\ 0 \;\;&\;\; \text{o.w.}\end{array} \right.\nonumber\\
%		\end{IEEEeqnarray}
\begin{align}
P_{X_s^n}(x^n)= \left\{ \begin{array}{cc}
\frac{1}{|\mathcal{M} |\cdot |\mathcal{K} |} & x^n \in\mathcal{C}_s \\
0 & \mbox{ else }
\end{array} \right.
\end{align}	
In the following, we first consider false alarm probability and next the missed detection probability. Consider the false alarm probability as follows:
\begin{align}
\alpha_n &=  \mathbb{P}_{Q_1^n}\left(\sum_{i=1}^n\mathcal{T}(Y_i)\leq  \tau\right)\\
&= \sum_{y^n} Q_1^n(y^n) \mathbbm{1} \left\{\sum_{i=1}^n\mathcal{T}(y_i)\leq  \tau\right\}\\
&= \sum_{y^n}\bigg( \sum_{x^n  \in \mathcal{C}_{1}} P_{X_1^n}(x^n)W_1^n(y^n|x^n) \bigg) \mathbbm{1} \left\{\sum_{i=1}^n\mathcal{T}(y_i)\leq \tau\right\}\\
&= \sum_{x^n\in\mathcal{C}_{1}}P_{X_1^n}(x^n) \sum_{y^n } W_1^n(y^n|x^n)  \mathbbm{1} \left\{\sum_{i=1}^n\mathcal{T}(y_i)\leq \tau\right\}\\
&= \sum_{\pi_X}\sum_{x^n\in\mathcal{C}_{1,\pi_X}}P_{{X_1^n}}(x^n) \mathbb{P}_{W_1^n(\cdot|x^n)}\left( \sum_{i=1}^n\mathcal{T}(Y_i)\leq \tau \right)
\end{align}
Since  $ \mathbb{P}_{W_1^n(\cdot|x^n)}\left(\sum_{i=1}^n\mathcal{T}(Y_i)\leq \tau \right) $ remains the same for each $x^n$ with the same type, and $\{P_{{X_1^n}} \{  \mathcal{C}_{1,\pi_X} \}\}_{\pi_X\in\mathcal{P}^n}$ is a probability distribution,  we have 
\begin{align}
\alpha_n \le \max_{\pi_X}\;  \mathbb{P}_{W_1^n(\cdot|x_1^n)}\left( \sum_{i=1}^n\mathcal{T}(Y_i)\leq \tau \right)\label{pr1}, 
\end{align}
where $x_1^n$ above refers to any vector with type $\pi_X$.  
%\blue{\begin{IEEEeqnarray}{rCl}
%	\beta_n &=& \mathbb{P}_{Q_2^n}\left[\sum_{i=1}^n\mathcal{T}_{\text{\test}}(Y_i)> \tau\right]\\
%	&\leq&\frac{1}{(n+1)^{|\mathcal{X}|}} \sum_{\pi_X\in\mathcal{P}^n}\;\; \nonumber\\&&\hspace{0.5cm}\mathbb{P}_{W_2^n(.|x^n)}\left[\sum_{i=1}^n\mathcal{T}_{\text{\test}}(Y_i)> \tau\Big| \text{tp}(X^n)=\pi_X\right]\\[1ex]
%	&\leq & \max_{\pi_{X}}\;\;\mathbb{P}_{W_2^n(.|X^n)}\left[\sum_{i=1}^n\mathcal{T}_{\text{\test}}(Y_i)> \tau\Big| The expectation and variance operators are written as $\mathbb{E}[.]$ and $\mathbb{V}[.]$, respectively. The notation $\mathbbm{1}\left\{ . \right\}$ denotes the indicator function.  The probability of an event $\mathcal{E}\subseteq \mathcal{X}$ is denoted by $\mathbb{P}(\mathcal{E})$. (X^n)=\pi_X\right].\nonumber\\\label{pr1}
%\end{IEEEeqnarray}
Denoting the maximizing type in \eqref{pr1} by $\pi^*$ and letting $x^{*n}$ be any vector with type $\pi^*$, \eqref{pr1} can be written as follows:
\begin{IEEEeqnarray}{rCl}
	\alpha_n %&\leq & \mathbb{P}_{W_2^n(.|X^n)}\left[\sum_{i=1}^n\mathcal{T}_{\text{\test}}(Y_i)> \tau\Big| \text{tp}(X^n)=\pi_2^*\right]\\
	&\le&\mathbb{P}_{W_1^n(\cdot |x^{*n})}\left(\sum_{i=1}^n\mathcal{T}(Y_i)\leq \tau \right).\label{alpha-upp}
	\end{IEEEeqnarray}
%where the last equality follows because all codewords $x_2^n$ of the same type $\pi_2^*$ have the same probability. 
  Notice that  $\pi^*$ relates to the Hamming weight of the codeword $x^{*n}$ as follows:
  \begin{IEEEeqnarray}{rCl}
  	\pi^*(x)=\left\{\begin{array}{ll} \frac{\w(x^{*n})}{n} & x=1\\ 1-\frac{\w(x^{*n})}{n} & x=0\end{array} \right.
  	\end{IEEEeqnarray}
 We also define
\begin{IEEEeqnarray}{rCl}
	\mu_{1,n}\triangleq \pi^*(1)= \frac{\w(x^{*n})}{n}.
\end{IEEEeqnarray}
Since the channel is memoryless, $\left\{\mathcal{T}(Y_i)\right\}_{i=1}^n$ are mutually independent so Theorem~\ref{BE-thm} can be applied to upper bound the probability in \eqref{alpha-upp}. For every $x^{*n}$ with type $\pi^*$, we calculate $\sum_{i=1}^n\mathbb{E}\left[\mathcal{T}(Y_i) \right]$ and $\sum_{i=1}^n\mathbb{V}\left[\mathcal{T}(Y_i) \right]$ in the following. First, we calculate the expectation as follows,
\begin{IEEEeqnarray}{rCl}
	%&=&\sum_{i=1}^n\mathbb{E}_{W_1^n(.|x^{*n})}\left[\mathcal{T}(Y_i)\right]\\
	\sum_{i=1}^n\mathbb{E}_{W_1(\cdot |x^*_{i})}\left[\mathcal{T}(Y_i)\right]
	&=&\sum_{i=1}^n\mathbb{E}_{W_1(\cdot |x_{i}^*)}\left[\frac{\varphi(\tilde{Q}_{1}(Y_i)-Q_0(Y_i))-(1-\varphi)(\tilde{Q}_{2}(Y_i)-Q_0(Y_i))}{Q_{0}(Y_i)}\right]\\
	&=&\sum_{\substack{i:x_{i}^*=1}}\mathbb{E}_{\tilde{Q}_{1}}\left[\frac{\varphi(\tilde{Q}_{1}(Y_i)-Q_0(Y_i))-(1-\varphi)(\tilde{Q}_{2}(Y_i)-Q_0(Y_i))}{Q_{0}(Y_i)}\right]\label{step31}\\
	&=&n\mu_{1,n}\sum_y \frac{\tilde{Q}_{1}(y)\left(\varphi(\tilde{Q}_{1}(y)-Q_0(y))-(1-\varphi)(\tilde{Q}_{2}(y)-Q_{0}(y))\right)}{Q_{0}(y)}\\
	&=&n\mu_{1,n}\sum_y \frac{\varphi(\tilde{Q}_{1}(y)-Q_0(y))^2+(1-\varphi)(\tilde{Q}_{2}(y)-Q_0(y))\cdot (\tilde{Q}_{1}(y)-Q_0(y))}{Q_{0}(y)}\\
	&=& n\mu_{1,n}\left(\varphi\chi_2(\tilde{Q}_1\|Q_0)-(1-\varphi) \rho(\tilde{Q}_1,\tilde{Q}_2,Q_0)\right)\\
	&=& n\mu_{1,n} \D_{1},\label{def1}
\end{IEEEeqnarray}
where \eqref{step31} follows because when $x_{i}^*=0$, we have $W_1(Y_i|x_{i}^*)=Q_0(Y_i)$ and the expectation in the summand is equal to zero. Next, we calculate the variance as follows
\begin{IEEEeqnarray}{rCl}
 \sum_{i=1}^n\mathbb{V}\left[\mathcal{T}(Y_i)\right]
	&=& \sum_{i=1}^n  \mathbb{E}\left[\left(\mathcal{T}(Y_i)\right)^2\right]-\sum_{i=1}^n\left(\mathbb{E}\left[\mathcal{T}(Y_i)\right]\right)^2 ,\label{step32}
\end{IEEEeqnarray}
Now, we calculate each expectation term in \eqref{step32} as follows,
 \begin{IEEEeqnarray}{rCl}
\sum_{i=1}^n \mathbb{E}\left[\left(\mathcal{T}(Y_i)\right)^2\right] &=&  \sum_{i=1}^n  \mathbb{E}_{W_1(\cdot |x_{i}^*)}\left[\left(\mathcal{T}(Y_i)\right)^2\right]	\\
&=& \sum_{i:x_{i}^*=1} \mathbb{E}_{\tilde{Q}_1}\left[\left(\mathcal{T}(Y_i)\right)^2\right]+\sum_{i:x_{i}^*=0} \mathbb{E}_{Q_0}\left[\left(\mathcal{T}(Y_i)\right)^2\right]\\
&=& n\mu_{1,n}\sum_y\frac{\tilde{Q}_1(y)\big(\varphi(\tilde{Q}_{1}(y)-Q_0(y))-(1-\varphi)(\tilde{Q}_{2}(y)-Q_{0}(y))\big)^2}{Q_{0}^2(y)}+n(1-\mu_{1,n})\Delta,\label{step102}
	\end{IEEEeqnarray}
and 
\begin{IEEEeqnarray}{rCl}
\sum_{i=1}^n\left(\mathbb{E}\left[\mathcal{T}(Y_i)\right]\right)^2&=& \sum_{i=1}^n\left(\mathbb{E}_{W_1(\cdot |x_{i}^*)}\left[\mathcal{T}(Y_i)\right]\right)^2\\
&=& \sum_{i=1}^n\left(\mathbb{E}_{W_1(\cdot|x_{i}^*)}\left[\frac{\varphi(\tilde{Q}_{1}(Y_i)-Q_0(Y_i))-(1-\varphi)(\tilde{Q}_{2}(Y_i)-Q_0(Y_i))}{Q_{0}(Y_i)}\right]\right)^2\\
&=&\sum_{i:x_{i}^*=1}\left(\mathbb{E}_{\tilde{Q}_1}\left[\frac{\varphi(\tilde{Q}_{1}(Y_i)-Q_0(Y_i))-(1-\varphi)(\tilde{Q}_{2}(Y_i)-Q_0(Y_i))}{Q_{0}(Y_i)}\right]\right)^2\nonumber\\&&\hspace{1cm}+ \sum_{i:x_{i}^*=0}\left(\mathbb{E}_{Q_0}\left[\frac{\varphi(\tilde{Q}_{1}(Y_i)-Q_0(Y_i))-(1-\varphi)(\tilde{Q}_{2}(Y_i)-Q_0(Y_i))}{Q_{0}(Y_i)}\right]\right)^2 \\[1ex]
&=&n\mu_{1,n}\left(\sum_y\frac{\tilde{Q}_1(y)\big(\varphi(\tilde{Q}_{1}(y)-Q_0(y))-(1-\varphi)(\tilde{Q}_{2}(y)-Q_{0}(y))\big)}{Q_0(y)}\right)^2\\
&=&n\mu_{1,n}\left(\varphi\chi_2(\tilde{Q}_1\|Q_0)-(1-\varphi) \rho(\tilde{Q}_1,\tilde{Q}_2,Q_0)\right)^2\\
&=&n\mu_{1,n}\D_1^2.\label{step103}
	\end{IEEEeqnarray}

Combining \eqref{step32}, \eqref{step102} and \eqref{step103}, we get 
\begin{IEEEeqnarray}{rCl}
	\sum_{i=1}^n\mathbb{V}\left[\mathcal{T}(Y_i) \right]
	&=&n\Bigg(\mu_{1,n} \sum_y \frac{\tilde{Q}_{1}(y)\big(\varphi(\tilde{Q}_{1}(y)-Q_0(y))-(1-\varphi)(\tilde{Q}_{2}(y)-Q_{0}(y))\big)^2}{Q_{0}^2(y)}+(1-\mu_{1,n})\Delta-\mu_{1,n}\D_{1}^2\Bigg)\\
	&\triangleq &n\V_{1,n}.\label{step41}
\end{IEEEeqnarray}
We now bound the sum of the  third absolute moments 
\begin{IEEEeqnarray}{rCl}
	\sum_{i=1}^n \mathbb{E}_{W_1(\cdot|x_i^{*})}\left[ \big|\mathcal{T}(Y_i) -\mathbb{E}_{W_1(\cdot|x_i^{*})} [ \mathcal{T}(Y_i) ] \big|^3 \right]\label{third1}.
\end{IEEEeqnarray}
We know that $\mathcal{T}(y_i)= \frac{\varphi(\tilde{Q}_{1}(y_i)-Q_0(y_i))-(1-\varphi)(\tilde{Q}_{2}(y_i)-Q_{0}(y_i))}{Q_0(y_i)}$. Hence,
\begin{align}|\mathcal{T}(y_i)|  \le \frac{2}{\min_{y_i} Q_0(y_i)} \triangleq \frac{2}{\eta} <\infty.\label{t-bound}\end{align}	
By the triangle inequality,
\begin{align}
\big|\mathcal{T}(Y_i) -\mathbb{E}  [ \mathcal{T}(Y_i) ] \big|\le \frac{4}{\eta},\quad\mbox{a.s.}\label{third2}
\end{align}
Combining \eqref{third1} and \eqref{third2}, we obtain
\begin{align}
&\sum_{i=1}^n \mathbb{E}_{W_1(\cdot|x_i^{*})}\left[ \big|\mathcal{T}(Y_i) -\mathbb{E}_{W_1(\cdot|x_i^{*})} [ \mathcal{T}(Y_i) ] \big|^3\right]\leq \frac{64n}{\eta^3}\triangleq n\T.\label{third-abs5}
\end{align}

Notice that $\V_{1,n}$ in \eqref{step41} is further upper bounded as follows:
\begin{IEEEeqnarray}{rCl}
	\V_{1,n}&\leq& \Bigg(\mu_{1,n} \sum_y \frac{\tilde{Q}_{1}(y)\big(\varphi(\tilde{Q}_{1}(y)-Q_0(y))-(1-\varphi)(\tilde{Q}_{2}(y)-Q_{0}(y))\big)^2}{Q_{0}^2(y)}+(1-\mu_{1,n})\Delta\Bigg)\\
	&=& \Delta+\mu_{1,n}\Gamma_1\\
	&\leq& \Delta+\mu_{\Hi,n}|\Gamma_1|\\
	&\triangleq &\V_{1,n}^{*}.
\end{IEEEeqnarray}
Thus, from Theorem~\ref{BE-thm}, we get the following:
\begin{IEEEeqnarray}{rCl}
	\hspace{0cm}\mathbb{P}\left(\sum_{i=1}^n\mathcal{T}(Y_i)\leq  \tau\right)&\leq& 	1-\Phi\left(\frac{-\tau+n \mu_{1,n}\D_{1}}{\sqrt{n \V_{1,n}}}\right)+\frac{6\T}{\V_{1,n}^{3/2}\sqrt{n}}.\label{step4}
\end{IEEEeqnarray} 
Combining \eqref{pr1} and \eqref{step4}, we get the following:
\begin{IEEEeqnarray}{rCl}
	\alpha_n \leq \max_{\mu_{1,n}} \;1-\Phi\left(\frac{-\tau+n \mu_{1,n}\D_{1}}{\sqrt{n \V_{1,n}}}\right)+O\left(\frac{1}{\sqrt{n}}\right),\;\;\;\;\;\;\label{step5}
\end{IEEEeqnarray}
where the maximization is over all $\mu_{1,n}$ such that $\mu_{\Lo,n}\leq \mu_{1,n}\leq \mu_{\Hi,n}$.
Now, consider the missed detection probability as follows:
\begin{IEEEeqnarray}{rCl}
	\beta_n &=& \mathbb{P}_{Q_2^n}\left( \sum_{i=1}^n\mathcal{T}(Y_i)> \tau \right).
\end{IEEEeqnarray}
Following similar steps leading to \eqref{pr1}, we can write the missed detection probability as:
\begin{IEEEeqnarray}{rCl}
	\beta_n 
	&\leq & \max_{\mu_{2,n}}\;1-\Phi\left(\frac{\tau-n\mu_{2,n}\D_{2}}{\sqrt{n \V_{2,n}}}\right)+\frac{6\T}{\V_{2,n}^{3/2}\sqrt{n}},
	\label{alpha-final}
\end{IEEEeqnarray}
where the maximization is over all $\mu_{2,n}$ such that $\mu_{\Lo,n}\leq \mu_{2,n}\leq \mu_{\Hi,n}$ and we define:
\begin{IEEEeqnarray}{rCl}
	\V_{2,n} &\triangleq& \Bigg(\mu_{2,n} \sum_y \frac{\tilde{Q}_{2}(y)\big(\varphi(\tilde{Q}_{1}(y)-Q_0(y))-(1-\varphi)(\tilde{Q}_{2}(y)-Q_{0}(y))\big)^2}{Q_{0}^2(y)}+(1-\mu_{2,n})\Delta-\mu_{2,n}\D_{2}^2\Bigg).
\end{IEEEeqnarray}
We can further upper bound $\V_{2,n}$ as follows:
\begin{align}
	\V_{2,n}&\leq  \mu_{2,n} \sum_y \frac{\tilde{Q}_{2}(y)\big(\varphi(\tilde{Q}_{1}(y)-Q_0(y))-(1-\varphi)(\tilde{Q}_{2}(y)-Q_{0}(y))\big)^2}{Q_{0}^2(y)}+(1-\mu_{2,n})\Delta \\
	&= \Delta+ \mu_{2,n}  \Gamma_2\\
	&\leq   \Delta+ \mu_{\Hi,n} |\Gamma_2| \\
	&\triangleq  \V^{*}_{2,n},
\end{align}
 The proof is followed by upper bounding the false alarm and missed detection probabilities using the choice of $\tau$ in \eqref{step15}:
\begin{IEEEeqnarray}{rCl}
	\alpha_n&\leq&\max_{\mu_{1,n}}\;1-\Phi\left(\frac{-\tau+n\mu_{1,n}\D_{1}}{\sqrt{n \V_{1,n}}}\right)+O\left(\frac{1}{\sqrt{n}}\right)\\
	&=& \max_{\mu_{1,n}}\;1-\Phi\left(\frac{-\frac{n\mu_{\Lo,n}}{2}\left(\D_{2}+\D_{1}\right)+n\mu_{1,n}\D_{1}}{\sqrt{n \V_{1,n}}}\right)+O\left(\frac{1}{\sqrt{n}}\right)\\
	&=& 1-\Phi\left(\frac{-\frac{n\mu_{\Lo,n}}{2}\left(\D_{2}+\D_{1}\right)+n\mu_{\Lo,n}\D_{1}}{\sqrt{n \V_{1,n}}}\right)+O\left(\frac{1}{\sqrt{n}}\right)\nonumber\\\\
	&\leq&	1-\Phi\left(\frac{\sqrt{n}\mu_{\Lo,n}\left(\D_{1}-\D_{2}\right)}{2\sqrt{ \V_{1,n}}}\right)+O\left(\frac{1}{\sqrt{n}}\right)\label{step301}\\[1ex]
	&\leq &1-\Phi\left(\frac{\sqrt{n}\mu_{\Lo,n} (\D_1-\D_2)}{2\sqrt{ \V^*_{1,n}}}\right)+O\left(\frac{1}{\sqrt{n}}\right)\label{step305}\\[1ex]
	&=& 1-\Phi\left(\frac{\sqrt{n} \mu_{\Lo,n} (\D_1-\D_2)}{2\sqrt{ \left(\Delta+\mu_{\Hi,n} |\Gamma_1|\right)}}\right)+O\left(\frac{1}{\sqrt{n}}\right)\\[1.5ex]
	&\leq & 1-\Phi\left(\frac{1}{2}\mu_{\Lo,n}\sqrt{n}\frac{(\D_1-\D_2)}{\sqrt{\Delta}} \cdot \left(1-\frac{\mu_{\Hi,n} |\Gamma_1|}{2\Delta}\right)\right)+O\left(\frac{1}{\sqrt{n}}\right)\nonumber\\\label{step6}\\[1ex]
	&\leq & 1-\Phi\left(\frac{1}{2}\mu_{\Lo,n}\sqrt{n}\frac{\D_1-\D_2}{\sqrt{\Delta}}  \right)+\frac{\sqrt{n}\mu_{\Lo,n}\mu_{\Hi,n}|\Gamma_1|(\D_1-\D_2)}{4\sqrt{2\pi\Delta^3}}+O\left(\frac{1}{\sqrt{n}}\right),\label{step16}
\end{IEEEeqnarray}
where
\begin{itemize}
	\item \eqref{step301} follows because assumption \eqref{conv-cons1} implies that $\D_1\geq 0$;
	\item \eqref{step305} follows because $\D_1\geq \D_2$ and $\V_{1,n}\leq \V^*_{1,n}$;
	\item \eqref{step6} follows from $\frac{1}{\sqrt{1+x}}\geq 1-\frac{x}{2}$; 
	\item \eqref{step16} follows because $1-\Phi(x-y)\leq 1-\Phi(x)+\frac{y}{\sqrt{2\pi}}$ for all $0<y<x$. 
\end{itemize}
With the choice of $\tau$ in \eqref{step15}, $\beta_n$ in \eqref{step5} can be upper bounded as follows:
\begin{IEEEeqnarray}{rCl}
	\beta_n&\leq & \max_{\mu_{2,n}}\; 1-\Phi\left(\frac{\tau-n \mu_{2,n}\D_{2}}{\sqrt{n \V_{2,n}}}\right)+O\left(\frac{1}{\sqrt{n}}\right)\label{step45}\\
	&=& \max_{\mu_{2,n}}\;1-\Phi\left(\frac{\frac{n\mu_{\Lo,n}}{2}\left(-\D_{2}+\D_{1}\right)+n( \mu_{\Lo,n}-\mu_{2,n})\D_{2}}{\sqrt{n \V_{2,n}}}\right)+O\left(\frac{1}{\sqrt{n}}\right)\\[1.5ex]
	&\leq &   1-\Phi\left(\frac{\frac{n\mu_{\Lo,n}}{2}\left(-\D_{2}+\D_{1}\right)}{\sqrt{n \V_{2,n}}}\right)+O\left(\frac{1}{\sqrt{n}}\right)\label{step300}\\[1.5ex]
	&\leq &1-\Phi\left(\frac{n\mu_{\Lo,n}(\D_1-\D_2)}{2\sqrt{n \V_{2,n}^*}}\right)+O\left(\frac{1}{\sqrt{n}}\right)\label{step302}\\[1.5ex]
	&\leq & 1-\Phi\left(\frac{1}{2}\mu_{\Lo,n}\sqrt{n}\frac{\D_1-\D_2}{\sqrt{\Delta}}\right)+\frac{\sqrt{n}\mu_{\Lo,n}\mu_{\Hi,n}|\Gamma_2|(\D_1-\D_2)}{4\sqrt{2\pi\Delta^3}}+O\left(\frac{1}{\sqrt{n}}\right),\label{step303}
\end{IEEEeqnarray} 
where 
\begin{itemize}
\item \eqref{step300} follows because assumption \eqref{conv-cons1} implies that $\D_2\leq 0$ and hence, $(\mu_{\Lo,n}-\mu_{2,n})\D_2\geq 0$;
\item \eqref{step302} follows because $\V_{2,n}\leq \V_{2,n}^*$;
\item \eqref{step303} follows because $\V_{2,n}^*=\Delta+\mu_{\Hi,n}|\Gamma_2|$ and also from the fact that $1-\Phi(x-y)\leq 1-\Phi(x)+\frac{y}{\sqrt{2\pi}}$ for all $0<y<x$.
\end{itemize}

This completes the proof of lemma.

\section{Conclusion of Proof of Theorem~\ref{thm-upper}}\label{existence-proof}
Just as in the previous section,  for the ease of notation, we denote $\D_s(\varphi)$, $\Gamma_s(\varphi)$, $\Delta(\varphi)$ and $\mathcal{T}(\cdot ;\varphi)$ by $\D_s$, $\Gamma_s$, $\Delta$ and $\mathcal{T}(\cdot )$, respectively.

 In the following, we first show that there exists a sub-codebook which satisfies \eqref{step42} and its size is at least $\frac{|\mathcal{M}|\cdot |\mathcal{K}|}{\sqrt{n}}$. Define the following set of codewords for some $\gamma>0$:
 \begin{IEEEeqnarray}{rCl}
	\mathcal{D}_s\triangleq\left\{ x_s^n\in\mathcal{C}_s\colon \w(x_s^n)\leq \frac{2\sqrt{n\Delta}}{\D_1-\D_2}\Phi^{-1}\left(\frac{1+\delta}{2}+\gamma+\frac{E}{\sqrt{n}}\right) \right\},\label{Ds-def}
 \end{IEEEeqnarray}
 where the constant $E$ is chosen such that
 \begin{IEEEeqnarray}{rCl}
 	E> \frac{(|\Gamma_1|+|\Gamma_2|)(\D_1-\D_2)}{8\sqrt{2\pi\Delta^3}}\cdot \left(\Phi^{-1}\left(\frac{1+\delta}{2}+\gamma\right)\right)^2.\label{step13}
 \end{IEEEeqnarray}

 Let $\widehat{Q}_{s}^n$ and $\overline{Q}_{s}^n$ be the induced output distributions for codes $\mathcal{D}_s$ and $\mathcal{C}_s \backslash \mathcal{D}_s$, respectively. The distribution $Q_s^n$ can be written as follows:
 \begin{IEEEeqnarray}{rCl}
 	Q_{s}^n= \xi_s\widehat{Q}_{s}^n+(1-\xi_s)\overline{Q}_{s}^n,
 \end{IEEEeqnarray}
 where $\xi_s\triangleq \frac{|\mathcal{D}_s|}{|\mathcal{M}|\cdot |\mathcal{K}|}$. Without loss of generality, assume that $\xi_1\geq \xi_2$.
 Then, we get the following:
\begin{align}
 	\delta &\geq d_{\TV}\left(Q_1^n,Q_2^n\right)=\frac{1}{2} \sum_{y^n}\left|Q_1^n(y^n)-Q_2^n(y^n)\right|\\
 	&=\frac{1}{2}\sum_{y^n}\left|\xi_1\widehat{Q}_1^n(y^n)+(1-\xi_1)\overline{Q}_1^n(y^n)-\xi_2\widehat{Q}_2^n(y^n)-(1-\xi_2)\overline{Q}_2^n(y^n)\right|\\
 	&\geq d_{\TV}\big(\overline{Q}_{1}^n,\overline{Q}_{2}^n\big) -\xi_1  d_{\TV}\big(\widehat{Q}_{1}^n,\overline{Q}_{1}^n\big)-\xi_2 d_{\TV}\big(\widehat{Q}_{2}^n,\overline{Q}_2^n\big)\label{step11}
\end{align}
 where the last inequality follows from the triangle inequality. 
 We know that for any $x_s^n\in \mathcal{C}_s\backslash \mathcal{D}_s$, 
 \begin{IEEEeqnarray}{rCl}
 	\w(x_s^n)\geq \frac{2\sqrt{n\Delta}}{\D_1-\D_2}\Phi^{-1}\left(\frac{1+\delta}{2}+ \gamma+\frac{E}{\sqrt{n}}\right),\label{step10}
 \end{IEEEeqnarray}
 Combining \eqref{step42} with \eqref{step10} and considering \eqref{step13}, we can further lower bound the total variation distance as follows:
 \begin{IEEEeqnarray}{rCl}
 	d_{\TV}\left(\overline{Q}_{1}^n,\overline{Q}_{2}^n\right)&\geq& \delta+\frac{2E}{\sqrt{n}}+2\gamma-\frac{\omega_{\Lo,n}\omega_{\Hi,n}(|\Gamma_1|+|\Gamma_2|)(\D_1-\D_2)}{4n\sqrt{2n\pi\Delta^3}} +O\left(\frac{1}{\sqrt{n}}\right)\\*
 	&\geq& \delta+2\gamma,
 	\label{step12}
 \end{IEEEeqnarray}
 where the last inequality follows from \eqref{step13}.
Uniting \eqref{step11} and \eqref{step12}   yields 
 \begin{IEEEeqnarray}{rCl}
 	\delta &\geq& \delta+2\gamma-\xi_1-\xi_2.
 \end{IEEEeqnarray}
 If we choose $\gamma=\frac{1}{\sqrt{n}}$, we obtain
 \begin{align}
 \xi_1 +\xi_2\geq \frac{2}{\sqrt{n}}.\label{step400}
 \end{align}
 In summary, from the assumption $\xi_1\geq \xi_2$, we obtain a set of codewords with size \begin{align}
 |\mathcal{D}_1| \geq \frac{|\mathcal{M}|\cdot |\mathcal{K}|}{\sqrt{n}},\label{ineq100}
 \end{align} and the Hamming weight of these codewords is given by 
 \begin{align}
 \w(x_1^n)&\leq \frac{2\sqrt{n\Delta}}{\D_1-\D_2}\Phi^{-1}\left(\frac{1+\delta}{2}+\frac{E}{\sqrt{n}}+\gamma\right)\\
 &=\frac{2\sqrt{n\Delta}}{\D_1-\D_2}\Phi^{-1}\left(\frac{1+\delta}{2}\right)+O\left(\frac{1}{\sqrt{n}}\right)\\
 &\triangleq \psi(n,\delta)+O\left(\frac{1}{\sqrt{n}}\right). \label{eqn:def_psi}
 \end{align}
Since we assumed that $\xi_1\geq \xi_2$, inequality \eqref{ineq100} can be equivalently written as:
\begin{IEEEeqnarray}{rCl}
	\max_{s}	|\mathcal{D}_s| \geq \frac{|\mathcal{M}|\cdot |\mathcal{K}|}{\sqrt{n}}\label{ineq101}
\end{IEEEeqnarray}
For each $k\in\mathcal{K}$, we denote the sub-codebook of all codewords with the same key $k$ by $\mathcal{C}_s^k$. From the pigeonhole principle, there exists a sub-codebook $\mathcal{C}_s^k$ such that $\max_s |\mathcal{D}_s\cap \mathcal{C}_s^k|\geq \frac{|\mathcal{M}|}{\sqrt{n}}$. Define:
\begin{IEEEeqnarray}{rCl}
	\mathcal{D}_{s,i}^k\triangleq\left\{ x_s^n\in\mathcal{D}_s\cap \mathcal{C}_s^k\colon \w(x_s^n)=i \right\}.\label{def500}
\end{IEEEeqnarray}
The codewords in the sub-codebook $|\mathcal{D}_{s,i}^k|$ have the same type which we denote by $\pi_s^i$. This sub-codebook is  with maximum probability of error not larger than $\epsilon$. Using Fano's inequality, we can write the following set of inequalities for $s\in\{1,2\}$:
\begin{IEEEeqnarray}{rCl}
	(1-\epsilon)\log |\mathcal{D}_{s,i}^k\cap \mathcal{C}_s^k|-1&\leq& I(X_s^n;Y^n)\\
	&\leq & \sum_{t=1}^nI(X_{s,t};Y_t)\\
	&\leq &nI(\pi_s^i,W_s)\\
	&\leq & iD(\tilde{Q}_{s}\|Q_0).\label{ineq401}
\end{IEEEeqnarray}
Next, we continue to upper bound the size of the message set as follows:
\begin{IEEEeqnarray}{rCl}
	\log \frac{|\mathcal{M}|}{\sqrt{n}}&\le& \max_s\;\log |\mathcal{D}_s\cap \mathcal{C}_s^k| \label{eqn:conv1}\\
	&=& \max_s\;\log\left(\sum_{i=0}^{\psi(n,\delta)}|\mathcal{D}_{s,i}^k\cap \mathcal{C}_s^k|\right)\label{eqn:conv2}\\
	&\le& \max_s\;\log\left( \sum_{i=0}^{\psi(n,\delta)} \exp(iD(\tilde{Q}_{s}\|Q_0))\right)\label{eqn:conv3}\\
	&\leq & \max_s\;\log\left( \psi(n,\delta) \exp\big(\psi(n,\delta)\cdot D(\tilde{Q}_{s}\|Q_0)\big)\right)\label{eqn:conv4}\\
	&\leq & \psi(n,\delta)\cdot  \max_s\; D(\tilde{Q}_{s}\|Q_0)+O(\log n )\label{eqn:conv5}\\
	&\leq & \frac{2\sqrt{n\Delta}}{\D_1-\D_2}\Phi^{-1}\left(\frac{1+\delta}{2}\right)\cdot  \max_s\, D(\tilde{Q}_{s}\|Q_0)+O(\log n ),\label{ineq700}
\end{IEEEeqnarray}
where \eqref{eqn:conv1} follows from \eqref{ineq101}, \eqref{eqn:conv2} follows from the definition of $\mathcal{D}_{s,i}^k$ in~\eqref{def500},  \eqref{eqn:conv3} follows from inequality \eqref{ineq401} and \eqref{eqn:conv5} follows from the definition of $\psi(n,\delta)$ in \eqref{eqn:def_psi}.
Dividing both sides of \eqref{ineq700} by $\sqrt{n}$ and taking limits in $n$ completes the proof.

\section{Proof of Corollary~\ref{Gaus-cor}}\label{Gaus-proof}
For the proposed Gaussian setup, the lower bound to $L_{\TV}^*(\epsilon,\delta)$ in Theorem~\ref{PPM-thm} can be generalized for a continuous alphabet. Several steps in the proof of upper bound to $L_{\TV}^*(0,\delta)$ in Appendix~\ref{lem2-proof} remain valid for the Gaussian distribution. The only step that should be refined is bounding the sum of third absolute moments in \eqref{t-bound}. The reason is as follows. In this step of the proof for the discrete memoryless channel, we know that $\eta=\min_y Q_0(y)$ is positive. However, for the Gaussian case, this statement does not hold and $\eta$ is zero. So, in the following, we first bound the sum of third absolute moments for the Gaussian setup. The evaluation of the relevant information quantities involving Gaussian distributions will be presented later.

\underline{\textit{Bounding the sum of third absolute moments}}:

Consider the following sum of third absolute moments:
\begin{IEEEeqnarray}{rCl}
	\sum_{i=1}^n \mathbb{E}_{W_1(\cdot|x_i^{*})}\left[ \big|\mathcal{T}(Y_i) -\mathbb{E}_{W_1(\cdot|x_i^{*})} [ \mathcal{T}(Y_i) ] \big|^3 \right] &=& \sum_{i:x_i^*=0} \mathbb{E}_{W_1(\cdot|x_i^{*})}\left[ \big|\mathcal{T}(Y_i) -\mathbb{E}_{W_1(\cdot|x_i^{*})} [ \mathcal{T}(Y_i) ] \big|^3 \right] \nonumber\\&&\hspace{1cm}+\sum_{i:x_i^*=1} \mathbb{E}_{W_1(\cdot|x_i^{*})}\left[ \big|\mathcal{T}(Y_i) -\mathbb{E}_{W_1(\cdot|x_i^{*})} [ \mathcal{T}(Y_i) ] \big|^3 \right]  \\
	&=&n(1-\mu_{1,n}) \mathbb{E}_{Q_0}\left[ \big|\mathcal{T}(Y) -\mathbb{E}_{Q_0} [ \mathcal{T}(Y) ] \big|^3 \right] \nonumber\\*&&\hspace{1cm}+n\mu_{1,n} \mathbb{E}_{\tilde{Q}_1}\left[ \big|\mathcal{T}(Y) -\mathbb{E}_{\tilde{Q}_1} [ \mathcal{T}(Y) ] \big|^3 \right]. \label{third-abs1}
\end{IEEEeqnarray}
We study each expectation term of \eqref{third-abs1}, separately, and show that each expectation term is a constant and does not depend on $n$. First, consider the following term:
\begin{IEEEeqnarray}{rCl}
	\mathbb{E}_{Q_0}\left[ \big|\mathcal{T}(Y) -\mathbb{E}_{Q_0} [ \mathcal{T}(Y) ] \big|^3 \right] &=& \mathbb{E}_{Q_0}\left[ \big|\mathcal{T}(Y)  \big|^3 \right] \label{step700}\\
	&=& \mathbb{E}_{Q_0}\left[ \Bigg| \frac{\varphi(\tilde{Q}_1(Y)-Q_0(Y))-(1-\varphi)(\tilde{Q}_2(Y)-Q_0(Y))}{Q_0(Y)}  \Bigg|^3 \right]\\
	&=& \frac{1}{\sigma\sqrt{2\pi}}\int_{-\infty}^{\infty} \Bigg|\varphi\exp\left(\frac{2y-1}{2\sigma^2}\right)-(1-\varphi)\exp\left(-\frac{2y+1}{2\sigma^2}\right)+1\Bigg|^3\cdot \exp\left(-\frac{y^2}{2\sigma^2}\right)\, \mathrm{d}y\nonumber\\\\
	&\leq & \frac{1}{\sigma\sqrt{2\pi}}\int_{-\infty}^{\infty} \Bigg(\varphi\exp\left(\frac{2y-1}{2\sigma^2}\right)+(1-\varphi)\exp\left(-\frac{2y+1}{2\sigma^2}\right)+1\Bigg)^3\cdot \exp\left(-\frac{y^2}{2\sigma^2}\right)\, \mathrm{d}y\nonumber\\
\label{step701}\\
	&=& \kappa_1,\label{third-abs2}
	%	&=& \frac{1}{\sqrt{2\pi}\sigma_0\sigma_1}\int_{-\infty}^{\infty} \exp \Bigg(-\frac{3\sigma_0^2-2\sigma_1^2}{2\sigma_0^2\sigma_1^2}y^2+\frac{3}{\sigma_1^2}y-\frac{9\sigma_0^2}{2\sigma_1^2\left(3\sigma_0^2-2\sigma_1^2\right)}\nonumber\\&&\hspace{3.5cm}+\frac{9\sigma_0^2}{2\sigma_1^2\left(3\sigma_0^2-2\sigma_1^2\right)}-\frac{3}{2\sigma_1^2}\Bigg)dy+O(1)\\
	%	&=& \frac{1}{\sqrt{2\pi}\sigma_0\sigma_1}\exp\left(\frac{9\sigma_0^2}{2\sigma_1^2\left(3\sigma_0^2-2\sigma_1^2\right)}-\frac{3}{2\sigma_1^2}\right)\times\nonumber\\&&\hspace{3cm}
	%	\int_{-\infty}^{\infty} \exp\left(- \left(\sqrt{\frac{3\sigma_0^2-2\sigma_1^2}{2\sigma_0^2\sigma_1^2}}y-\frac{3}{\sqrt{\frac{2\sigma_1^2\left(3\sigma_0^2-2\sigma_1^2\right)}{\sigma_0^2}}}\right)^2\right)dy+O(1)\nonumber\\\\
	%	&=&\frac{1}{\sqrt{2\pi}\sigma_0\sigma_1}\exp\left(\frac{9\sigma_0^2}{2\sigma_1^2\left(3\sigma_0^2-2\sigma_1^2\right)}-\frac{3}{2\sigma_1^2}\right)\sqrt{\frac{2\sigma_0^2\sigma_1^2}{3\sigma_0^2-2\sigma_1^2}}\int_{-\infty}^{\infty}e^{-u^2}du+O(1)\\
	%	&=& \frac{1}{\sigma_0\sigma_1}\exp\left(\frac{9\sigma_0^2}{2\sigma_1^2\left(3\sigma_0^2-2\sigma_1^2\right)}-\frac{3}{2\sigma_1^2}\right)\sqrt{\frac{\sigma_0^2\sigma_1^2}{3\sigma_0^2-2\sigma_1^2}}+O(1)
\end{IEEEeqnarray}
where $\kappa_1$ is a positive constant. Here, \eqref{step700} follows because $\mathbb{E}_{Q_0}\left[\mathcal{T}(Y)\right]=0$, \eqref{step701} follows from the triangle inequality, and~\eqref{third-abs2} follows because calculation of all involved integrals yields a finite value.

Next, we analyze the second term of \eqref{third-abs1}:
\begin{IEEEeqnarray}{rCl}
	\mathbb{E}_{\tilde{Q}_1}\left[ \big|\mathcal{T}(Y) -\mathbb{E}_{\tilde{Q}_1} [ \mathcal{T}(Y) ] \big|^3 \right] &\leq & \mathbb{E}_{\tilde{Q}_1}\left[ \big|\mathcal{T}(Y)\big|^3\right]+3\mathbb{E}_{\tilde{Q}_1}\left[ \big|\mathcal{T}(Y)\big|^2\right]\cdot \big| \mathbb{E}_{\tilde{Q}_1} [ \mathcal{T}(Y) ]\big|\\
	&&\hspace{1cm}+3\mathbb{E}_{\tilde{Q}_1}\left[ \big|\mathcal{T}(Y)\big|\right]\cdot \big| \mathbb{E}_{\tilde{Q}_1} [ \mathcal{T}(Y) ]\big|^2+\big| \mathbb{E}_{\tilde{Q}_1} [ \mathcal{T}(Y) ]\big|^3\\
	&\leq & \mathbb{E}_{\tilde{Q}_1}\left[ \big|\mathcal{T}(Y)\big|^3\right]+3\mathbb{E}_{\tilde{Q}_1}\left[ \big|\mathcal{T}(Y)\big|^3\right]^{\frac{2}{3}}\cdot \big| \mathbb{E}_{\tilde{Q}_1} [ \mathcal{T}(Y) ]\big|\\
	&&\hspace{1cm}+3\mathbb{E}_{\tilde{Q}_1}\left[ \big|\mathcal{T}(Y)\big|^3\right]^{\frac{1}{3}}\cdot \big| \mathbb{E}_{\tilde{Q}_1} [ \mathcal{T}(Y) ]\big|^2+\big| \mathbb{E}_{\tilde{Q}_1} [ \mathcal{T}(Y) ]\big|^3,\label{step500}
\end{IEEEeqnarray}
where inequality \eqref{step500} follows because $\mathbb{E}\left[|\mathcal{T}(Y)|^2\right]\leq \mathbb{E}\left[|\mathcal{T}(Y)|^3\right]^{\frac{2}{3}}$ and $\mathbb{E}\left[|\mathcal{T}(Y)|\right]\leq \mathbb{E}\left[|\mathcal{T}(Y)|^3\right]^{\frac{1}{3}}$. Thus, it remains to prove that $\mathbb{E}_{\tilde{Q}_1}\left[ \big|\mathcal{T}(Y)\big|^3\right]$ and $\big| \mathbb{E}_{\tilde{Q}_1} [ \mathcal{T}(Y) ]\big|$ are bounded. We first write $\mathbb{E}_{\tilde{Q}_1}\left[\mathcal{T}(Y)\right]$ as follows: 
\begin{IEEEeqnarray}{rCl}
	\mathbb{E}_{\tilde{Q}_1}\left[\mathcal{T}(Y)\right]&=&\mathbb{E}_{\tilde{Q}_1}\left[\frac{\varphi(\tilde{Q}_1(Y)-Q_0(Y))-(1-\varphi)(\tilde{Q}_2(Y)-Q_0(Y))}{Q_0(Y)}\right]\\
	&=&\mathbb{E}_{Q_0}\left[\frac{\varphi(\tilde{Q}_1(Y)-Q_0(Y))-(1-\varphi)(\tilde{Q}_2(Y)-Q_0(Y))}{Q_0(Y)}.\frac{\tilde{Q}_1(Y)}{Q_0(Y)}\right]\label{third-abs3-1}\\
	&\le& \sqrt{\mathbb{E}_{Q_0}\left[\left(\frac{\varphi(\tilde{Q}_1(Y)-Q_0(Y))-(1-\varphi)(\tilde{Q}_2(Y)-Q_0(Y))}{Q_0(Y)}\right)^2\right]\mathbb{E}_{Q_0}\left[\left(\frac{\tilde{Q}_1(Y)}{Q_0(Y)}\right)^2\right]}\label{third-abs3-2}\\
	&=& \sqrt{(\varphi^2 \chi_2(\tilde{Q}_1\|Q_0)+(1-\varphi)^2\chi_2(\tilde{Q}_2\|Q_0)-2\varphi(1-\varphi)\rho(\tilde{Q}_1,\tilde{Q}_2,Q_0))(\chi_2(\tilde{Q}_1\|Q_0)+1)}\label{third-abs3-3}\\
&:=&\kappa_2, \label{third-abs3}
\end{IEEEeqnarray}
where  \eqref{third-abs3-1} follows by a change of measure argument, \eqref{third-abs3-2} follows from the Cauchy-Schwartz inequality, \eqref{third-abs3-3} follows from the definitions of $\rho$ and $\chi_2$ and finally, the the right-hand-side of~\eqref{third-abs3-3} is evaluated in the sequel and is finite. Thus,  we denote it by $\kappa_2$.

Now, it remains to show that $\mathbb{E}_{\tilde{Q}_1}\left[ \big|\mathcal{T}(Y)\big|^3\right]$ is also bounded.
\begin{IEEEeqnarray}{rCl}
	\mathbb{E}_{\tilde{Q}_1}\left[ \big|\mathcal{T}(Y)  \big|^3 \right] 
	&=& \mathbb{E}_{\tilde{Q}_1}\left[ \Bigg| \frac{\varphi(\tilde{Q}_1(Y)-Q_0(Y))-(1-\varphi)(\tilde{Q}_2(Y)-Q_0(Y))}{Q_0(Y)}  \Bigg|^3 \right]\\
	&=& \frac{1}{\sigma\sqrt{2\pi}}\int_{-\infty}^{\infty} \Bigg|\varphi\exp\left(\frac{2y-1}{2\sigma^2}\right)-(1-\varphi)\exp\left(-\frac{2y+1}{2\sigma^2}\right)+1\Bigg|^3\cdot \exp\left(-\frac{(y-1)^2}{2\sigma^2}\right)\, \mathrm{d}y\\
	&\leq & \frac{1}{\sigma\sqrt{2\pi}}\int_{-\infty}^{\infty} \Bigg(\varphi\exp\left(\frac{2y-1}{2\sigma^2}\right)+(1-\varphi)\exp\left(-\frac{2y+1}{2\sigma^2}\right)+1\Bigg)^3\cdot \exp\left(-\frac{(y-1)^2}{2\sigma^2}\right)\, \mathrm{d}y\label{step7001}\\
%	&=& \frac{1}{\sqrt{2\pi}}\int_{-\infty}^{\infty} \Bigg|\frac{1}{\sigma}\exp\left(-\frac{(y-1)^2}{2\sigma^2}\right)-\frac{1}{\sigma}\exp\left(-\frac{(y+1)^2}{2\sigma^2}\right)\Bigg|^3\cdot \sigma^3\exp\left(\frac{3y^2}{2\sigma^2}\right)\cdot \frac{1}{\sigma}\exp\left(-\frac{(y-1)^2}{2\sigma^2}\right)  dy\nonumber\\\\
%	&\leq& \frac{1}{\sqrt{2\pi}}\int_{-\infty}^{\infty} \frac{1}{\sigma^3}\exp\left(-\frac{3(y-1)^2}{2\sigma^2}\right)\cdot \sigma^3\exp\left(\frac{3y^2}{2\sigma^2}\right)\cdot \frac{1}{\sigma}\exp\left(-\frac{(y-1)^2}{2\sigma^2}\right)  dy\nonumber\\
%	&&\hspace{1cm}+\frac{1}{\sqrt{2\pi}}\int_{-\infty}^{\infty} \frac{1}{\sigma^3}\exp\left(-\frac{3(y+1)^2}{2\sigma^2}\right)\cdot \sigma^3\exp\left(\frac{3y^2}{2\sigma^2}\right)\cdot \frac{1}{\sigma}\exp\left(-\frac{(y-1)^2}{2\sigma^2}\right)  dy\nonumber\\
%	&&\hspace{1cm}+\frac{1}{\sqrt{2\pi}}\int_{-\infty}^{\infty} \frac{3}{\sigma^3}\exp\left(-\frac{2(y+1)^2+(y-1)^2}{2\sigma^2}\right)\cdot \sigma^3\exp\left(\frac{3y^2}{2\sigma^2}\right)\cdot \frac{1}{\sigma}\exp\left(-\frac{(y-1)^2}{2\sigma^2}\right)  dy\nonumber\\
%	&&\hspace{1cm}+\frac{1}{\sqrt{2\pi}}\int_{-\infty}^{\infty} \frac{3}{\sigma^3}\exp\left(-\frac{(y+1)^2+2(y-1)^2}{2\sigma^2}\right)\cdot \sigma^3\exp\left(\frac{3y^2}{2\sigma^2}\right)\cdot \frac{1}{\sigma}\exp\left(-\frac{(y-1)^2}{2\sigma^2}\right)  dy\\
	&=& \kappa_3,\label{third-abs4}
\end{IEEEeqnarray}
where $\kappa_3$  is a positive constant. The last equality follows because calculation of all involved integrals yields a finite value. Combining \eqref{third-abs1}, \eqref{third-abs2}, \eqref{step500}, \eqref{third-abs3} and \eqref{third-abs4}, we get:
\begin{IEEEeqnarray}{rCl}
	&&\sum_{i=1}^n \mathbb{E}_{W_1(\cdot|x_i^{*})}\left[ \big|\mathcal{T}(Y_i) -\mathbb{E}_{W_1(\cdot|x_i^{*})} [ \mathcal{T}(Y_i) ] \big|^3 \right]\\
	&&\hspace{1cm}=n(1-\mu_{1,n}) \mathbb{E}_{Q_0}\left[ \big|\mathcal{T}(Y) -\mathbb{E}_{Q_0} [ \mathcal{T}(Y) ] \big|^3 \right] +n\mu_{1,n} \mathbb{E}_{\tilde{Q}_1}\left[ \big|\mathcal{T}(Y) -\mathbb{E}_{\tilde{Q}_1} [ \mathcal{T}(Y) ] \big|^3 \right]\nonumber\\
	&&\hspace{1cm}\leq n(1-\mu_{1,n})\kappa_1+n\mu_{1,n}(\kappa_3+3\kappa_3^{2/3}\kappa_2+3\kappa_3^{1/3}\kappa_2^2+\kappa_2^2)\\
	&&\hspace{1cm}\leq n(\kappa_1+\kappa_3+3\kappa_3^{2/3}\kappa_2+3\kappa_3^{1/3}\kappa_2^2+\kappa_2^2)\\
	&&\hspace{1cm}\triangleq n\mathbf{T}.
	\end{IEEEeqnarray}
Thus, the above replaces \eqref{third-abs5} and the rest of the proof in Appendix~\ref{lem2-proof} holds.

\underline{\textit{Evaluation of the relevant information quantities}}:

We now evaluate the information quantities involved in Theorem~\ref{opt-thm}.
First, consider the KL-divergence term $D(\tilde{Q}_s\|Q_0)$ as follows:
\begin{IEEEeqnarray}{rCl}
	D(\tilde{Q}_s\|Q_0) &=& -h_{\tilde{Q}_s}(Y)+\mathbb{E}_{\tilde{Q}_s}\left[\log\frac{1}{Q_0(Y)}\right]\\
	&=&-\frac{1}{2}\log\left( 2\pi e \sigma^2 \right)+\mathbb{E}_{\tilde{Q}_s}\left[\log\frac{1}{Q_0(Y)}\right]\\
		&=&-\frac{1}{2}\log\left( 2\pi e \sigma^2 \right)+\mathbb{E}_{\tilde{Q}_s}\left[\frac{1}{2}\log (2\pi \sigma^2)+\frac{Y^2}{2\sigma^2}\right]\\
		&=&\frac{1}{2\sigma^2}\label{DG-ineq1}   
	\end{IEEEeqnarray}
From \eqref{DG-ineq1}, the assumption $D(\tilde{Q}_1\|Q_0)=D(\tilde{Q}_2\|Q_0)$ is satisfied.
Next, consider the chi-squared distance $\chi_2(\tilde{Q}_1\|Q_0)$ as follows:
\begin{IEEEeqnarray}{rCl}
	\chi_2 \big(\tilde{Q}_1\|Q_0\big) &=& \mathbb{E}_{Q_0}\left[\left(\frac{\tilde{Q}_1(Y)}{Q_0(Y)}\right)^2\right]-1\\
	&=&\mathbb{E}_{Q_0}\left[ \exp \left(-\frac{2(Y-1)^2}{2\sigma^2}+\frac{2Y^2}{2\sigma^2}\right) \right]-1\\
	&=&\frac{1}{\sqrt{2\pi}\sigma}\;\int_{-\infty}^{\infty}\exp\left(-\frac{2(y-1)^2}{2\sigma^2}+\frac{y^2}{2\sigma^2}\right)\, \mathrm{d}y-1\\
	&=&\frac{1}{\sqrt{2\pi}\sigma}\;\int_{-\infty}^{\infty}\exp\left(-\frac{1}{2\sigma^2}y^2+\frac{2}{\sigma^2}y-\frac{1}{\sigma^2}\right)\, \mathrm{d}y-1 \\
	&=&\frac{1}{\sqrt{2\pi}\sigma}\;\exp\left(\frac{1}{\sigma^2}\right)\int_{-\infty}^{\infty} \exp \left( -\left(\sqrt{\frac{1}{2\sigma^2}}y-\sqrt{\frac{2}{\sigma^2}}\right)^2 \right)\, \mathrm{d}y-1\\
	&=&\frac{1}{\sqrt{2\pi}\sigma}\;\exp\left(\frac{1}{\sigma^2}\right)\frac{1}{\sqrt{\frac{1}{2\sigma^2}}}\int_{-\infty}^{\infty} \exp \left( -u^2 \right)\, \mathrm{d}u-1\\
	&=&\exp\left(\frac{1}{\sigma^2}\right)-1.\label{chi-term}
	\end{IEEEeqnarray}
Similary, it can be shown that $\chi_2(\tilde{Q}_2\|Q_0)=\chi_2 (\tilde{Q}_1\|Q_0)$.
Finally, we evaluate the term $\rho(\tilde{Q}_1,\tilde{Q}_2,Q_0)$ as the following:
\begin{IEEEeqnarray}{rCl}
	\rho(\tilde{Q}_1,\tilde{Q}_2,Q_0) &=& \mathbb{E}_{Q_0}\left[\frac{\tilde{Q}_1(Y)}{Q_0(Y)}\cdot \frac{\tilde{Q}_2(Y)}{Q_0(Y)}\right]-1\\
	&=& \frac{1}{\sqrt{2\pi}\sigma}\int_{-\infty}^{\infty}\exp\left(-\frac{(y-1)^2}{2\sigma^2}-\frac{(y+1)^2}{2\sigma^2}+\frac{y^2}{2\sigma^2}\right)\, \mathrm{d}y-1\\
	&=&\frac{1}{\sqrt{2\pi}\sigma}\int_{-\infty}^{\infty}\exp\left(-\frac{1}{2\sigma^2}y^2-\frac{1}{\sigma^2}\right)\, \mathrm{d}y-1\\
	&=&\frac{1}{\sqrt{2\pi}\sigma}\exp\left(-\frac{1}{\sigma^2}\right)\int_{-\infty}^{\infty}\exp\left(-\frac{1}{2\sigma^2}y^2\right)\, \mathrm{d}y-1\\
	&=&\exp\left(-\frac{1}{\sigma^2}\right)-1.\label{rho-term}
	\end{IEEEeqnarray}
Moreover, consider the fact that 
\begin{IEEEeqnarray}{rCl}
	\Delta = \chi_2(\tilde{Q}_1\|Q_0)+\chi_2(\tilde{Q}_2\|Q_0)-2\rho(\tilde{Q}_1,\tilde{Q}_2,Q_0).\label{DeltaG}
	\end{IEEEeqnarray}
The proof is concluded by combining \eqref{DG-ineq1} with \eqref{DeltaG} and \eqref{chi-term} with \eqref{rho-term}.

\subsubsection*{Acknowledgements} The authors acknowledge the constructive  comments from the reviewers and the associate editor Prof.\ Matthieu Bloch. 
\bibliographystyle{IEEEtran}
\bibliography{references}

% that's all folks
\end{document}